\documentclass[aps,prl,reprint,amsmath,amssymb,superscriptaddress, 10pt]{revtex4-2}

\usepackage[usenames,dvipsnames]{xcolor}
\usepackage{amsmath,amssymb}
\usepackage{graphicx}
\usepackage{hyperref}
\usepackage{braket}
\usepackage[normalem]{ulem}
\usepackage{enumerate}

\newcommand{\be}{\begin{equation}}
\newcommand{\ee}{\end{equation}}
\def\bes{\begin{subequations}}
\def\esu{\end{subequations}}

\newcommand{\sm}{s_{\text{max}}}

\newcommand{\dd}{{\rm d}}

\newcommand{\cdr}{\text{dr}}
\newcommand{\p}{\partial}

\DeclareMathOperator{\sign}{sign}

\begin{document}

\newcommand{\titleinfo}{
Exact large-scale fluctuations of the phase field in the sine-Gordon model}

\title{\titleinfo}

\author{Giuseppe Del Vecchio Del Vecchio}
\affiliation{Department of Mathematics, King's College London, Strand, London WC2R 2LS, United Kingdom}
\author{M\'arton Kormos}
\affiliation{Department of Theoretical Physics, Institute of Physics, Budapest University of Technology and Economics, M\H uegyetem rkp. 3., H-1111 Budapest, Hungary}
\affiliation{MTA-BME Quantum Dynamics and Correlations Research Group, Budapest University of Technology and Economics, M\H uegyetem rkp. 3., H-1111 Budapest, Hungary}
\author{Benjamin Doyon}
\affiliation{Department of Mathematics, King's College London, Strand, London WC2R 2LS, United Kingdom}
\author{Alvise Bastianello}
\affiliation{Technical University of Munich, TUM School of Natural Sciences, Physics Department, 85748 Garching, Germany}
\affiliation{Munich Center for Quantum Science and Technology (MCQST), Schellingstr. 4, 80799 M{\"u}nchen, Germany}

\begin{abstract}
We present the first exact theory and analytical formulas for the large-scale phase fluctuations in the sine-Gordon model, valid in all regimes of the field theory, for arbitrary temperatures and interaction strengths. Our result is based on the Ballistic Fluctuation Theory combined with Generalized Hydrodynamics, and can be seen as an exact ``dressing" of the phenomenological soliton-gas
picture first introduced by Sachdev and Young [S. Sachdev and A. P. Young, PRL 78, 2220 (1997)], to the modes of Generalised Hydrodynamics. The resulting physics of phase fluctuations in the sine-Gordon model is qualitatively different, as the stable quasi-particles of integrability give coherent ballistic  propagation instead of diffusive spreading. We provide extensive numerical checks of our analytical predictions within the classical regime of the field theory by using Monte Carlo methods. We discuss how our results are of ready applicability to experiments on tunnel-coupled quasicondensates.
\end{abstract}

\maketitle

\paragraph{\textbf{Introduction ---}}

Understanding correlations and fluctuations in quantum and classical interacting many-body systems is a crucial problem of theoretical physics. 
Needless to say, in strongly interacting models this is a daunting task, too complicated to be carried out in the most general setting. However, at large scales {\em universality} emerges \cite{Fischer1998,spohn2012large,giamarchi2003quantum,francesco2012conformal,Bernard_2016}: microscopic details are unimportant and information is carried only by slowly decaying modes, coupled to the local conservation laws of the underlying Hamiltonian. This is the hallmark of hydrodynamics.

The most significant correlations of a given observable are due to its coupling with hydrodynamic modes (sound, heat, etc.) associated with conservation laws \cite{DeNardis_2022}.
Along the velocities of such modes, power-law behavior is observed instead of exponential decay. But some observables do not couple to hydrodynamic modes, such as those sensitive to topological excitations, because of the intrinsically non-local nature of the latter. For instance, correlation functions of order parameters often show exponential decay throughout space-time. Is there a general theory for understanding such behavior? How do hydrodynamic modes interact with order parameters and what information can be extracted from their correlation functions?
Even with the mathematical tools of integrability, computing correlations of order parameters from a microscopic analysis is challenging in non-interacting cases \cite{Pfeuty1970,PERK19801,PERK19803,PERK19841,Perk2009,Doyon_2005,DoyonFF_2007,Calabrese2011,Calabrese_2012,Chen_2014,2018Groha} and unpractical in the presence of interactions \cite{Altshuler2006,LeClair1999,Pozsgay2018,Collura_2017}. This calls for a more universal hydrodynamic approach.
The relation between order parameters and hydrodynamic modes was recently addressed \cite{delvecchio2022,delvecchio2023} in the XX quantum chain using free-fermionic techniques; and a general, but phenomenological, picture for the influence of topological excitations on correlations was proposed by Sachdev and Young (SY) \cite{Sachdev1997}. However, to our knowledge, there are no results beyond free excitations or extremely dilute gases.

A paradigmatic model where these questions are of central relevance is the sine-Gordon model 
\be\label{eq_H}
H=\int \dd x \Big[\frac{c^2 g^2 }{2}\Pi^2+\frac{1}{2g^2}(\partial_x\phi)^2-\frac{c^2 m^2}{g^2}\cos(\phi) \Big]\, ,
\ee
that manifests itself in the most diverse contexts \cite{giamarchi2003quantum,Lomdahl1985,Davidson1985,Roy2019,Roy2021,Zvyagin2021}. Above, the field $\Pi(x)$ is conjugated to the phase $\phi(x)$, $g$ tunes the interactions, $c$ is the ``velocity of light", and $m$ is a mass scale. In the following, we measure lengths in units of $[mc]^{-1}$.
The low-energy sector of many systems is well described by the sine-Gordon model, as perturbations can induce Berezinskii-Kosterlitz-Thouless transitions in the ubiquitous Luttinger Liquid \cite{giamarchi2003quantum} ($m=0$) field theory, but it has many applications in high-energy physics as well \cite{cuevassinegordon}.
Notably, the sine-Gordon model is a paradigmatic example of {\em integrable field theory} \cite{Smirnov1992}, hence it is amenable to nonperturbative analysis, and shows peculiar thermalization \cite{Calabrese_2016} and transport \cite{Bertini2021,specialissueGHD} properties. In this model, fluctuations create topological excitations of the phase field $\phi$: phase-slips of $2\pi$ interpolating between the degenerate ground states $\phi\in 2\pi \mathbb{Z}$. The non-locality of these excitations with respect to the model's order parameter $\phi$ places it in the general category we outlined, leading us to the central question of this Letter: can we build a general framework able to capture the large scale fluctuations of the phase field?

In addition to being a long-standing unsolved problem of mathematical physics, this question is of central experimental relevance. Correlation functions of vertex operators $e^{i\lambda \phi}$ capture response functions at low energy in certain materials \cite{Zvyagin2004,Umegaki2009,Essler1998,Essler1999} and in multi-species cold atomic gases \cite{giamarchi2003quantum}, and order-parameter correlations functions in spin chains 
\cite{Essler2005,Zvyagin2021}. Moreover, recent experimental advances probe fluctuations of phase differences $\Delta \phi(t,x) = \phi(t,x)-\phi(0,0)$ themselves.
This is possible in tabletop quantum simulators of the sine-Gordon model realized by tunnel-coupled condensates \cite{Gritsev2007,Gritsev2007a}. Matter-wave interferometry gives access to projective measurements of phase differences in both equilibrium \cite{Schweigler2017,Zache2020} and nonequilibrium settings \cite{Pigneur2018}: any analytical insight would be of utmost interest not only from a theoretical point of view, but also in very concrete experiments.
In this Letter, we solve this problem.

We show that the probability distribution of the phase differences, $P\left[\frac{\Delta\phi(t,x)}{2\pi}=\delta\right]$, obeys a large deviation principle $P[\delta]\asymp e^{-\ell I_{\alpha}(\delta)}$,
where $\ell=\sqrt{x^2+c^2t^2}$ and the large-deviation function (LDF) $I_\alpha$ is fully determined by hydrodynamic modes and depends only on the ``ray" $x/(ct)=\tan\alpha$. 
Our theory fully corrects the SY picture  \cite{DamleSachdev2005, Kormos2016, Moca2019, Bertini2019, PhysRevB.106.205151} by accounting for quantum distributions and coherence, and gives {\em qualitatively different, analytical and exact} results that are valid in the scaling limit of large $\ell$, and are applicable at arbitrary interactions, finite temperature and even on Generalized Gibbs Ensembles \cite{Calabrese_2016}.

\paragraph{\textbf{The sine-Gordon field theory ---}}
The sine-Gordon model is integrable both in the classical \cite{Faddeev:1987ph} and quantum \cite{ZAMOLODCHIKOV1979253} regimes. 
The fundamental excitations are relativistic topological solitons interpolating between the valleys of the periodic potential and parameterized by their rapidity $\theta$. Hereafter, we refer to ``kinks" (``antikinks") when they cause a positive (negative) phase slip $+2\pi$ from left to right. A kink-antikink pair can form a stable bound state called breather \cite{Faddeev:1987ph,ZAMOLODCHIKOV1979253}.
In the quantum case, breathers are absent in the repulsive regime $cg^2>4\pi$; for smaller interactions, breathers appear in the spectrum, with masses $m_n=2M \sin(\frac{\pi}{2}\xi n)$ for integers $n< \xi^{-1}=\frac{8\pi}{cg^2}-1$.
The classical soliton mass $M=8m/(cg^2)$ is renormalized upon quantization \cite{ZAMOLODCHIKOV1995}. At weak interactions $g\to 0$, the breather masses collapse to a continuum, and classical physics is recovered \cite{koch2023exact}.

\begin{figure}[t!]
\centering
	\includegraphics[width=0.99\columnwidth]{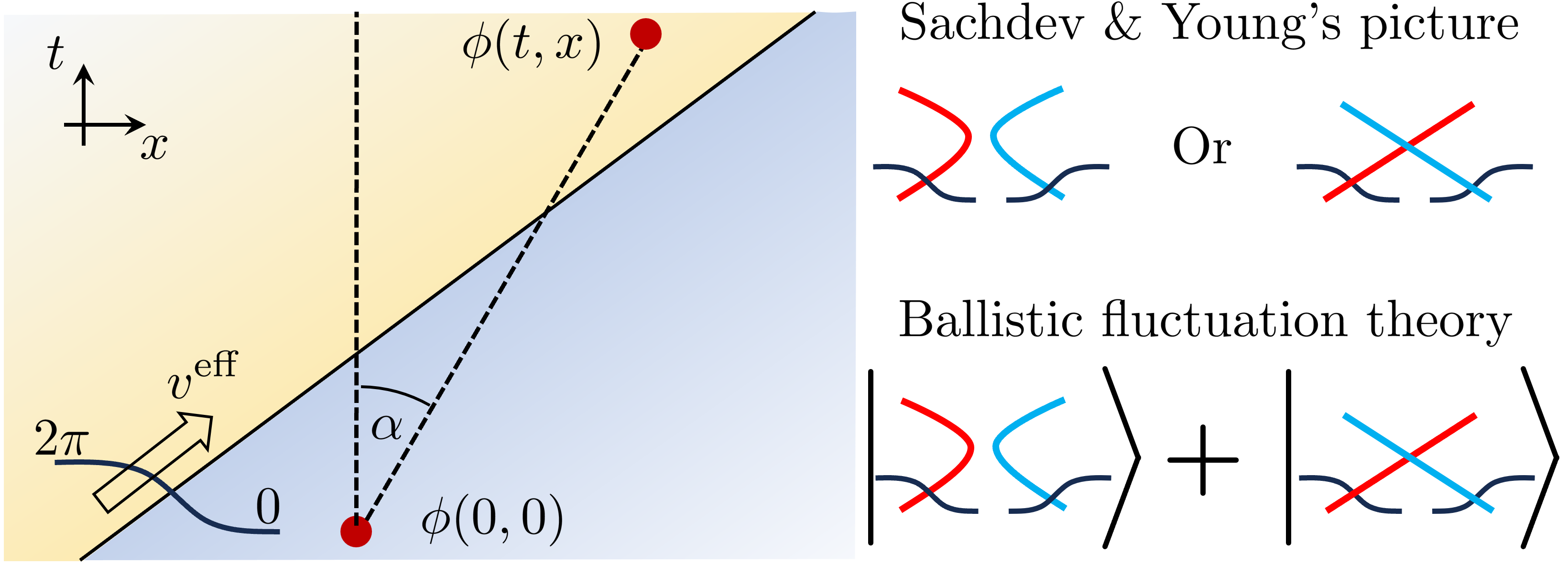}
	\caption{\textbf{Phenomenology of phase fluctuations and topological solitons.---} Pictorial representation of the phase fluctuation induced by a traveling soliton, moving with a velocity $v^{\text{eff}}$. Whenever the (anti-)kink worldline intersects the segment connecting $(t,x)$ with the origin, the phase difference jumps $\pm2\pi$. Ballistic Fluctuation Theory exactly captures  coherence that causes ballistic fluctuations and exponential decay of vertex operator correlations, neglected by the SY phenomenological approach which instead gives diffusion around the space-time ray at $x=0$.
}
	\label{fig_cartoon}
\end{figure}

The Hilbert space is described in terms of the asymptotic scattering states of these stable excitations. In integrable models, the interactions are fully encoded within the two-body scattering matrix  \cite{Smirnov1992}. 
For example, two wave packets of colliding breathers are transmitted through each other, experiencing in the meanwhile a non-trivial displacement  (see e.g.~\cite{PhysRevB.92.214427}) akin to classical soliton gases \cite{Doyon2018}. Kink-kink and breather-kink scattering behaves similarly, but reflection is generally possible in kink-antikink scattering (except for the reflectionless points $\xi^{-1}\in \mathbb{N}$). The task of diagonalizing this quantum process in terms of appropriate coherent combinations of scattering states, is accomplished by the Thermodynamic Bethe Ansatz (TBA)\cite{10.1063/1.1664947,ZAMOLODCHIKOV1990695,takahashi2005thermodynamics,Mossel_2012,caux2013,caux2016} and Generalized Hydrodynamics (GHD) \cite{Doyon2016,Bertini2016,specialissueGHD,Doyon2020Notes}, worked out in the sine-Gordon model in \cite{Bertini2019,Nagy_2023}. 
We summarize some aspects in the Supplementary Material (SM)\footnote{See supplementary material for \emph{(i)} summary of SY approach, \emph{(ii)} overview of the Ballistic Fluctuation Theory, \emph{(iii)} details on the sine-Gordon thermodynamics, and \emph{(iv)} details on coupled-condensates experiment.}.

In the SY phenomenological approach applied to sine-Gordon \cite{DamleSachdev2005, Kormos2016, Moca2019, Bertini2019, PhysRevB.106.205151}, phase fluctuations are assumed to come solely from a dilute gas of (anti-)kinks with Maxwell-Boltzmann statistics, justified at low temperatures. Whenever the trajectory of an (anti-)kink intersects the ray connecting $(t,x)$ with the origin, it causes the phase difference $\phi(t,x)-\phi(0,0)$ to jump, see Fig. \ref{fig_cartoon}: hence, the statistics of phase differences is intimately connected with that of traveling solitons. Damle and Sachdev \cite{DamleSachdev2005} considered the repulsive regime, where only (anti)kinks are present, and assumed a fully-reflective scattering, as justified by the universal low-energy limit of the scattering matrix, leading to a diffusive behavior of the vertex operator correlation function in space-time \cite{DamleSachdev2005}. At finite temperature, transmission is possible, but a hybrid semiclassical picture of incoherent processes with finite transmission probability is still diffusive as reflection eventually dominates.
We find that these conclusions do not hold in the sine-Gordon theory, where integrability plays a pivotal role in preserving coherent scattering, giving rise to ballistic transport {\em at all temperatures and coupling strengths} and exponential decay of correlation functions everywhere in space-time.

\paragraph{\textbf{Large scale correlation functions and full counting statistics ---}} The topological charge is defined as $Q_{\rm top}=\int \dd x\, q_{\rm top}(x)$, where $q_{\rm top}=\frac{1}{2\pi}\partial_x \phi$ is its density.
It is an extensive conserved quantity: since the cosine-potential does not confine the field, $\phi(x)-\phi(0)$ can grow indefinitely. The associated continuity equation is $\partial_t q_{\rm top}+\partial_x j_{\rm top}=0$, with the current $j_{\rm top}=-\frac{1}{2\pi}\partial_t \phi$. Integrating $q_{\rm top}(x)$ on a finite interval, we recover the difference of phases at the interval's endpoints.
The Ballistic Fluctuation Theorey (BFT) provides general formulae for the Full Counting Statistics (FCS) of total charges and currents on large intervals of space-time solely from hydrodynamic data \cite{10.21468/SciPostPhys.8.1.007, Doyon2020}. For a generic density $q(x,t)$ and current $j(x,t)$, and a thermal or or Generalized Gibbs Ensemble (GGE) \cite{Calabrese_2016} $\braket{\cdots}$,
the theory predicts that for $(x,c t)=(\ell \sin\alpha,\ell \cos\alpha)$ and large $\ell$ one has
\begin{equation}\label{expF}
    \braket{e^{\lambda \int_0^1 \dd s\,\big(\dot t_s j(x_s,t_s)-\dot x_s q(x_s,t_s)\big)} }\asymp
    e^{\ell F_{\alpha}(\lambda)}\,,
\end{equation}
where $s\mapsto (x_s,t_s)$ is a path in space-time from $(x_0,t_0)=(0,0)$ to $(x_1,t_1)=(x,t)$. The FCS $F_{\alpha}(\lambda)$, a ``dynamical specific free energy", is the main result of the BFT. It is expressed in terms of the current  $\mathtt j_\lambda = \braket{j}_\lambda$ and density $\mathtt q_\lambda = \braket{q}_\lambda$ evaluated in a $\lambda$-dependent GGE $\braket{\cdots}_\lambda$ as
\begin{equation}\label{eq:scaled_cumulant_generating_function}
    F_\alpha(\lambda)
    = \int_0^\lambda \dd \lambda'\,\big( c^{-1}\cos\alpha \,\mathtt j_{\lambda'} - \sin\alpha\, \mathtt q_{\lambda'}\big).
\end{equation}
The $\lambda$-dependent GGE is fixed by a  {\em flow equation} from $\braket{\cdots}$ at $\lambda=0$,
that describes the deformation of the state by the exponential operator on the l.h.s.~of Eq.~\eqref{expF}.

\begin{figure}[t!]
\centering
	\includegraphics[width=0.95\columnwidth]{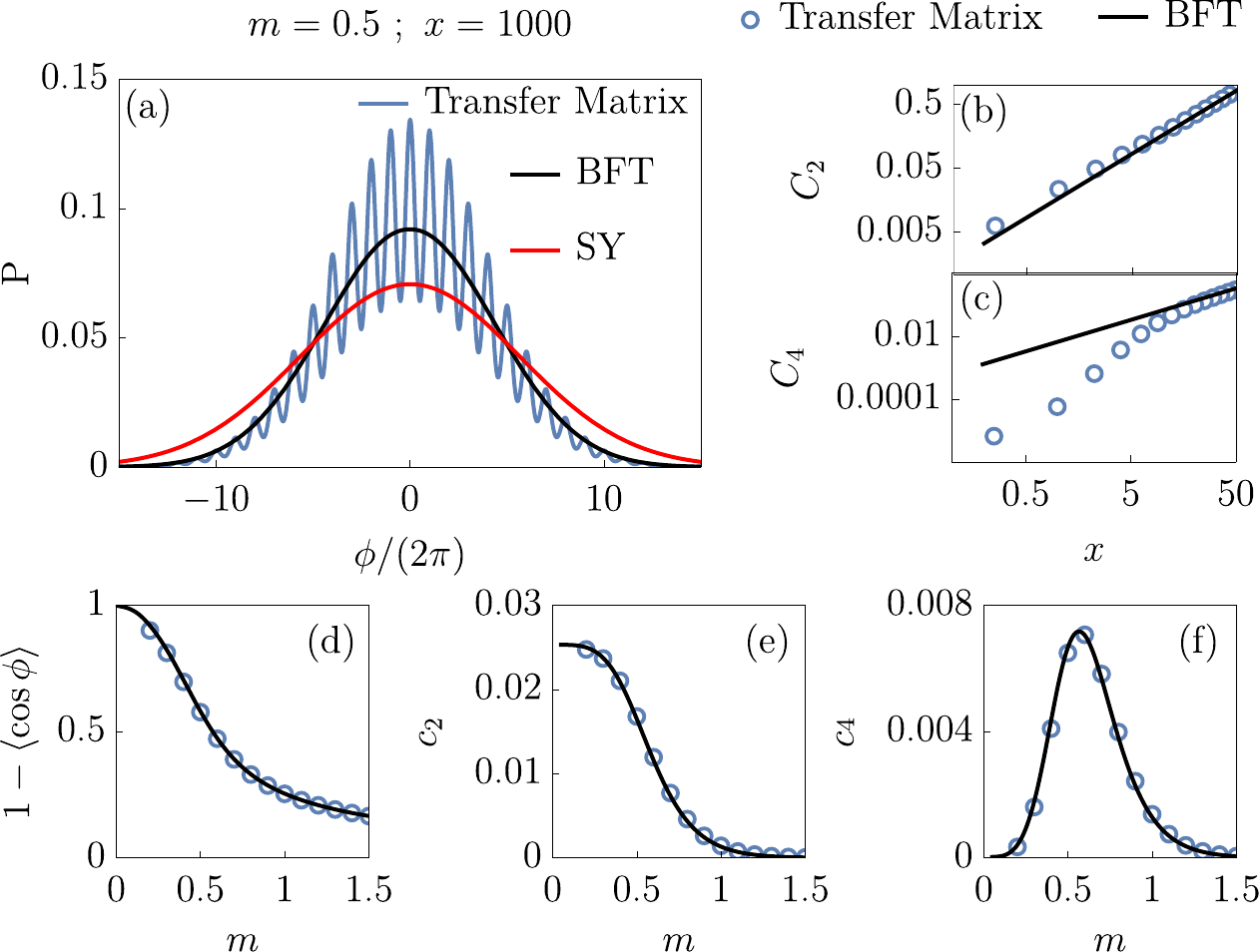}
	\caption{\textbf{Equal-time probability \& cumulants.---} We compare analytic predictions (BFT) [black line] in the classical regime of sine-Gordon with numerical results from Transfer Matrix [blue line and symbols], and with predictions from the SY picture of a gas of (anti)kinks with Maxwell-Boltzmann statistics [red line], in order to illustrate the neglected dressing effects in this picture, in the chosen regime of parameters. The bare mass $m$ is tuned while keeping $\beta=g=c=1$. (a) The probability of phase differences is reported for a typical mass scale and distance, showing the convergence to the scaling behavior. (b-c) The convergence of cumulants upon increasing the relative distance is shown. (d-f) We scan different values of the bare mass: the vertex operator (d) helps to identify the strongly-interacting regimes away from the massless limit $\langle \cos\phi\rangle \simeq 1$ and the large-mass non-interacting regime $1-\langle\cos\phi\rangle \simeq (4 m)^{-1}$ \cite{koch2023exact}. (e-f) The large-distance scaling of the 2$^{\rm nd}$ and 4$^{\rm th}$ cumulants is shown, clearly non-Gaussian and in perfect agreement with numerics.
	}
	\label{fig_1}
\end{figure}

Taking $Q=Q_{\rm top}$ we have $\int_0^1 \dd s\,\big(\dot t_s j_{\rm top}(x_s,t_s)-\dot x_s q_{\rm top}(x_s,t_s)\big) = -\frac{\Delta\phi(x,t)}{2\pi}$, thus the l.h.s.~of Eq.~\eqref{expF} is $\int \dd \delta\,P\left[\frac{\Delta\phi(t,x)}{2\pi}=\delta\right] e^{-\lambda \delta}$. The theory predicts that all cumulants $C_n = \left\langle\left(\frac{\Delta \phi}{2\pi}\right)^n\right\rangle_c$ of phase differences scale extensively $C_n \sim \ell c_n$ as $\ell\to\infty$, with $F_\alpha(\lambda)=  \sum_{n=1}^\infty c_n (-\lambda)^n/n!$ the Legendre-Fenchel transform of $I_\alpha(\delta)$. 
In the SM \cite{Note1} we review  the BFT and apply it to the topological charge of the sine-Gordon model.
Remarkably simple is the closed expression, valid at reflectionless points and in the classical regime, for the second cumulant whenever the average topological charge is zero 
\be\label{eq_c2}
c_2(\alpha)=2\int\dd\theta\,  \rho_K(\theta)f(\theta)|c^{-1}v^\text{eff}_K(\theta)\cos\alpha-\sin\alpha|\,.
\ee
Here kinks and antikinks have the same GGE distribution $\rho_K(\theta)$ and (dressed) velocity $v^\text{eff}_K(\theta)$, and $f(\theta)$ is a state dependent statistical factor ($f\to 1$ in the semiclassical limit).
In practice, $c_2(\alpha)$ is the scaled variance for the number of solitons whose wordline intersects the segment connecting $(t,x)$ and the origin, see Fig. \ref{fig_cartoon}.
All the terms in Eq. \eqref{eq_c2} are exactly known from TBA and GHD. The full second $c_2$ and fourth $c_4$ cumulants are reported in the SM \cite{Note1} at the reflectionless points and classical regime, and can be obtained for arbitrary coupling from the sine-Gordon TBA \cite{Bertini2019,Nagy_2023}.
\begin{figure}[t!]
\centering
	\includegraphics[width=0.95\columnwidth]{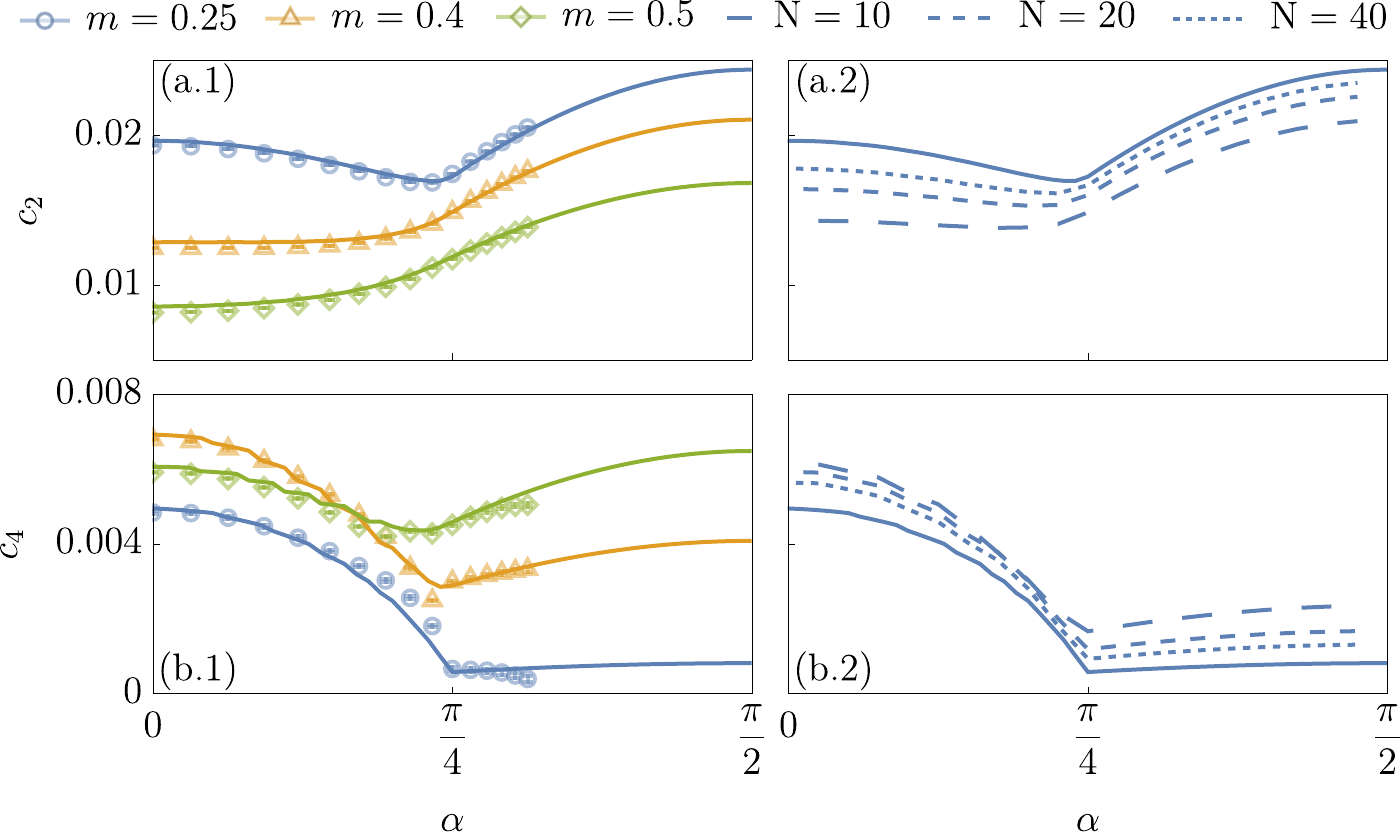}
	\caption{\textbf{Cumulants with space-time separation.---} Scaling behavior of the second (a.1) and fourth (b.1) cumulant as function of the ray $x/(ct) =\tan\alpha$ for representative choices of the mass scale $m$ ($\beta=g=c=1$) in the classical regime. Numerical values obtained with Monte Carlo (symbols) closely follow the analytic BFT prediction (solid lines). In Figure (a.2) and (b.2) we show the approach of the quantum prediction (dashed lines) to the classical limit (solid line) for the $c_2$ and $c_4$, respectively. We take $m=0.25$ and increase the number of breathers $N$, while tuning the quantum soliton mass according to the semiclassical limit \cite{koch2023exact,Note1}.
	}
	\label{fig_2}
\end{figure}

The BFT results should be contrasted with the SY picture \cite{DamleSachdev2005, Kormos2016, Moca2019, Bertini2019, PhysRevB.106.205151}. 
Clearly, the latter picture neglects dressing effects by setting $\rho_K(\theta)=\frac{Mc \cosh\theta}{2\pi}\exp(-\beta Mc^2\cosh\theta)$ in \eqref{eq_c2}
(see Fig. \ref{fig_1}(a)). But most importantly, at unequal times, the resulting physics is qualitatively different: fully reflective scattering makes the topological charge an isolated hydrodynamic mode with zero velocity, and indeed Ref. \cite{DamleSachdev2005} predicts a diffusive, instead of ballistic, behavior, with power-law instead of exponential decay of vertex operator correlations at $\alpha=0$.
The BFT captures the resulting coherent scattering and shows that a ballistic behavior and exponential decay is generic in the sine-Gordon model.
Note that taking purely transmissive scattering in the SY picture \cite{DamleSachdev2005}, one obtains the correct ballistic low-temperature behavior at reflectionless points \cite{Note1}.

\paragraph{\textbf{The semiclassical limit and numerical benchmarks ---}} 
Strongly interacting systems at finite temperature are extremely challenging to simulate \cite{Schollwock2011}.
Hence, we now focus on the classical regime, which is amenable to efficient numerical benchmarks \cite{Note1}.

The exact thermodynamics of the classical sine-Gordon model has recently been developed in Ref. \cite{koch2023exact} building on classical limits \cite{DeLuca2016,Bastianello2018,DelVecchio2020,Koch_2022,Bezzaz2023} of quantum integrability; we apply these to the BFT framework \cite{Note1}. 
In equilibrium, one can set interaction, temperature and velocity to $1$ upon a length scale renormalization: we opt for this choice and use the mass $m$ as a tunable parameter.
In Fig. \ref{fig_1}, we compare equal-time phase fluctuations derived from \emph{i)} our result, \emph{ii)} SY classical picture (see the SM \cite{Note1}) and \emph{iii)} numerical results obtained with the Transfer Matrix method \cite{Scalapino1972,Castin2000,Note1}.
In Fig. \ref{fig_1}(a) we show the full distribution of the phase for a typical example. 
Notably, numerics shows ``spikes", reminiscent of the fact that the number of solitons comprised in an interval $[0,x]$ is an integer number; for lower temperature (larger mass scales) the spikes are more peaked. The BFT prediction, substantially different from SY, captures the smoothed probability distribution: convergence at large separation holds in a weak sense. The BFT scaling is clearer for the cumulants, see Fig. \ref{fig_1}(b-c); it becomes slower for higher cumulants.
In Fig.~\ref{fig_1}(d-e-f), we scan a wide parametric regime finding excellent agreement between our analytical result and numerics. In Fig.~\ref{fig_2}, we analyze unequal-times phase fluctuations: we observed spikes (not shown), but the cumulants quickly reach their asymptotic scaling. For representative values of the mass, we compare the ray-dependent growth of cumulants predicted by BFT against Monte Carlo simulations \cite{Metropolis1953,Hasting1970} with good agreement, and show the convergence of quantum predictions at the reflectionless points to the semiclassical ones. 
Further analysis is left to the SM \cite{Note1}.

\begin{figure}[t!]
\centering
	\includegraphics[width=0.99\columnwidth]{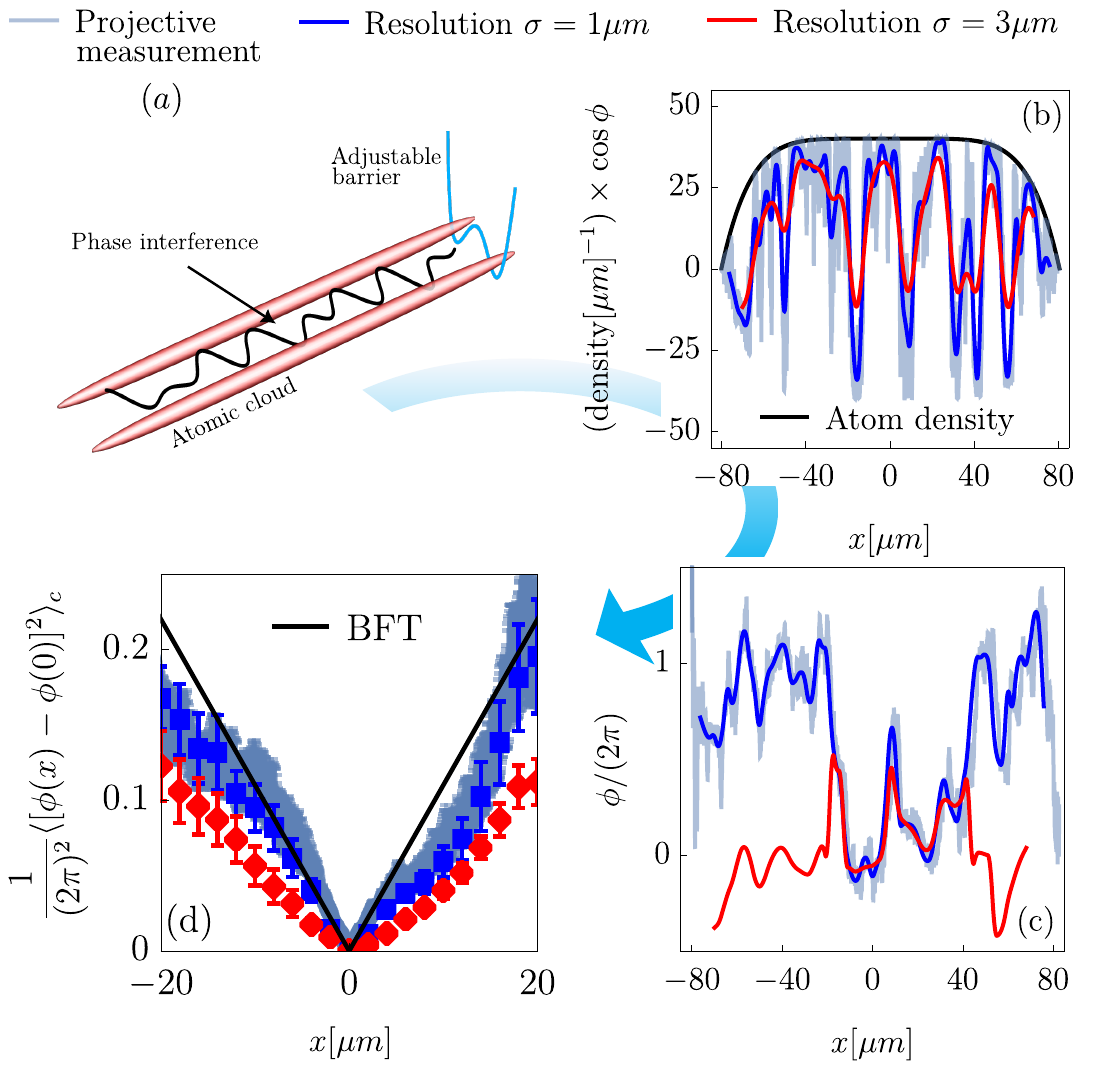}
	\caption{\textbf{The sine-Gordon model from coupled condensates.---}(a) Sketch of the experimental setup. (b-c) Example of phase-reconstruction protocol from the outcome of a single projective measurement for different pixel resolutions $\sigma$ (see main text). (d) Statistics built on $100$ samples already shows the scaling behavior of the equal-time second cumulant stemming from the center of the trap. The effect of a low resolution $\sigma$ is to ``miss" kinks (see also (c)) and underestimate phase fluctuations. A good-quality measurement is already obtained with $\sigma=1\text{$\mu$m}$. See main text for discussion of parameters, and the SM \cite{Note1} for further details and data.
	}
	\label{fig_3}
\end{figure}

\paragraph{\textbf{Experimental feasibility ---}}
A versatile tabletop simulator of the sine-Gordon model is realized by the experimental group in Vienna via two tunnel-coupled quasicondensates \cite{Gritsev2007}, see Fig.~\ref{fig_3}(a);  phase  fluctuations are probed by matter-wave interferometry measurements \cite{Schumm2005,Hofferberth2007,Nieuwkerk2018}. Our result can arguably give quantitative predictions for such experimental data and may be useful in state characterization, both in equilibrium \cite{Schweigler2017} and nonequilibrium setups \cite{langen2016}. 
However, imperfections and finite resolution may undermine a correct phase measurement: to show that faithful phase tomography is within the reach of current experimental capability, we analyze a toy model of the measurement process. 
Due to the weak interactions of the atoms, sine-Gordon is realized close to its semiclassical regime \cite{Blakie2008,Horvath2019} and Monte Carlo accounts for experimental observations \cite{Schweigler2017,Mennemann2021}.
We use typical experimental parameters, see the SM \cite{Note1}. Atoms are trapped in a smoothed box potential of length $\sim 160\text{$\mu$m}$, and the transverse trap frequency \cite{Olshanii1998} tunes interactions. The inhomogeneous density profile $n(x)$ is well described by the Thomas-Fermi approximation and causes weak inhomogeneities in the sine-Gordon coupling \cite{tajik2022}. The mass scale, which is changed by adjusting the strength of the tunneling between the tubes, affects the spatial extent of the kinks, and the overall population. We choose a temperature $60\text{nK}$ and a bulk atom density $40 \text{atm/$\mu$m}$ to retain an appreciable kink density, and a mass scale such that $\langle \cos\phi\rangle\simeq 0.32$ in the bulk. Further discussion is left to the SM \cite{Note1}.

Matter-wave interferometry gives access to spatially-resolved projective measurements of trigonometric functions of the phase $n(x)\cos\phi(x)$ and $n(x)\sin\phi(x)$ \cite{Kuhnert2013}. 
In an ideal scenario, the phase itself can then be recovered, but the finite imaging resolution causes a detrimental coarse graining (see Fig. \ref{fig_3}(b)). The latter is modeled by convolving $n(x)\cos(\phi(x))$ with Gaussians with standard deviation $\sigma$ and the phase is then reconstructed from these coarse grained data \cite{schweigler2019,Note1}.
Depending on the resolution, the phase profile may be correctly recovered or kinks may be washed out by the local coarse graining, see Fig. \ref{fig_3}(c).
Finally, by building statistics over many measurements, phase correlations are obtained. In Fig. \ref{fig_3}(d), we show the outcome of $100$ independent samples: a resolution $\sigma=1\text{$\mu$m}$ (a slight improvement on the current experimental resolution $\sigma\simeq 3\text{$\mu$m}$ \cite{schweigler2019}) is enough to capture microscopic phase fluctuations that compare well with analytical results. Large fluctuations of the second cumulant are due to the relatively small number of samples: we used $100$ as a typical experimental situation.

\paragraph{\textbf{Conclusion and outlook ---}}
Exact results on correlation functions in interacting field theories are scarce: we give analytical predictions for the large scale phase-fluctuations in the sine-Gordon model, valid at any temperature and interactions, and in Generalized Gibbs Ensembles. We discuss how our results are of ready applicability in experiments on coupled condensates, where equal-time correlations are accessed.
Unequal-time phase differences may be accessible by locally exciting the topological charge via Raman-coupling \cite{Kasper2020}.
The main appeal of our results is its applicability to the quantum regime: sine-Gordon simulators in the quantum regime may be within reach of Quantum Gas Microscopes \cite{Wybo2022,wybo2023}.
One can analyse the full range of couplings using \cite{Bertini2019,Nagy_2023} (work in preparation) and integrability-breaking perturbations \cite{Friedman2020,Piqueres2021,Durnin2021b,Bastianello_2021} within the BFT. It will be important to include diffusive corrections \cite{denardis2018,denardis2019}: a possible scenario at low temperatures is that the diffusive behaviour predicted by Damle and Sachdev \cite{DamleSachdev2005} is seen at early times, with a slow exponential decay at later times as predicted by the BFT. In contrast, if integrability is broken, isolated hydrodynamic modes are present and the diffusive SY picture should hold at all times and temperatures \cite{PhysRevB.106.205151}. Studying the time scales of the various crossovers implied is of utmost interest for future studies.

\paragraph{\textbf{Data and code availability---}} Raw data and working codes are available on Zenodo \cite{Zenodo}.

\paragraph{\textbf{Acknowledgments---}} We are indebted to Frederik M{\o}ller, Mohammadamin Tajik, and Joerg Schmiedmayer for valuable discussions on the experimental setup. We thank Fabian Essler, Michael Knap and Rebekka Koch for useful discussion.
AB acknowledges support from the Deutsche Forschungsgemeinschaft (DFG, German Research Foundation) under Germany’s Excellence Strategy–EXC–2111–390814868. The work of BD was supported by the Engineering and Physical Sciences Research Council (EPSRC) under grant EP/W010194/1. MK acknowledges support from the National Research, Development and Innovation Office (NKFIH) through research Grants No. K138606 and No. ANN142584, and from the QuantERA grant QuSiED.

\bibliography{biblio}

\begin{thebibliography}{99}%
\makeatletter
\providecommand \@ifxundefined [1]{%
 \@ifx{#1\undefined}
}%
\providecommand \@ifnum [1]{%
 \ifnum #1\expandafter \@firstoftwo
 \else \expandafter \@secondoftwo
 \fi
}%
\providecommand \@ifx [1]{%
 \ifx #1\expandafter \@firstoftwo
 \else \expandafter \@secondoftwo
 \fi
}%
\providecommand \natexlab [1]{#1}%
\providecommand \enquote  [1]{``#1''}%
\providecommand \bibnamefont  [1]{#1}%
\providecommand \bibfnamefont [1]{#1}%
\providecommand \citenamefont [1]{#1}%
\providecommand \href@noop [0]{\@secondoftwo}%
\providecommand \href [0]{\begingroup \@sanitize@url \@href}%
\providecommand \@href[1]{\@@startlink{#1}\@@href}%
\providecommand \@@href[1]{\endgroup#1\@@endlink}%
\providecommand \@sanitize@url [0]{\catcode `\\12\catcode `\$12\catcode
  `\&12\catcode `\#12\catcode `\^12\catcode `\_12\catcode `\%12\relax}%
\providecommand \@@startlink[1]{}%
\providecommand \@@endlink[0]{}%
\providecommand \url  [0]{\begingroup\@sanitize@url \@url }%
\providecommand \@url [1]{\endgroup\@href {#1}{\urlprefix }}%
\providecommand \urlprefix  [0]{URL }%
\providecommand \Eprint [0]{\href }%
\providecommand \doibase [0]{https://doi.org/}%
\providecommand \selectlanguage [0]{\@gobble}%
\providecommand \bibinfo  [0]{\@secondoftwo}%
\providecommand \bibfield  [0]{\@secondoftwo}%
\providecommand \translation [1]{[#1]}%
\providecommand \BibitemOpen [0]{}%
\providecommand \bibitemStop [0]{}%
\providecommand \bibitemNoStop [0]{.\EOS\space}%
\providecommand \EOS [0]{\spacefactor3000\relax}%
\providecommand \BibitemShut  [1]{\csname bibitem#1\endcsname}%
\let\auto@bib@innerbib\@empty
\bibitem [{\citenamefont {Fisher}(1998)}]{Fischer1998}%
  \BibitemOpen
  \bibfield  {author} {\bibinfo {author} {\bibfnamefont {M.~E.}\ \bibnamefont
  {Fisher}},\ }\bibfield  {title} {\bibinfo {title} {Renormalization group
  theory: Its basis and formulation in statistical physics},\ }\href
  {https://doi.org/10.1103/RevModPhys.70.653} {\bibfield  {journal} {\bibinfo
  {journal} {Rev. Mod. Phys.}\ }\textbf {\bibinfo {volume} {70}},\ \bibinfo
  {pages} {653} (\bibinfo {year} {1998})}\BibitemShut {NoStop}%
\bibitem [{\citenamefont {Spohn}(2012)}]{spohn2012large}%
  \BibitemOpen
  \bibfield  {author} {\bibinfo {author} {\bibfnamefont {H.}~\bibnamefont
  {Spohn}},\ }\href@noop {} {\emph {\bibinfo {title} {Large scale dynamics of
  interacting particles}}}\ (\bibinfo  {publisher} {Springer Science \&
  Business Media},\ \bibinfo {year} {2012})\BibitemShut {NoStop}%
\bibitem [{\citenamefont {Giamarchi}(2003)}]{giamarchi2003quantum}%
  \BibitemOpen
  \bibfield  {author} {\bibinfo {author} {\bibfnamefont {T.}~\bibnamefont
  {Giamarchi}},\ }\href@noop {} {\emph {\bibinfo {title} {Quantum physics in
  one dimension}}},\ Vol.\ \bibinfo {volume} {121}\ (\bibinfo  {publisher}
  {Clarendon press},\ \bibinfo {year} {2003})\BibitemShut {NoStop}%
\bibitem [{\citenamefont {Francesco}\ \emph {et~al.}(2012)\citenamefont
  {Francesco}, \citenamefont {Mathieu},\ and\ \citenamefont
  {S{\'e}n{\'e}chal}}]{francesco2012conformal}%
  \BibitemOpen
  \bibfield  {author} {\bibinfo {author} {\bibfnamefont {P.}~\bibnamefont
  {Francesco}}, \bibinfo {author} {\bibfnamefont {P.}~\bibnamefont {Mathieu}},\
  and\ \bibinfo {author} {\bibfnamefont {D.}~\bibnamefont {S{\'e}n{\'e}chal}},\
  }\href@noop {} {\emph {\bibinfo {title} {Conformal field theory}}}\ (\bibinfo
   {publisher} {Springer Science \& Business Media},\ \bibinfo {year}
  {2012})\BibitemShut {NoStop}%
\bibitem [{\citenamefont {Bernard}\ and\ \citenamefont
  {Doyon}(2016)}]{Bernard_2016}%
  \BibitemOpen
  \bibfield  {author} {\bibinfo {author} {\bibfnamefont {D.}~\bibnamefont
  {Bernard}}\ and\ \bibinfo {author} {\bibfnamefont {B.}~\bibnamefont
  {Doyon}},\ }\bibfield  {title} {\bibinfo {title} {Conformal field theory out
  of equilibrium: a review},\ }\href
  {https://doi.org/10.1088/1742-5468/2016/06/064005} {\bibfield  {journal}
  {\bibinfo  {journal} {Journal of Statistical Mechanics: Theory and
  Experiment}\ }\textbf {\bibinfo {volume} {2016}},\ \bibinfo {pages} {064005}
  (\bibinfo {year} {2016})}\BibitemShut {NoStop}%
\bibitem [{\citenamefont {Nardis}\ \emph {et~al.}(2022)\citenamefont {Nardis},
  \citenamefont {Doyon}, \citenamefont {Medenjak},\ and\ \citenamefont
  {Panfil}}]{DeNardis_2022}%
  \BibitemOpen
  \bibfield  {author} {\bibinfo {author} {\bibfnamefont {J.~D.}\ \bibnamefont
  {Nardis}}, \bibinfo {author} {\bibfnamefont {B.}~\bibnamefont {Doyon}},
  \bibinfo {author} {\bibfnamefont {M.}~\bibnamefont {Medenjak}},\ and\
  \bibinfo {author} {\bibfnamefont {M.}~\bibnamefont {Panfil}},\ }\bibfield
  {title} {\bibinfo {title} {Correlation functions and transport coefficients
  in generalised hydrodynamics},\ }\href
  {https://doi.org/10.1088/1742-5468/ac3658} {\bibfield  {journal} {\bibinfo
  {journal} {Journal of Statistical Mechanics: Theory and Experiment}\ }\textbf
  {\bibinfo {volume} {2022}},\ \bibinfo {pages} {014002} (\bibinfo {year}
  {2022})}\BibitemShut {NoStop}%
\bibitem [{\citenamefont {Pfeuty}(1970)}]{Pfeuty1970}%
  \BibitemOpen
  \bibfield  {author} {\bibinfo {author} {\bibfnamefont {P.}~\bibnamefont
  {Pfeuty}},\ }\bibfield  {title} {\bibinfo {title} {The one-dimensional ising
  model with a transverse field},\ }\href
  {https://doi.org/https://doi.org/10.1016/0003-4916(70)90270-8} {\bibfield
  {journal} {\bibinfo  {journal} {Annals of Physics}\ }\textbf {\bibinfo
  {volume} {57}},\ \bibinfo {pages} {79} (\bibinfo {year} {1970})}\BibitemShut
  {NoStop}%
\bibitem [{\citenamefont {Perk}(1980{\natexlab{a}})}]{PERK19801}%
  \BibitemOpen
  \bibfield  {author} {\bibinfo {author} {\bibfnamefont {J.}~\bibnamefont
  {Perk}},\ }\bibfield  {title} {\bibinfo {title} {Equations of motion for the
  transverse correlations of the one-dimensional xy-model at finite
  temperature},\ }\href
  {https://doi.org/https://doi.org/10.1016/0375-9601(80)90298-4} {\bibfield
  {journal} {\bibinfo  {journal} {Physics Letters A}\ }\textbf {\bibinfo
  {volume} {79}},\ \bibinfo {pages} {1} (\bibinfo {year}
  {1980}{\natexlab{a}})}\BibitemShut {NoStop}%
\bibitem [{\citenamefont {Perk}(1980{\natexlab{b}})}]{PERK19803}%
  \BibitemOpen
  \bibfield  {author} {\bibinfo {author} {\bibfnamefont {J.}~\bibnamefont
  {Perk}},\ }\bibfield  {title} {\bibinfo {title} {Quadratic identities for
  ising model correlations},\ }\href
  {https://doi.org/https://doi.org/10.1016/0375-9601(80)90299-6} {\bibfield
  {journal} {\bibinfo  {journal} {Physics Letters A}\ }\textbf {\bibinfo
  {volume} {79}},\ \bibinfo {pages} {3} (\bibinfo {year}
  {1980}{\natexlab{b}})}\BibitemShut {NoStop}%
\bibitem [{\citenamefont {Perk}\ \emph {et~al.}(1984)\citenamefont {Perk},
  \citenamefont {Capel}, \citenamefont {Quispel},\ and\ \citenamefont
  {Nijhoff}}]{PERK19841}%
  \BibitemOpen
  \bibfield  {author} {\bibinfo {author} {\bibfnamefont {J.}~\bibnamefont
  {Perk}}, \bibinfo {author} {\bibfnamefont {H.}~\bibnamefont {Capel}},
  \bibinfo {author} {\bibfnamefont {G.}~\bibnamefont {Quispel}},\ and\ \bibinfo
  {author} {\bibfnamefont {F.}~\bibnamefont {Nijhoff}},\ }\bibfield  {title}
  {\bibinfo {title} {Finite-temperature correlations for the ising chain in a
  transverse field},\ }\href
  {https://doi.org/https://doi.org/10.1016/0378-4371(84)90102-X} {\bibfield
  {journal} {\bibinfo  {journal} {Physica A: Statistical Mechanics and its
  Applications}\ }\textbf {\bibinfo {volume} {123}},\ \bibinfo {pages} {1}
  (\bibinfo {year} {1984})}\BibitemShut {NoStop}%
\bibitem [{\citenamefont {Perk}\ and\ \citenamefont
  {Au-Yang}(2009)}]{Perk2009}%
  \BibitemOpen
  \bibfield  {author} {\bibinfo {author} {\bibfnamefont {J.~H.~H.}\
  \bibnamefont {Perk}}\ and\ \bibinfo {author} {\bibfnamefont {H.}~\bibnamefont
  {Au-Yang}},\ }\bibfield  {title} {\bibinfo {title} {New results for the
  correlation functions of the ising model and the transverse ising chain},\
  }\href {https://doi.org/10.1007/s10955-009-9758-5} {\bibfield  {journal}
  {\bibinfo  {journal} {Journal of Statistical Physics}\ }\textbf {\bibinfo
  {volume} {135}},\ \bibinfo {pages} {599} (\bibinfo {year}
  {2009})}\BibitemShut {NoStop}%
\bibitem [{\citenamefont {Doyon}(2005)}]{Doyon_2005}%
  \BibitemOpen
  \bibfield  {author} {\bibinfo {author} {\bibfnamefont {B.}~\bibnamefont
  {Doyon}},\ }\bibfield  {title} {\bibinfo {title} {Finite-temperature form
  factors in the free majorana theory},\ }\href
  {https://doi.org/10.1088/1742-5468/2005/11/P11006} {\bibfield  {journal}
  {\bibinfo  {journal} {Journal of Statistical Mechanics: Theory and
  Experiment}\ }\textbf {\bibinfo {volume} {2005}},\ \bibinfo {pages} {P11006}
  (\bibinfo {year} {2005})}\BibitemShut {NoStop}%
\bibitem [{\citenamefont {Doyon}(2007)}]{DoyonFF_2007}%
  \BibitemOpen
  \bibfield  {author} {\bibinfo {author} {\bibfnamefont {B.}~\bibnamefont
  {Doyon}},\ }\bibfield  {title} {\bibinfo {title} {Finite-temperature form
  factors: a review},\ }\href {https://doi.org/10.3842/SIGMA.2007.011}
  {\bibfield  {journal} {\bibinfo  {journal} {SIGMA}\ }\textbf {\bibinfo
  {volume} {3}},\ \bibinfo {pages} {011} (\bibinfo {year} {2007})}\BibitemShut
  {NoStop}%
\bibitem [{\citenamefont {Calabrese}\ \emph {et~al.}(2011)\citenamefont
  {Calabrese}, \citenamefont {Essler},\ and\ \citenamefont
  {Fagotti}}]{Calabrese2011}%
  \BibitemOpen
  \bibfield  {author} {\bibinfo {author} {\bibfnamefont {P.}~\bibnamefont
  {Calabrese}}, \bibinfo {author} {\bibfnamefont {F.~H.~L.}\ \bibnamefont
  {Essler}},\ and\ \bibinfo {author} {\bibfnamefont {M.}~\bibnamefont
  {Fagotti}},\ }\bibfield  {title} {\bibinfo {title} {Quantum quench in the
  transverse-field ising chain},\ }\href
  {https://doi.org/10.1103/PhysRevLett.106.227203} {\bibfield  {journal}
  {\bibinfo  {journal} {Phys. Rev. Lett.}\ }\textbf {\bibinfo {volume} {106}},\
  \bibinfo {pages} {227203} (\bibinfo {year} {2011})}\BibitemShut {NoStop}%
\bibitem [{\citenamefont {Calabrese}\ \emph {et~al.}(2012)\citenamefont
  {Calabrese}, \citenamefont {Essler},\ and\ \citenamefont
  {Fagotti}}]{Calabrese_2012}%
  \BibitemOpen
  \bibfield  {author} {\bibinfo {author} {\bibfnamefont {P.}~\bibnamefont
  {Calabrese}}, \bibinfo {author} {\bibfnamefont {F.~H.~L.}\ \bibnamefont
  {Essler}},\ and\ \bibinfo {author} {\bibfnamefont {M.}~\bibnamefont
  {Fagotti}},\ }\bibfield  {title} {\bibinfo {title} {Quantum quench in the
  transverse field ising chain: I. time evolution of order parameter
  correlators},\ }\href {https://doi.org/10.1088/1742-5468/2012/07/P07016}
  {\bibfield  {journal} {\bibinfo  {journal} {Journal of Statistical Mechanics:
  Theory and Experiment}\ }\textbf {\bibinfo {volume} {2012}},\ \bibinfo
  {pages} {P07016} (\bibinfo {year} {2012})}\BibitemShut {NoStop}%
\bibitem [{\citenamefont {Chen}\ and\ \citenamefont {Doyon}(2014)}]{Chen_2014}%
  \BibitemOpen
  \bibfield  {author} {\bibinfo {author} {\bibfnamefont {Y.}~\bibnamefont
  {Chen}}\ and\ \bibinfo {author} {\bibfnamefont {B.}~\bibnamefont {Doyon}},\
  }\bibfield  {title} {\bibinfo {title} {Form factors in equilibrium and
  non-equilibrium mixed states of the ising model},\ }\href
  {https://doi.org/10.1088/1742-5468/2014/09/P09021} {\bibfield  {journal}
  {\bibinfo  {journal} {Journal of Statistical Mechanics: Theory and
  Experiment}\ }\textbf {\bibinfo {volume} {2014}},\ \bibinfo {pages} {P09021}
  (\bibinfo {year} {2014})}\BibitemShut {NoStop}%
\bibitem [{\citenamefont {Groha}\ \emph {et~al.}(2018)\citenamefont {Groha},
  \citenamefont {Essler},\ and\ \citenamefont {Calabrese}}]{2018Groha}%
  \BibitemOpen
  \bibfield  {author} {\bibinfo {author} {\bibfnamefont {S.}~\bibnamefont
  {Groha}}, \bibinfo {author} {\bibfnamefont {F.~H.~L.}\ \bibnamefont
  {Essler}},\ and\ \bibinfo {author} {\bibfnamefont {P.}~\bibnamefont
  {Calabrese}},\ }\bibfield  {title} {\bibinfo {title} {{Full counting
  statistics in the transverse field Ising chain}},\ }\href
  {https://doi.org/10.21468/SciPostPhys.4.6.043} {\bibfield  {journal}
  {\bibinfo  {journal} {SciPost Phys.}\ }\textbf {\bibinfo {volume} {4}},\
  \bibinfo {pages} {043} (\bibinfo {year} {2018})}\BibitemShut {NoStop}%
\bibitem [{\citenamefont {Altshuler}\ and\ \citenamefont
  {Tsvelik}(2006)}]{Altshuler2006}%
  \BibitemOpen
  \bibfield  {author} {\bibinfo {author} {\bibfnamefont {B.~L.}\ \bibnamefont
  {Altshuler}}\ and\ \bibinfo {author} {\bibfnamefont {A.~M.}\ \bibnamefont
  {Tsvelik}},\ }\href@noop {} {\bibinfo {title} {Finite temperature correlation
  functions in integrable models: derivation of the semiclassical limit from
  the formfactor expansion}} (\bibinfo {year} {2006}),\ \Eprint
  {https://arxiv.org/abs/cond-mat/0505367} {arXiv:cond-mat/0505367
  [cond-mat.stat-mech]} \BibitemShut {NoStop}%
\bibitem [{\citenamefont {LeClair}\ and\ \citenamefont
  {Mussardo}(1999)}]{LeClair1999}%
  \BibitemOpen
  \bibfield  {author} {\bibinfo {author} {\bibfnamefont {A.}~\bibnamefont
  {LeClair}}\ and\ \bibinfo {author} {\bibfnamefont {G.}~\bibnamefont
  {Mussardo}},\ }\bibfield  {title} {\bibinfo {title} {Finite temperature
  correlation functions in integrable qft},\ }\href
  {https://doi.org/https://doi.org/10.1016/S0550-3213(99)00280-1} {\bibfield
  {journal} {\bibinfo  {journal} {Nuclear Physics B}\ }\textbf {\bibinfo
  {volume} {552}},\ \bibinfo {pages} {624} (\bibinfo {year}
  {1999})}\BibitemShut {NoStop}%
\bibitem [{\citenamefont {Pozsgay}\ and\ \citenamefont
  {Sz{\'e}cs{\'e}nyi}(2018)}]{Pozsgay2018}%
  \BibitemOpen
  \bibfield  {author} {\bibinfo {author} {\bibfnamefont {B.}~\bibnamefont
  {Pozsgay}}\ and\ \bibinfo {author} {\bibfnamefont {I.~M.}\ \bibnamefont
  {Sz{\'e}cs{\'e}nyi}},\ }\bibfield  {title} {\bibinfo {title}
  {Leclair-mussardo series for two-point functions in integrable qft},\ }\href
  {https://doi.org/10.1007/JHEP05(2018)170} {\bibfield  {journal} {\bibinfo
  {journal} {Journal of High Energy Physics}\ }\textbf {\bibinfo {volume}
  {2018}},\ \bibinfo {pages} {170} (\bibinfo {year} {2018})}\BibitemShut
  {NoStop}%
\bibitem [{\citenamefont {Collura}\ \emph {et~al.}(2017)\citenamefont
  {Collura}, \citenamefont {Essler},\ and\ \citenamefont
  {Groha}}]{Collura_2017}%
  \BibitemOpen
  \bibfield  {author} {\bibinfo {author} {\bibfnamefont {M.}~\bibnamefont
  {Collura}}, \bibinfo {author} {\bibfnamefont {F.~H.~L.}\ \bibnamefont
  {Essler}},\ and\ \bibinfo {author} {\bibfnamefont {S.}~\bibnamefont
  {Groha}},\ }\bibfield  {title} {\bibinfo {title} {Full counting statistics in
  the spin-1/2 heisenberg xxz chain},\ }\href
  {https://doi.org/10.1088/1751-8121/aa87dd} {\bibfield  {journal} {\bibinfo
  {journal} {Journal of Physics A: Mathematical and Theoretical}\ }\textbf
  {\bibinfo {volume} {50}},\ \bibinfo {pages} {414002} (\bibinfo {year}
  {2017})}\BibitemShut {NoStop}%
\bibitem [{\citenamefont {Vecchio}\ and\ \citenamefont
  {Doyon}(2022)}]{delvecchio2022}%
  \BibitemOpen
  \bibfield  {author} {\bibinfo {author} {\bibfnamefont {G.~D. V.~D.}\
  \bibnamefont {Vecchio}}\ and\ \bibinfo {author} {\bibfnamefont
  {B.}~\bibnamefont {Doyon}},\ }\bibfield  {title} {\bibinfo {title} {The
  hydrodynamic theory of dynamical correlation functions in the xx chain},\
  }\href {https://doi.org/10.1088/1742-5468/ac6667} {\bibfield  {journal}
  {\bibinfo  {journal} {Journal of Statistical Mechanics: Theory and
  Experiment}\ }\textbf {\bibinfo {volume} {2022}},\ \bibinfo {pages} {053102}
  (\bibinfo {year} {2022})}\BibitemShut {NoStop}%
\bibitem [{\citenamefont {Vecchio}\ \emph {et~al.}(2023)\citenamefont
  {Vecchio}, \citenamefont {Doyon},\ and\ \citenamefont
  {Ruggiero}}]{delvecchio2023}%
  \BibitemOpen
  \bibfield  {author} {\bibinfo {author} {\bibfnamefont {G.~D. V.~D.}\
  \bibnamefont {Vecchio}}, \bibinfo {author} {\bibfnamefont {B.}~\bibnamefont
  {Doyon}},\ and\ \bibinfo {author} {\bibfnamefont {P.}~\bibnamefont
  {Ruggiero}},\ }\href@noop {} {\bibinfo {title} {Entanglement r\'enyi
  entropies from ballistic fluctuation theory: the free fermionic case}}
  (\bibinfo {year} {2023}),\ \Eprint {https://arxiv.org/abs/2301.02326}
  {arXiv:2301.02326 [quant-ph]} \BibitemShut {NoStop}%
\bibitem [{\citenamefont {Sachdev}\ and\ \citenamefont
  {Young}(1997)}]{Sachdev1997}%
  \BibitemOpen
  \bibfield  {author} {\bibinfo {author} {\bibfnamefont {S.}~\bibnamefont
  {Sachdev}}\ and\ \bibinfo {author} {\bibfnamefont {A.~P.}\ \bibnamefont
  {Young}},\ }\bibfield  {title} {\bibinfo {title} {Low temperature
  relaxational dynamics of the ising chain in a transverse field},\ }\href
  {https://doi.org/10.1103/PhysRevLett.78.2220} {\bibfield  {journal} {\bibinfo
   {journal} {Phys. Rev. Lett.}\ }\textbf {\bibinfo {volume} {78}},\ \bibinfo
  {pages} {2220} (\bibinfo {year} {1997})}\BibitemShut {NoStop}%
\bibitem [{\citenamefont {Lomdahl}(1985)}]{Lomdahl1985}%
  \BibitemOpen
  \bibfield  {author} {\bibinfo {author} {\bibfnamefont {P.~S.}\ \bibnamefont
  {Lomdahl}},\ }\bibfield  {title} {\bibinfo {title} {Solitons in josephson
  junctions: An overview},\ }\href {https://doi.org/10.1007/BF01008351}
  {\bibfield  {journal} {\bibinfo  {journal} {Journal of Statistical Physics}\
  }\textbf {\bibinfo {volume} {39}},\ \bibinfo {pages} {551} (\bibinfo {year}
  {1985})}\BibitemShut {NoStop}%
\bibitem [{\citenamefont {Davidson}\ \emph {et~al.}(1985)\citenamefont
  {Davidson}, \citenamefont {Dueholm}, \citenamefont {Kryger},\ and\
  \citenamefont {Pedersen}}]{Davidson1985}%
  \BibitemOpen
  \bibfield  {author} {\bibinfo {author} {\bibfnamefont {A.}~\bibnamefont
  {Davidson}}, \bibinfo {author} {\bibfnamefont {B.}~\bibnamefont {Dueholm}},
  \bibinfo {author} {\bibfnamefont {B.}~\bibnamefont {Kryger}},\ and\ \bibinfo
  {author} {\bibfnamefont {N.~F.}\ \bibnamefont {Pedersen}},\ }\bibfield
  {title} {\bibinfo {title} {Experimental investigation of trapped sine-gordon
  solitons},\ }\href {https://doi.org/10.1103/PhysRevLett.55.2059} {\bibfield
  {journal} {\bibinfo  {journal} {Phys. Rev. Lett.}\ }\textbf {\bibinfo
  {volume} {55}},\ \bibinfo {pages} {2059} (\bibinfo {year}
  {1985})}\BibitemShut {NoStop}%
\bibitem [{\citenamefont {Roy}\ and\ \citenamefont {Saleur}(2019)}]{Roy2019}%
  \BibitemOpen
  \bibfield  {author} {\bibinfo {author} {\bibfnamefont {A.}~\bibnamefont
  {Roy}}\ and\ \bibinfo {author} {\bibfnamefont {H.}~\bibnamefont {Saleur}},\
  }\bibfield  {title} {\bibinfo {title} {Quantum electronic circuit simulation
  of generalized sine-gordon models},\ }\href
  {https://doi.org/10.1103/PhysRevB.100.155425} {\bibfield  {journal} {\bibinfo
   {journal} {Phys. Rev. B}\ }\textbf {\bibinfo {volume} {100}},\ \bibinfo
  {pages} {155425} (\bibinfo {year} {2019})}\BibitemShut {NoStop}%
\bibitem [{\citenamefont {Roy}\ \emph {et~al.}(2021)\citenamefont {Roy},
  \citenamefont {Schuricht}, \citenamefont {Hauschild}, \citenamefont
  {Pollmann},\ and\ \citenamefont {Saleur}}]{Roy2021}%
  \BibitemOpen
  \bibfield  {author} {\bibinfo {author} {\bibfnamefont {A.}~\bibnamefont
  {Roy}}, \bibinfo {author} {\bibfnamefont {D.}~\bibnamefont {Schuricht}},
  \bibinfo {author} {\bibfnamefont {J.}~\bibnamefont {Hauschild}}, \bibinfo
  {author} {\bibfnamefont {F.}~\bibnamefont {Pollmann}},\ and\ \bibinfo
  {author} {\bibfnamefont {H.}~\bibnamefont {Saleur}},\ }\bibfield  {title}
  {\bibinfo {title} {The quantum sine-gordon model with quantum circuits},\
  }\href {https://doi.org/https://doi.org/10.1016/j.nuclphysb.2021.115445}
  {\bibfield  {journal} {\bibinfo  {journal} {Nuclear Physics B}\ }\textbf
  {\bibinfo {volume} {968}},\ \bibinfo {pages} {115445} (\bibinfo {year}
  {2021})}\BibitemShut {NoStop}%
\bibitem [{\citenamefont {Zvyagin}(2021)}]{Zvyagin2021}%
  \BibitemOpen
  \bibfield  {author} {\bibinfo {author} {\bibfnamefont {S.}~\bibnamefont
  {Zvyagin}},\ }\bibfield  {title} {\bibinfo {title} {Spin dynamics in quantum
  sine-gordon spin chains: High-field esr studies},\ }\href
  {https://doi.org/10.1007/s00723-021-01310-9} {\bibfield  {journal} {\bibinfo
  {journal} {Applied Magnetic Resonance}\ }\textbf {\bibinfo {volume} {52}},\
  \bibinfo {pages} {337} (\bibinfo {year} {2021})}\BibitemShut {NoStop}%
\bibitem [{\citenamefont {Cuevas-Maraver}\ \emph {et~al.}(2014)\citenamefont
  {Cuevas-Maraver}, \citenamefont {Kevrekidis},\ and\ \citenamefont
  {Williams}}]{cuevassinegordon}%
  \BibitemOpen
  \bibfield  {author} {\bibinfo {author} {\bibfnamefont {J.}~\bibnamefont
  {Cuevas-Maraver}}, \bibinfo {author} {\bibfnamefont {P.~G.}\ \bibnamefont
  {Kevrekidis}},\ and\ \bibinfo {author} {\bibfnamefont {F.}~\bibnamefont
  {Williams}},\ }\href@noop {} {\emph {\bibinfo {title} {The sine-Gordon Model
  and its Applications. From Pendula and Josephson Junctions to Gravity and
  High-Energy Physics}}}\ (\bibinfo  {publisher} {Springer},\ \bibinfo {year}
  {2014})\BibitemShut {NoStop}%
\bibitem [{\citenamefont {Smirnov}(1992)}]{Smirnov1992}%
  \BibitemOpen
  \bibfield  {author} {\bibinfo {author} {\bibfnamefont {F.~A.}\ \bibnamefont
  {Smirnov}},\ }\href {https://doi.org/10.1142/1115} {\emph {\bibinfo {title}
  {Form Factors in Completely Integrable Models of Quantum Field Theory}}}\
  (\bibinfo  {publisher} {World Scientific},\ \bibinfo {year}
  {1992})\BibitemShut {NoStop}%
\bibitem [{\citenamefont {Calabrese}\ \emph {et~al.}(2016)\citenamefont
  {Calabrese}, \citenamefont {Essler},\ and\ \citenamefont
  {Mussardo}}]{Calabrese_2016}%
  \BibitemOpen
  \bibfield  {author} {\bibinfo {author} {\bibfnamefont {P.}~\bibnamefont
  {Calabrese}}, \bibinfo {author} {\bibfnamefont {F.~H.~L.}\ \bibnamefont
  {Essler}},\ and\ \bibinfo {author} {\bibfnamefont {G.}~\bibnamefont
  {Mussardo}},\ }\bibfield  {title} {\bibinfo {title} {Introduction to
  ‘quantum integrability in out of equilibrium systems’},\ }\href
  {https://doi.org/10.1088/1742-5468/2016/06/064001} {\bibfield  {journal}
  {\bibinfo  {journal} {Journal of Statistical Mechanics: Theory and
  Experiment}\ }\textbf {\bibinfo {volume} {2016}},\ \bibinfo {pages} {064001}
  (\bibinfo {year} {2016})}\BibitemShut {NoStop}%
\bibitem [{\citenamefont {Bertini}\ \emph {et~al.}(2021)\citenamefont
  {Bertini}, \citenamefont {Heidrich-Meisner}, \citenamefont {Karrasch},
  \citenamefont {Prosen}, \citenamefont {Steinigeweg},\ and\ \citenamefont
  {\ifmmode \check{Z}\else \v{Z}\fi{}nidari\ifmmode~\check{c}\else
  \v{c}\fi{}}}]{Bertini2021}%
  \BibitemOpen
  \bibfield  {author} {\bibinfo {author} {\bibfnamefont {B.}~\bibnamefont
  {Bertini}}, \bibinfo {author} {\bibfnamefont {F.}~\bibnamefont
  {Heidrich-Meisner}}, \bibinfo {author} {\bibfnamefont {C.}~\bibnamefont
  {Karrasch}}, \bibinfo {author} {\bibfnamefont {T.}~\bibnamefont {Prosen}},
  \bibinfo {author} {\bibfnamefont {R.}~\bibnamefont {Steinigeweg}},\ and\
  \bibinfo {author} {\bibfnamefont {M.}~\bibnamefont {\ifmmode \check{Z}\else
  \v{Z}\fi{}nidari\ifmmode~\check{c}\else \v{c}\fi{}}},\ }\bibfield  {title}
  {\bibinfo {title} {Finite-temperature transport in one-dimensional quantum
  lattice models},\ }\href {https://doi.org/10.1103/RevModPhys.93.025003}
  {\bibfield  {journal} {\bibinfo  {journal} {Rev. Mod. Phys.}\ }\textbf
  {\bibinfo {volume} {93}},\ \bibinfo {pages} {025003} (\bibinfo {year}
  {2021})}\BibitemShut {NoStop}%
\bibitem [{\citenamefont {Bastianello}\ \emph {et~al.}(2022)\citenamefont
  {Bastianello}, \citenamefont {Bertini}, \citenamefont {Doyon},\ and\
  \citenamefont {Vasseur}}]{specialissueGHD}%
  \BibitemOpen
  \bibfield  {author} {\bibinfo {author} {\bibfnamefont {A.}~\bibnamefont
  {Bastianello}}, \bibinfo {author} {\bibfnamefont {B.}~\bibnamefont
  {Bertini}}, \bibinfo {author} {\bibfnamefont {B.}~\bibnamefont {Doyon}},\
  and\ \bibinfo {author} {\bibfnamefont {R.}~\bibnamefont {Vasseur}},\
  }\bibfield  {title} {\bibinfo {title} {Introduction to the special issue on
  emergent hydrodynamics in integrable many-body systems},\ }\href
  {https://doi.org/10.1088/1742-5468/ac3e6a} {\bibfield  {journal} {\bibinfo
  {journal} {Journal of Statistical Mechanics: Theory and Experiment}\ }\textbf
  {\bibinfo {volume} {2022}},\ \bibinfo {pages} {014001} (\bibinfo {year}
  {2022})}\BibitemShut {NoStop}%
\bibitem [{\citenamefont {Zvyagin}\ \emph {et~al.}(2004)\citenamefont
  {Zvyagin}, \citenamefont {Kolezhuk}, \citenamefont {Krzystek},\ and\
  \citenamefont {Feyerherm}}]{Zvyagin2004}%
  \BibitemOpen
  \bibfield  {author} {\bibinfo {author} {\bibfnamefont {S.~A.}\ \bibnamefont
  {Zvyagin}}, \bibinfo {author} {\bibfnamefont {A.~K.}\ \bibnamefont
  {Kolezhuk}}, \bibinfo {author} {\bibfnamefont {J.}~\bibnamefont {Krzystek}},\
  and\ \bibinfo {author} {\bibfnamefont {R.}~\bibnamefont {Feyerherm}},\
  }\bibfield  {title} {\bibinfo {title} {Excitation hierarchy of the quantum
  sine-gordon spin chain in a strong magnetic field},\ }\href
  {https://doi.org/10.1103/PhysRevLett.93.027201} {\bibfield  {journal}
  {\bibinfo  {journal} {Phys. Rev. Lett.}\ }\textbf {\bibinfo {volume} {93}},\
  \bibinfo {pages} {027201} (\bibinfo {year} {2004})}\BibitemShut {NoStop}%
\bibitem [{\citenamefont {Umegaki}\ \emph {et~al.}(2009)\citenamefont
  {Umegaki}, \citenamefont {Tanaka}, \citenamefont {Ono}, \citenamefont
  {Uekusa},\ and\ \citenamefont {Nojiri}}]{Umegaki2009}%
  \BibitemOpen
  \bibfield  {author} {\bibinfo {author} {\bibfnamefont {I.}~\bibnamefont
  {Umegaki}}, \bibinfo {author} {\bibfnamefont {H.}~\bibnamefont {Tanaka}},
  \bibinfo {author} {\bibfnamefont {T.}~\bibnamefont {Ono}}, \bibinfo {author}
  {\bibfnamefont {H.}~\bibnamefont {Uekusa}},\ and\ \bibinfo {author}
  {\bibfnamefont {H.}~\bibnamefont {Nojiri}},\ }\bibfield  {title} {\bibinfo
  {title} {Elementary excitations of the $s=\frac{1}{2}$ one-dimensional
  antiferromagnet ${\text{kcugaf}}_{6}$ in a magnetic field and quantum
  sine-gordon model},\ }\href {https://doi.org/10.1103/PhysRevB.79.184401}
  {\bibfield  {journal} {\bibinfo  {journal} {Phys. Rev. B}\ }\textbf {\bibinfo
  {volume} {79}},\ \bibinfo {pages} {184401} (\bibinfo {year}
  {2009})}\BibitemShut {NoStop}%
\bibitem [{\citenamefont {Essler}\ and\ \citenamefont
  {Tsvelik}(1998)}]{Essler1998}%
  \BibitemOpen
  \bibfield  {author} {\bibinfo {author} {\bibfnamefont {F.~H.~L.}\
  \bibnamefont {Essler}}\ and\ \bibinfo {author} {\bibfnamefont {A.~M.}\
  \bibnamefont {Tsvelik}},\ }\bibfield  {title} {\bibinfo {title} {Dynamical
  magnetic susceptibilities in copper benzoate},\ }\href
  {https://doi.org/10.1103/PhysRevB.57.10592} {\bibfield  {journal} {\bibinfo
  {journal} {Phys. Rev. B}\ }\textbf {\bibinfo {volume} {57}},\ \bibinfo
  {pages} {10592} (\bibinfo {year} {1998})}\BibitemShut {NoStop}%
\bibitem [{\citenamefont {E\ss{}ler}(1999)}]{Essler1999}%
  \BibitemOpen
  \bibfield  {author} {\bibinfo {author} {\bibfnamefont {F.~H.~L.}\
  \bibnamefont {E\ss{}ler}},\ }\bibfield  {title} {\bibinfo {title}
  {Sine-gordon low-energy effective theory for copper benzoate},\ }\href
  {https://doi.org/10.1103/PhysRevB.59.14376} {\bibfield  {journal} {\bibinfo
  {journal} {Phys. Rev. B}\ }\textbf {\bibinfo {volume} {59}},\ \bibinfo
  {pages} {14376} (\bibinfo {year} {1999})}\BibitemShut {NoStop}%
\bibitem [{\citenamefont {Essler}\ and\ \citenamefont
  {Konik}(2005)}]{Essler2005}%
  \BibitemOpen
  \bibfield  {author} {\bibinfo {author} {\bibfnamefont {F.~H.}\ \bibnamefont
  {Essler}}\ and\ \bibinfo {author} {\bibfnamefont {R.~M.}\ \bibnamefont
  {Konik}},\ }\bibinfo {title} {Application of massive integrable quantum field
  theories to problems in condensed matter physics}\ (\bibinfo  {publisher}
  {World Scientific},\ \bibinfo {year} {2005})\ pp.\ \bibinfo {pages}
  {684--830},\ \bibinfo {note} {0}\BibitemShut {NoStop}%
\bibitem [{\citenamefont {Gritsev}\ \emph
  {et~al.}(2007{\natexlab{a}})\citenamefont {Gritsev}, \citenamefont
  {Polkovnikov},\ and\ \citenamefont {Demler}}]{Gritsev2007}%
  \BibitemOpen
  \bibfield  {author} {\bibinfo {author} {\bibfnamefont {V.}~\bibnamefont
  {Gritsev}}, \bibinfo {author} {\bibfnamefont {A.}~\bibnamefont
  {Polkovnikov}},\ and\ \bibinfo {author} {\bibfnamefont {E.}~\bibnamefont
  {Demler}},\ }\bibfield  {title} {\bibinfo {title} {Linear response theory for
  a pair of coupled one-dimensional condensates of interacting atoms},\ }\href
  {https://doi.org/10.1103/PhysRevB.75.174511} {\bibfield  {journal} {\bibinfo
  {journal} {Phys. Rev. B}\ }\textbf {\bibinfo {volume} {75}},\ \bibinfo
  {pages} {174511} (\bibinfo {year} {2007}{\natexlab{a}})}\BibitemShut
  {NoStop}%
\bibitem [{\citenamefont {Gritsev}\ \emph
  {et~al.}(2007{\natexlab{b}})\citenamefont {Gritsev}, \citenamefont {Demler},
  \citenamefont {Lukin},\ and\ \citenamefont {Polkovnikov}}]{Gritsev2007a}%
  \BibitemOpen
  \bibfield  {author} {\bibinfo {author} {\bibfnamefont {V.}~\bibnamefont
  {Gritsev}}, \bibinfo {author} {\bibfnamefont {E.}~\bibnamefont {Demler}},
  \bibinfo {author} {\bibfnamefont {M.}~\bibnamefont {Lukin}},\ and\ \bibinfo
  {author} {\bibfnamefont {A.}~\bibnamefont {Polkovnikov}},\ }\bibfield
  {title} {\bibinfo {title} {Spectroscopy of collective excitations in
  interacting low-dimensional many-body systems using quench dynamics},\ }\href
  {https://doi.org/10.1103/PhysRevLett.99.200404} {\bibfield  {journal}
  {\bibinfo  {journal} {Phys. Rev. Lett.}\ }\textbf {\bibinfo {volume} {99}},\
  \bibinfo {pages} {200404} (\bibinfo {year} {2007}{\natexlab{b}})}\BibitemShut
  {NoStop}%
\bibitem [{\citenamefont {Schweigler}\ \emph {et~al.}(2017)\citenamefont
  {Schweigler}, \citenamefont {Kasper}, \citenamefont {Erne}, \citenamefont
  {Mazets}, \citenamefont {Rauer}, \citenamefont {Cataldini}, \citenamefont
  {Langen}, \citenamefont {Gasenzer}, \citenamefont {Berges},\ and\
  \citenamefont {Schmiedmayer}}]{Schweigler2017}%
  \BibitemOpen
  \bibfield  {author} {\bibinfo {author} {\bibfnamefont {T.}~\bibnamefont
  {Schweigler}}, \bibinfo {author} {\bibfnamefont {V.}~\bibnamefont {Kasper}},
  \bibinfo {author} {\bibfnamefont {S.}~\bibnamefont {Erne}}, \bibinfo {author}
  {\bibfnamefont {I.}~\bibnamefont {Mazets}}, \bibinfo {author} {\bibfnamefont
  {B.}~\bibnamefont {Rauer}}, \bibinfo {author} {\bibfnamefont
  {F.}~\bibnamefont {Cataldini}}, \bibinfo {author} {\bibfnamefont
  {T.}~\bibnamefont {Langen}}, \bibinfo {author} {\bibfnamefont
  {T.}~\bibnamefont {Gasenzer}}, \bibinfo {author} {\bibfnamefont
  {J.}~\bibnamefont {Berges}},\ and\ \bibinfo {author} {\bibfnamefont
  {J.}~\bibnamefont {Schmiedmayer}},\ }\bibfield  {title} {\bibinfo {title}
  {Experimental characterization of a quantum many-body system via higher-order
  correlations},\ }\href {https://doi.org/10.1038/nature22310} {\bibfield
  {journal} {\bibinfo  {journal} {Nature}\ }\textbf {\bibinfo {volume} {545}},\
  \bibinfo {pages} {323} (\bibinfo {year} {2017})}\BibitemShut {NoStop}%
\bibitem [{\citenamefont {Zache}\ \emph {et~al.}(2020)\citenamefont {Zache},
  \citenamefont {Schweigler}, \citenamefont {Erne}, \citenamefont
  {Schmiedmayer},\ and\ \citenamefont {Berges}}]{Zache2020}%
  \BibitemOpen
  \bibfield  {author} {\bibinfo {author} {\bibfnamefont {T.~V.}\ \bibnamefont
  {Zache}}, \bibinfo {author} {\bibfnamefont {T.}~\bibnamefont {Schweigler}},
  \bibinfo {author} {\bibfnamefont {S.}~\bibnamefont {Erne}}, \bibinfo {author}
  {\bibfnamefont {J.}~\bibnamefont {Schmiedmayer}},\ and\ \bibinfo {author}
  {\bibfnamefont {J.}~\bibnamefont {Berges}},\ }\bibfield  {title} {\bibinfo
  {title} {Extracting the field theory description of a quantum many-body
  system from experimental data},\ }\href
  {https://doi.org/10.1103/PhysRevX.10.011020} {\bibfield  {journal} {\bibinfo
  {journal} {Phys. Rev. X}\ }\textbf {\bibinfo {volume} {10}},\ \bibinfo
  {pages} {011020} (\bibinfo {year} {2020})}\BibitemShut {NoStop}%
\bibitem [{\citenamefont {Pigneur}\ \emph {et~al.}(2018)\citenamefont
  {Pigneur}, \citenamefont {Berrada}, \citenamefont {Bonneau}, \citenamefont
  {Schumm}, \citenamefont {Demler},\ and\ \citenamefont
  {Schmiedmayer}}]{Pigneur2018}%
  \BibitemOpen
  \bibfield  {author} {\bibinfo {author} {\bibfnamefont {M.}~\bibnamefont
  {Pigneur}}, \bibinfo {author} {\bibfnamefont {T.}~\bibnamefont {Berrada}},
  \bibinfo {author} {\bibfnamefont {M.}~\bibnamefont {Bonneau}}, \bibinfo
  {author} {\bibfnamefont {T.}~\bibnamefont {Schumm}}, \bibinfo {author}
  {\bibfnamefont {E.}~\bibnamefont {Demler}},\ and\ \bibinfo {author}
  {\bibfnamefont {J.}~\bibnamefont {Schmiedmayer}},\ }\bibfield  {title}
  {\bibinfo {title} {Relaxation to a phase-locked equilibrium state in a
  one-dimensional bosonic josephson junction},\ }\href
  {https://doi.org/10.1103/PhysRevLett.120.173601} {\bibfield  {journal}
  {\bibinfo  {journal} {Phys. Rev. Lett.}\ }\textbf {\bibinfo {volume} {120}},\
  \bibinfo {pages} {173601} (\bibinfo {year} {2018})}\BibitemShut {NoStop}%
\bibitem [{\citenamefont {Damle}\ and\ \citenamefont
  {Sachdev}(2005)}]{DamleSachdev2005}%
  \BibitemOpen
  \bibfield  {author} {\bibinfo {author} {\bibfnamefont {K.}~\bibnamefont
  {Damle}}\ and\ \bibinfo {author} {\bibfnamefont {S.}~\bibnamefont
  {Sachdev}},\ }\bibfield  {title} {\bibinfo {title} {Universal relaxational
  dynamics of gapped one-dimensional models in the quantum sine-gordon
  universality class},\ }\href {https://doi.org/10.1103/PhysRevLett.95.187201}
  {\bibfield  {journal} {\bibinfo  {journal} {Phys. Rev. Lett.}\ }\textbf
  {\bibinfo {volume} {95}},\ \bibinfo {pages} {187201} (\bibinfo {year}
  {2005})}\BibitemShut {NoStop}%
\bibitem [{\citenamefont {Kormos}\ and\ \citenamefont
  {Zar\'and}(2016)}]{Kormos2016}%
  \BibitemOpen
  \bibfield  {author} {\bibinfo {author} {\bibfnamefont {M.}~\bibnamefont
  {Kormos}}\ and\ \bibinfo {author} {\bibfnamefont {G.}~\bibnamefont
  {Zar\'and}},\ }\bibfield  {title} {\bibinfo {title} {Quantum quenches in the
  sine-gordon model: A semiclassical approach},\ }\href
  {https://doi.org/10.1103/PhysRevE.93.062101} {\bibfield  {journal} {\bibinfo
  {journal} {Phys. Rev. E}\ }\textbf {\bibinfo {volume} {93}},\ \bibinfo
  {pages} {062101} (\bibinfo {year} {2016})}\BibitemShut {NoStop}%
\bibitem [{\citenamefont {Moca}\ \emph {et~al.}(2017)\citenamefont {Moca},
  \citenamefont {Kormos},\ and\ \citenamefont {Zar\'and}}]{Moca2019}%
  \BibitemOpen
  \bibfield  {author} {\bibinfo {author} {\bibfnamefont {C.~P.}\ \bibnamefont
  {Moca}}, \bibinfo {author} {\bibfnamefont {M.}~\bibnamefont {Kormos}},\ and\
  \bibinfo {author} {\bibfnamefont {G.}~\bibnamefont {Zar\'and}},\ }\bibfield
  {title} {\bibinfo {title} {Hybrid semiclassical theory of quantum quenches in
  one-dimensional systems},\ }\href
  {https://doi.org/10.1103/PhysRevLett.119.100603} {\bibfield  {journal}
  {\bibinfo  {journal} {Phys. Rev. Lett.}\ }\textbf {\bibinfo {volume} {119}},\
  \bibinfo {pages} {100603} (\bibinfo {year} {2017})}\BibitemShut {NoStop}%
\bibitem [{\citenamefont {Bertini}\ \emph {et~al.}(2019)\citenamefont
  {Bertini}, \citenamefont {Piroli},\ and\ \citenamefont
  {Kormos}}]{Bertini2019}%
  \BibitemOpen
  \bibfield  {author} {\bibinfo {author} {\bibfnamefont {B.}~\bibnamefont
  {Bertini}}, \bibinfo {author} {\bibfnamefont {L.}~\bibnamefont {Piroli}},\
  and\ \bibinfo {author} {\bibfnamefont {M.}~\bibnamefont {Kormos}},\
  }\bibfield  {title} {\bibinfo {title} {Transport in the sine-gordon field
  theory: From generalized hydrodynamics to semiclassics},\ }\href
  {https://doi.org/10.1103/PhysRevB.100.035108} {\bibfield  {journal} {\bibinfo
   {journal} {Phys. Rev. B}\ }\textbf {\bibinfo {volume} {100}},\ \bibinfo
  {pages} {035108} (\bibinfo {year} {2019})}\BibitemShut {NoStop}%
\bibitem [{\citenamefont {Kormos}\ \emph {et~al.}(2022)\citenamefont {Kormos},
  \citenamefont {V\"or\"os},\ and\ \citenamefont
  {Zar\'and}}]{PhysRevB.106.205151}%
  \BibitemOpen
  \bibfield  {author} {\bibinfo {author} {\bibfnamefont {M.}~\bibnamefont
  {Kormos}}, \bibinfo {author} {\bibfnamefont {D.}~\bibnamefont {V\"or\"os}},\
  and\ \bibinfo {author} {\bibfnamefont {G.}~\bibnamefont {Zar\'and}},\
  }\bibfield  {title} {\bibinfo {title} {Finite-temperature dynamics in gapped
  one-dimensional models in the sine-gordon family},\ }\href
  {https://doi.org/10.1103/PhysRevB.106.205151} {\bibfield  {journal} {\bibinfo
   {journal} {Phys. Rev. B}\ }\textbf {\bibinfo {volume} {106}},\ \bibinfo
  {pages} {205151} (\bibinfo {year} {2022})}\BibitemShut {NoStop}%
\bibitem [{\citenamefont {Faddeev}\ and\ \citenamefont
  {Takhtajan}(1987)}]{Faddeev:1987ph}%
  \BibitemOpen
  \bibfield  {author} {\bibinfo {author} {\bibfnamefont {L.~D.}\ \bibnamefont
  {Faddeev}}\ and\ \bibinfo {author} {\bibfnamefont {L.~A.}\ \bibnamefont
  {Takhtajan}},\ }\href@noop {} {\emph {\bibinfo {title} {{Hamltonian methods
  in the theory of solitons}}}}\ (\bibinfo {year} {1987})\BibitemShut {NoStop}%
\bibitem [{\citenamefont {Zamolodchikov}\ and\ \citenamefont
  {Zamolodchikov}(1979)}]{ZAMOLODCHIKOV1979253}%
  \BibitemOpen
  \bibfield  {author} {\bibinfo {author} {\bibfnamefont {A.~B.}\ \bibnamefont
  {Zamolodchikov}}\ and\ \bibinfo {author} {\bibfnamefont {A.~B.}\ \bibnamefont
  {Zamolodchikov}},\ }\bibfield  {title} {\bibinfo {title} {Factorized
  s-matrices in two dimensions as the exact solutions of certain relativistic
  quantum field theory models},\ }\href
  {https://doi.org/https://doi.org/10.1016/0003-4916(79)90391-9} {\bibfield
  {journal} {\bibinfo  {journal} {Annals of Physics}\ }\textbf {\bibinfo
  {volume} {120}},\ \bibinfo {pages} {253} (\bibinfo {year}
  {1979})}\BibitemShut {NoStop}%
\bibitem [{\citenamefont {Zamolodchikov}(1995)}]{ZAMOLODCHIKOV1995}%
  \BibitemOpen
  \bibfield  {author} {\bibinfo {author} {\bibfnamefont {A.~B.}\ \bibnamefont
  {Zamolodchikov}},\ }\bibfield  {title} {\bibinfo {title} {Mass scale in the
  sine-gordon model and its reductions},\ }\href
  {https://doi.org/10.1142/S0217751X9500053X} {\bibfield  {journal} {\bibinfo
  {journal} {International Journal of Modern Physics A}\ }\textbf {\bibinfo
  {volume} {10}},\ \bibinfo {pages} {1125} (\bibinfo {year} {1995})},\ \Eprint
  {https://arxiv.org/abs/https://doi.org/10.1142/S0217751X9500053X}
  {https://doi.org/10.1142/S0217751X9500053X} \BibitemShut {NoStop}%
\bibitem [{\citenamefont {Koch}\ and\ \citenamefont
  {Bastianello}(2023)}]{koch2023exact}%
  \BibitemOpen
  \bibfield  {author} {\bibinfo {author} {\bibfnamefont {R.}~\bibnamefont
  {Koch}}\ and\ \bibinfo {author} {\bibfnamefont {A.}~\bibnamefont
  {Bastianello}},\ }\bibfield  {title} {\bibinfo {title} {{Exact thermodynamics
  and transport in the classical sine-Gordon model}},\ }\href
  {https://doi.org/10.21468/SciPostPhys.15.4.140} {\bibfield  {journal}
  {\bibinfo  {journal} {SciPost Phys.}\ }\textbf {\bibinfo {volume} {15}},\
  \bibinfo {pages} {140} (\bibinfo {year} {2023})}\BibitemShut {NoStop}%
\bibitem [{\citenamefont {Vlijm}\ \emph {et~al.}(2015)\citenamefont {Vlijm},
  \citenamefont {Ganahl}, \citenamefont {Fioretto}, \citenamefont {Brockmann},
  \citenamefont {Haque}, \citenamefont {Evertz},\ and\ \citenamefont
  {Caux}}]{PhysRevB.92.214427}%
  \BibitemOpen
  \bibfield  {author} {\bibinfo {author} {\bibfnamefont {R.}~\bibnamefont
  {Vlijm}}, \bibinfo {author} {\bibfnamefont {M.}~\bibnamefont {Ganahl}},
  \bibinfo {author} {\bibfnamefont {D.}~\bibnamefont {Fioretto}}, \bibinfo
  {author} {\bibfnamefont {M.}~\bibnamefont {Brockmann}}, \bibinfo {author}
  {\bibfnamefont {M.}~\bibnamefont {Haque}}, \bibinfo {author} {\bibfnamefont
  {H.~G.}\ \bibnamefont {Evertz}},\ and\ \bibinfo {author} {\bibfnamefont
  {J.-S.}\ \bibnamefont {Caux}},\ }\bibfield  {title} {\bibinfo {title}
  {Quasi-soliton scattering in quantum spin chains},\ }\href
  {https://doi.org/10.1103/PhysRevB.92.214427} {\bibfield  {journal} {\bibinfo
  {journal} {Phys. Rev. B}\ }\textbf {\bibinfo {volume} {92}},\ \bibinfo
  {pages} {214427} (\bibinfo {year} {2015})}\BibitemShut {NoStop}%
\bibitem [{\citenamefont {Doyon}\ \emph {et~al.}(2018)\citenamefont {Doyon},
  \citenamefont {Yoshimura},\ and\ \citenamefont {Caux}}]{Doyon2018}%
  \BibitemOpen
  \bibfield  {author} {\bibinfo {author} {\bibfnamefont {B.}~\bibnamefont
  {Doyon}}, \bibinfo {author} {\bibfnamefont {T.}~\bibnamefont {Yoshimura}},\
  and\ \bibinfo {author} {\bibfnamefont {J.-S.}\ \bibnamefont {Caux}},\
  }\bibfield  {title} {\bibinfo {title} {Soliton gases and generalized
  hydrodynamics},\ }\href {https://doi.org/10.1103/PhysRevLett.120.045301}
  {\bibfield  {journal} {\bibinfo  {journal} {Phys. Rev. Lett.}\ }\textbf
  {\bibinfo {volume} {120}},\ \bibinfo {pages} {045301} (\bibinfo {year}
  {2018})}\BibitemShut {NoStop}%
\bibitem [{\citenamefont {Yang}\ and\ \citenamefont
  {Yang}(2003)}]{10.1063/1.1664947}%
  \BibitemOpen
  \bibfield  {author} {\bibinfo {author} {\bibfnamefont {C.~N.}\ \bibnamefont
  {Yang}}\ and\ \bibinfo {author} {\bibfnamefont {C.~P.}\ \bibnamefont
  {Yang}},\ }\bibfield  {title} {\bibinfo {title} {{Thermodynamics of a
  One‐Dimensional System of Bosons with Repulsive Delta‐Function
  Interaction}},\ }\href {https://doi.org/10.1063/1.1664947} {\bibfield
  {journal} {\bibinfo  {journal} {Journal of Mathematical Physics}\ }\textbf
  {\bibinfo {volume} {10}},\ \bibinfo {pages} {1115} (\bibinfo {year}
  {2003})},\ \Eprint
  {https://arxiv.org/abs/https://pubs.aip.org/aip/jmp/article-pdf/10/7/1115/8144272/1115\_1\_online.pdf}
  {https://pubs.aip.org/aip/jmp/article-pdf/10/7/1115/8144272/1115\_1\_online.pdf}
  \BibitemShut {NoStop}%
\bibitem [{\citenamefont {Zamolodchikov}(1990)}]{ZAMOLODCHIKOV1990695}%
  \BibitemOpen
  \bibfield  {author} {\bibinfo {author} {\bibfnamefont {A.}~\bibnamefont
  {Zamolodchikov}},\ }\bibfield  {title} {\bibinfo {title} {Thermodynamic bethe
  ansatz in relativistic models: Scaling 3-state potts and lee-yang models},\
  }\href {https://doi.org/https://doi.org/10.1016/0550-3213(90)90333-9}
  {\bibfield  {journal} {\bibinfo  {journal} {Nuclear Physics B}\ }\textbf
  {\bibinfo {volume} {342}},\ \bibinfo {pages} {695} (\bibinfo {year}
  {1990})}\BibitemShut {NoStop}%
\bibitem [{\citenamefont {Takahashi}(2005)}]{takahashi2005thermodynamics}%
  \BibitemOpen
  \bibfield  {author} {\bibinfo {author} {\bibfnamefont {M.}~\bibnamefont
  {Takahashi}},\ }\href@noop {} {\emph {\bibinfo {title} {Thermodynamics of
  one-dimensional solvable models}}}\ (\bibinfo  {publisher} {Cambridge
  University Press},\ \bibinfo {year} {2005})\BibitemShut {NoStop}%
\bibitem [{\citenamefont {Mossel}\ and\ \citenamefont
  {Caux}(2012)}]{Mossel_2012}%
  \BibitemOpen
  \bibfield  {author} {\bibinfo {author} {\bibfnamefont {J.}~\bibnamefont
  {Mossel}}\ and\ \bibinfo {author} {\bibfnamefont {J.-S.}\ \bibnamefont
  {Caux}},\ }\bibfield  {title} {\bibinfo {title} {Generalized tba and
  generalized gibbs},\ }\href {https://doi.org/10.1088/1751-8113/45/25/255001}
  {\bibfield  {journal} {\bibinfo  {journal} {Journal of Physics A:
  Mathematical and Theoretical}\ }\textbf {\bibinfo {volume} {45}},\ \bibinfo
  {pages} {255001} (\bibinfo {year} {2012})}\BibitemShut {NoStop}%
\bibitem [{\citenamefont {Caux}\ and\ \citenamefont {Essler}(2013)}]{caux2013}%
  \BibitemOpen
  \bibfield  {author} {\bibinfo {author} {\bibfnamefont {J.-S.}\ \bibnamefont
  {Caux}}\ and\ \bibinfo {author} {\bibfnamefont {F.~H.~L.}\ \bibnamefont
  {Essler}},\ }\bibfield  {title} {\bibinfo {title} {Time evolution of local
  observables after quenching to an integrable model},\ }\href
  {https://doi.org/10.1103/PhysRevLett.110.257203} {\bibfield  {journal}
  {\bibinfo  {journal} {Phys. Rev. Lett.}\ }\textbf {\bibinfo {volume} {110}},\
  \bibinfo {pages} {257203} (\bibinfo {year} {2013})}\BibitemShut {NoStop}%
\bibitem [{\citenamefont {Caux}(2016)}]{caux2016}%
  \BibitemOpen
  \bibfield  {author} {\bibinfo {author} {\bibfnamefont {J.-S.}\ \bibnamefont
  {Caux}},\ }\bibfield  {title} {\bibinfo {title} {The quench action},\ }\href
  {https://doi.org/10.1088/1742-5468/2016/06/064006} {\bibfield  {journal}
  {\bibinfo  {journal} {Journal of Statistical Mechanics: Theory and
  Experiment}\ }\textbf {\bibinfo {volume} {2016}},\ \bibinfo {pages} {064006}
  (\bibinfo {year} {2016})}\BibitemShut {NoStop}%
\bibitem [{\citenamefont {Castro-Alvaredo}\ \emph {et~al.}(2016)\citenamefont
  {Castro-Alvaredo}, \citenamefont {Doyon},\ and\ \citenamefont
  {Yoshimura}}]{Doyon2016}%
  \BibitemOpen
  \bibfield  {author} {\bibinfo {author} {\bibfnamefont {O.~A.}\ \bibnamefont
  {Castro-Alvaredo}}, \bibinfo {author} {\bibfnamefont {B.}~\bibnamefont
  {Doyon}},\ and\ \bibinfo {author} {\bibfnamefont {T.}~\bibnamefont
  {Yoshimura}},\ }\bibfield  {title} {\bibinfo {title} {Emergent hydrodynamics
  in integrable quantum systems out of equilibrium},\ }\href
  {https://doi.org/10.1103/PhysRevX.6.041065} {\bibfield  {journal} {\bibinfo
  {journal} {Phys. Rev. X}\ }\textbf {\bibinfo {volume} {6}},\ \bibinfo {pages}
  {041065} (\bibinfo {year} {2016})}\BibitemShut {NoStop}%
\bibitem [{\citenamefont {Bertini}\ \emph {et~al.}(2016)\citenamefont
  {Bertini}, \citenamefont {Collura}, \citenamefont {De~Nardis},\ and\
  \citenamefont {Fagotti}}]{Bertini2016}%
  \BibitemOpen
  \bibfield  {author} {\bibinfo {author} {\bibfnamefont {B.}~\bibnamefont
  {Bertini}}, \bibinfo {author} {\bibfnamefont {M.}~\bibnamefont {Collura}},
  \bibinfo {author} {\bibfnamefont {J.}~\bibnamefont {De~Nardis}},\ and\
  \bibinfo {author} {\bibfnamefont {M.}~\bibnamefont {Fagotti}},\ }\bibfield
  {title} {\bibinfo {title} {Transport in out-of-equilibrium $xxz$ chains:
  Exact profiles of charges and currents},\ }\href
  {https://doi.org/10.1103/PhysRevLett.117.207201} {\bibfield  {journal}
  {\bibinfo  {journal} {Phys. Rev. Lett.}\ }\textbf {\bibinfo {volume} {117}},\
  \bibinfo {pages} {207201} (\bibinfo {year} {2016})}\BibitemShut {NoStop}%
\bibitem [{\citenamefont {Doyon}(2020)}]{Doyon2020Notes}%
  \BibitemOpen
  \bibfield  {author} {\bibinfo {author} {\bibfnamefont {B.}~\bibnamefont
  {Doyon}},\ }\bibfield  {title} {\bibinfo {title} {{Lecture notes on
  Generalised Hydrodynamics}},\ }\href
  {https://doi.org/10.21468/SciPostPhysLectNotes.18} {\bibfield  {journal}
  {\bibinfo  {journal} {SciPost Phys. Lect. Notes}\ ,\ \bibinfo {pages} {18}}
  (\bibinfo {year} {2020})}\BibitemShut {NoStop}%
\bibitem [{\citenamefont {Nagy}\ \emph {et~al.}(2023)\citenamefont {Nagy},
  \citenamefont {Takács},\ and\ \citenamefont {Kormos}}]{Nagy_2023}%
  \BibitemOpen
  \bibfield  {author} {\bibinfo {author} {\bibfnamefont {B.~C.}\ \bibnamefont
  {Nagy}}, \bibinfo {author} {\bibfnamefont {G.}~\bibnamefont {Takács}},\ and\
  \bibinfo {author} {\bibfnamefont {M.}~\bibnamefont {Kormos}},\ }\href@noop {}
  {\bibinfo {title} {Thermodynamic bethe ansatz and generalised hydrodynamics
  in the sine-gordon model}} (\bibinfo {year} {2023}),\ \Eprint
  {https://arxiv.org/abs/2312.03909} {arXiv:2312.03909 [cond-mat.str-el]}
  \BibitemShut {NoStop}%
\bibitem [{Note1()}]{Note1}%
  \BibitemOpen
  \bibinfo {note} {See supplementary material for \protect \emph {(i)} summary
  of SY approach, \protect \emph {(ii)} overview of the Ballistic Fluctuation
  Theory, \protect \emph {(iii)} details on the sine-Gordon thermodynamics, and
  \protect \emph {(iv)} details on coupled-condensates experiment.}\BibitemShut
  {Stop}%
\bibitem [{\citenamefont {Myers}\ \emph {et~al.}(2020)\citenamefont {Myers},
  \citenamefont {Bhaseen}, \citenamefont {Harris},\ and\ \citenamefont
  {Doyon}}]{10.21468/SciPostPhys.8.1.007}%
  \BibitemOpen
  \bibfield  {author} {\bibinfo {author} {\bibfnamefont {J.}~\bibnamefont
  {Myers}}, \bibinfo {author} {\bibfnamefont {M.~J.}\ \bibnamefont {Bhaseen}},
  \bibinfo {author} {\bibfnamefont {R.~J.}\ \bibnamefont {Harris}},\ and\
  \bibinfo {author} {\bibfnamefont {B.}~\bibnamefont {Doyon}},\ }\bibfield
  {title} {\bibinfo {title} {{Transport fluctuations in integrable models out
  of equilibrium}},\ }\href {https://doi.org/10.21468/SciPostPhys.8.1.007}
  {\bibfield  {journal} {\bibinfo  {journal} {SciPost Phys.}\ }\textbf
  {\bibinfo {volume} {8}},\ \bibinfo {pages} {007} (\bibinfo {year}
  {2020})}\BibitemShut {NoStop}%
\bibitem [{\citenamefont {Doyon}\ and\ \citenamefont
  {Myers}(2020)}]{Doyon2020}%
  \BibitemOpen
  \bibfield  {author} {\bibinfo {author} {\bibfnamefont {B.}~\bibnamefont
  {Doyon}}\ and\ \bibinfo {author} {\bibfnamefont {J.}~\bibnamefont {Myers}},\
  }\bibfield  {title} {\bibinfo {title} {Fluctuations in ballistic transport
  from euler hydrodynamics},\ }\href
  {https://doi.org/10.1007/s00023-019-00860-w} {\bibfield  {journal} {\bibinfo
  {journal} {Annales Henri Poincar{\'e}}\ }\textbf {\bibinfo {volume} {21}},\
  \bibinfo {pages} {255} (\bibinfo {year} {2020})}\BibitemShut {NoStop}%
\bibitem [{\citenamefont {Schollw\"ock}(2011)}]{Schollwock2011}%
  \BibitemOpen
  \bibfield  {author} {\bibinfo {author} {\bibfnamefont {U.}~\bibnamefont
  {Schollw\"ock}},\ }\bibfield  {title} {\bibinfo {title} {The density-matrix
  renormalization group in the age of matrix product states},\ }\href
  {https://doi.org/https://doi.org/10.1016/j.aop.2010.09.012} {\bibfield
  {journal} {\bibinfo  {journal} {Annals of Physics}\ }\textbf {\bibinfo
  {volume} {326}},\ \bibinfo {pages} {96} (\bibinfo {year} {2011})},\ \bibinfo
  {note} {january 2011 Special Issue}\BibitemShut {NoStop}%
\bibitem [{\citenamefont {Luca}\ and\ \citenamefont
  {Giuseppe}(2016)}]{DeLuca2016}%
  \BibitemOpen
  \bibfield  {author} {\bibinfo {author} {\bibfnamefont {A.~D.}\ \bibnamefont
  {Luca}}\ and\ \bibinfo {author} {\bibnamefont {Giuseppe}},\ }\bibfield
  {title} {\bibinfo {title} {Equilibration properties of classical integrable
  field theories},\ }\href {https://doi.org/10.1088/1742-5468/2016/06/064011}
  {\bibfield  {journal} {\bibinfo  {journal} {Journal of Statistical Mechanics:
  Theory and Experiment}\ }\textbf {\bibinfo {volume} {2016}},\ \bibinfo
  {pages} {064011} (\bibinfo {year} {2016})}\BibitemShut {NoStop}%
\bibitem [{\citenamefont {Bastianello}\ \emph {et~al.}(2018)\citenamefont
  {Bastianello}, \citenamefont {Doyon}, \citenamefont {Watts},\ and\
  \citenamefont {Yoshimura}}]{Bastianello2018}%
  \BibitemOpen
  \bibfield  {author} {\bibinfo {author} {\bibfnamefont {A.}~\bibnamefont
  {Bastianello}}, \bibinfo {author} {\bibfnamefont {B.}~\bibnamefont {Doyon}},
  \bibinfo {author} {\bibfnamefont {G.}~\bibnamefont {Watts}},\ and\ \bibinfo
  {author} {\bibfnamefont {T.}~\bibnamefont {Yoshimura}},\ }\bibfield  {title}
  {\bibinfo {title} {{Generalized hydrodynamics of classical integrable field
  theory: the sinh-Gordon model}},\ }\href
  {https://doi.org/10.21468/SciPostPhys.4.6.045} {\bibfield  {journal}
  {\bibinfo  {journal} {SciPost Phys.}\ }\textbf {\bibinfo {volume} {4}},\
  \bibinfo {pages} {045} (\bibinfo {year} {2018})}\BibitemShut {NoStop}%
\bibitem [{\citenamefont {Vecchio}\ \emph {et~al.}(2020)\citenamefont
  {Vecchio}, \citenamefont {Bastianello}, \citenamefont {Luca},\ and\
  \citenamefont {Mussardo}}]{DelVecchio2020}%
  \BibitemOpen
  \bibfield  {author} {\bibinfo {author} {\bibfnamefont {G.~D. V.~D.}\
  \bibnamefont {Vecchio}}, \bibinfo {author} {\bibfnamefont {A.}~\bibnamefont
  {Bastianello}}, \bibinfo {author} {\bibfnamefont {A.~D.}\ \bibnamefont
  {Luca}},\ and\ \bibinfo {author} {\bibfnamefont {G.}~\bibnamefont
  {Mussardo}},\ }\bibfield  {title} {\bibinfo {title} {{Exact
  out-of-equilibrium steady states in the semiclassical limit of the
  interacting Bose gas}},\ }\href
  {https://doi.org/10.21468/SciPostPhys.9.1.002} {\bibfield  {journal}
  {\bibinfo  {journal} {SciPost Phys.}\ }\textbf {\bibinfo {volume} {9}},\
  \bibinfo {pages} {002} (\bibinfo {year} {2020})}\BibitemShut {NoStop}%
\bibitem [{\citenamefont {Koch}\ \emph {et~al.}(2022)\citenamefont {Koch},
  \citenamefont {Caux},\ and\ \citenamefont {Bastianello}}]{Koch_2022}%
  \BibitemOpen
  \bibfield  {author} {\bibinfo {author} {\bibfnamefont {R.}~\bibnamefont
  {Koch}}, \bibinfo {author} {\bibfnamefont {J.-S.}\ \bibnamefont {Caux}},\
  and\ \bibinfo {author} {\bibfnamefont {A.}~\bibnamefont {Bastianello}},\
  }\bibfield  {title} {\bibinfo {title} {Generalized hydrodynamics of the
  attractive non-linear schr\"odinger equation},\ }\href
  {https://doi.org/10.1088/1751-8121/ac53c3} {\bibfield  {journal} {\bibinfo
  {journal} {Journal of Physics A: Mathematical and Theoretical}\ }\textbf
  {\bibinfo {volume} {55}},\ \bibinfo {pages} {134001} (\bibinfo {year}
  {2022})}\BibitemShut {NoStop}%
\bibitem [{\citenamefont {Bezzaz}\ \emph {et~al.}(2023)\citenamefont {Bezzaz},
  \citenamefont {Dubois},\ and\ \citenamefont {Bouchoule}}]{Bezzaz2023}%
  \BibitemOpen
  \bibfield  {author} {\bibinfo {author} {\bibfnamefont {Y.}~\bibnamefont
  {Bezzaz}}, \bibinfo {author} {\bibfnamefont {L.}~\bibnamefont {Dubois}},\
  and\ \bibinfo {author} {\bibfnamefont {I.}~\bibnamefont {Bouchoule}},\
  }\href@noop {} {\bibinfo {title} {Rapidity distribution within the defocusing
  non-linear schr\"odinger equation model}} (\bibinfo {year} {2023}),\ \Eprint
  {https://arxiv.org/abs/2301.11098} {arXiv:2301.11098 [cond-mat.quant-gas]}
  \BibitemShut {NoStop}%
\bibitem [{\citenamefont {Scalapino}\ \emph {et~al.}(1972)\citenamefont
  {Scalapino}, \citenamefont {Sears},\ and\ \citenamefont
  {Ferrell}}]{Scalapino1972}%
  \BibitemOpen
  \bibfield  {author} {\bibinfo {author} {\bibfnamefont {D.~J.}\ \bibnamefont
  {Scalapino}}, \bibinfo {author} {\bibfnamefont {M.}~\bibnamefont {Sears}},\
  and\ \bibinfo {author} {\bibfnamefont {R.~A.}\ \bibnamefont {Ferrell}},\
  }\bibfield  {title} {\bibinfo {title} {Statistical mechanics of
  one-dimensional ginzburg-landau fields},\ }\href
  {https://doi.org/10.1103/PhysRevB.6.3409} {\bibfield  {journal} {\bibinfo
  {journal} {Phys. Rev. B}\ }\textbf {\bibinfo {volume} {6}},\ \bibinfo {pages}
  {3409} (\bibinfo {year} {1972})}\BibitemShut {NoStop}%
\bibitem [{\citenamefont {Castin}\ \emph {et~al.}(2000)\citenamefont {Castin},
  \citenamefont {Dum}, \citenamefont {Mandonnet}, \citenamefont {Minguzzi},\
  and\ \citenamefont {Carusotto}}]{Castin2000}%
  \BibitemOpen
  \bibfield  {author} {\bibinfo {author} {\bibfnamefont {Y.}~\bibnamefont
  {Castin}}, \bibinfo {author} {\bibfnamefont {R.}~\bibnamefont {Dum}},
  \bibinfo {author} {\bibfnamefont {E.}~\bibnamefont {Mandonnet}}, \bibinfo
  {author} {\bibfnamefont {A.}~\bibnamefont {Minguzzi}},\ and\ \bibinfo
  {author} {\bibfnamefont {I.}~\bibnamefont {Carusotto}},\ }\bibfield  {title}
  {\bibinfo {title} {Coherence properties of a continuous atom laser},\ }\href
  {https://doi.org/10.1080/09500340008232189} {\bibfield  {journal} {\bibinfo
  {journal} {Journal of Modern Optics}\ }\textbf {\bibinfo {volume} {47}},\
  \bibinfo {pages} {2671} (\bibinfo {year} {2000})}\BibitemShut {NoStop}%
\bibitem [{\citenamefont {Metropolis}\ \emph {et~al.}(1953)\citenamefont
  {Metropolis}, \citenamefont {Rosenbluth}, \citenamefont {Rosenbluth},
  \citenamefont {Teller},\ and\ \citenamefont {Teller}}]{Metropolis1953}%
  \BibitemOpen
  \bibfield  {author} {\bibinfo {author} {\bibfnamefont {N.}~\bibnamefont
  {Metropolis}}, \bibinfo {author} {\bibfnamefont {A.~W.}\ \bibnamefont
  {Rosenbluth}}, \bibinfo {author} {\bibfnamefont {M.~N.}\ \bibnamefont
  {Rosenbluth}}, \bibinfo {author} {\bibfnamefont {A.~H.}\ \bibnamefont
  {Teller}},\ and\ \bibinfo {author} {\bibfnamefont {E.}~\bibnamefont
  {Teller}},\ }\bibfield  {title} {\bibinfo {title} {Equation of state
  calculations by fast computing machines},\ }\href
  {https://doi.org/10.1063/1.1699114} {\bibfield  {journal} {\bibinfo
  {journal} {The Journal of Chemical Physics}\ }\textbf {\bibinfo {volume}
  {21}},\ \bibinfo {pages} {1087} (\bibinfo {year} {1953})},\ \Eprint
  {https://arxiv.org/abs/https://doi.org/10.1063/1.1699114}
  {https://doi.org/10.1063/1.1699114} \BibitemShut {NoStop}%
\bibitem [{\citenamefont {Hastings}(1970)}]{Hasting1970}%
  \BibitemOpen
  \bibfield  {author} {\bibinfo {author} {\bibfnamefont {W.~K.}\ \bibnamefont
  {Hastings}},\ }\bibfield  {title} {\bibinfo {title} {{Monte Carlo sampling
  methods using Markov chains and their applications}},\ }\href
  {https://doi.org/10.1093/biomet/57.1.97} {\bibfield  {journal} {\bibinfo
  {journal} {Biometrika}\ }\textbf {\bibinfo {volume} {57}},\ \bibinfo {pages}
  {97} (\bibinfo {year} {1970})}\BibitemShut {NoStop}%
\bibitem [{\citenamefont {Schumm}\ \emph {et~al.}(2005)\citenamefont {Schumm},
  \citenamefont {Hofferberth}, \citenamefont {Andersson}, \citenamefont
  {Wildermuth}, \citenamefont {Groth}, \citenamefont {Bar-Joseph},
  \citenamefont {Schmiedmayer},\ and\ \citenamefont {Kr{\"u}ger}}]{Schumm2005}%
  \BibitemOpen
  \bibfield  {author} {\bibinfo {author} {\bibfnamefont {T.}~\bibnamefont
  {Schumm}}, \bibinfo {author} {\bibfnamefont {S.}~\bibnamefont {Hofferberth}},
  \bibinfo {author} {\bibfnamefont {L.~M.}\ \bibnamefont {Andersson}}, \bibinfo
  {author} {\bibfnamefont {S.}~\bibnamefont {Wildermuth}}, \bibinfo {author}
  {\bibfnamefont {S.}~\bibnamefont {Groth}}, \bibinfo {author} {\bibfnamefont
  {I.}~\bibnamefont {Bar-Joseph}}, \bibinfo {author} {\bibfnamefont
  {J.}~\bibnamefont {Schmiedmayer}},\ and\ \bibinfo {author} {\bibfnamefont
  {P.}~\bibnamefont {Kr{\"u}ger}},\ }\bibfield  {title} {\bibinfo {title}
  {Matter-wave interferometry in a double well on an atom chip},\ }\href
  {https://doi.org/10.1038/nphys125} {\bibfield  {journal} {\bibinfo  {journal}
  {Nature Physics}\ }\textbf {\bibinfo {volume} {1}},\ \bibinfo {pages} {57}
  (\bibinfo {year} {2005})}\BibitemShut {NoStop}%
\bibitem [{\citenamefont {Hofferberth}\ \emph {et~al.}(2007)\citenamefont
  {Hofferberth}, \citenamefont {Lesanovsky}, \citenamefont {Fischer},
  \citenamefont {Schumm},\ and\ \citenamefont
  {Schmiedmayer}}]{Hofferberth2007}%
  \BibitemOpen
  \bibfield  {author} {\bibinfo {author} {\bibfnamefont {S.}~\bibnamefont
  {Hofferberth}}, \bibinfo {author} {\bibfnamefont {I.}~\bibnamefont
  {Lesanovsky}}, \bibinfo {author} {\bibfnamefont {B.}~\bibnamefont {Fischer}},
  \bibinfo {author} {\bibfnamefont {T.}~\bibnamefont {Schumm}},\ and\ \bibinfo
  {author} {\bibfnamefont {J.}~\bibnamefont {Schmiedmayer}},\ }\bibfield
  {title} {\bibinfo {title} {Non-equilibrium coherence dynamics in
  one-dimensional bose gases},\ }\href {https://doi.org/10.1038/nature06149}
  {\bibfield  {journal} {\bibinfo  {journal} {Nature}\ }\textbf {\bibinfo
  {volume} {449}},\ \bibinfo {pages} {324} (\bibinfo {year}
  {2007})}\BibitemShut {NoStop}%
\bibitem [{\citenamefont {van Nieuwkerk}\ \emph {et~al.}(2018)\citenamefont
  {van Nieuwkerk}, \citenamefont {Schmiedmayer},\ and\ \citenamefont
  {Essler}}]{Nieuwkerk2018}%
  \BibitemOpen
  \bibfield  {author} {\bibinfo {author} {\bibfnamefont {Y.~D.}\ \bibnamefont
  {van Nieuwkerk}}, \bibinfo {author} {\bibfnamefont {J.}~\bibnamefont
  {Schmiedmayer}},\ and\ \bibinfo {author} {\bibfnamefont {F.~H.~L.}\
  \bibnamefont {Essler}},\ }\bibfield  {title} {\bibinfo {title} {{Projective
  phase measurements in one-dimensional Bose gases}},\ }\href
  {https://doi.org/10.21468/SciPostPhys.5.5.046} {\bibfield  {journal}
  {\bibinfo  {journal} {SciPost Phys.}\ }\textbf {\bibinfo {volume} {5}},\
  \bibinfo {pages} {046} (\bibinfo {year} {2018})}\BibitemShut {NoStop}%
\bibitem [{\citenamefont {Langen}\ \emph {et~al.}(2016)\citenamefont {Langen},
  \citenamefont {Gasenzer},\ and\ \citenamefont {Schmiedmayer}}]{langen2016}%
  \BibitemOpen
  \bibfield  {author} {\bibinfo {author} {\bibfnamefont {T.}~\bibnamefont
  {Langen}}, \bibinfo {author} {\bibfnamefont {T.}~\bibnamefont {Gasenzer}},\
  and\ \bibinfo {author} {\bibfnamefont {J.}~\bibnamefont {Schmiedmayer}},\
  }\bibfield  {title} {\bibinfo {title} {Prethermalization and universal
  dynamics in near-integrable quantum systems},\ }\href
  {https://doi.org/10.1088/1742-5468/2016/06/064009} {\bibfield  {journal}
  {\bibinfo  {journal} {Journal of Statistical Mechanics: Theory and
  Experiment}\ }\textbf {\bibinfo {volume} {2016}},\ \bibinfo {pages} {064009}
  (\bibinfo {year} {2016})}\BibitemShut {NoStop}%
\bibitem [{\citenamefont {Blakie}\ \emph {et~al.}(2008)\citenamefont {Blakie},
  \citenamefont {Bradley}, \citenamefont {Davis}, \citenamefont {Ballagh},\
  and\ \citenamefont {Gardiner}}]{Blakie2008}%
  \BibitemOpen
  \bibfield  {author} {\bibinfo {author} {\bibfnamefont {P.}~\bibnamefont
  {Blakie}}, \bibinfo {author} {\bibfnamefont {A.}~\bibnamefont {Bradley}},
  \bibinfo {author} {\bibfnamefont {M.}~\bibnamefont {Davis}}, \bibinfo
  {author} {\bibfnamefont {R.}~\bibnamefont {Ballagh}},\ and\ \bibinfo {author}
  {\bibfnamefont {C.}~\bibnamefont {Gardiner}},\ }\bibfield  {title} {\bibinfo
  {title} {Dynamics and statistical mechanics of ultra-cold bose gases using
  c-field techniques},\ }\href {https://doi.org/10.1080/00018730802564254}
  {\bibfield  {journal} {\bibinfo  {journal} {Advances in Physics}\ }\textbf
  {\bibinfo {volume} {57}},\ \bibinfo {pages} {363} (\bibinfo {year} {2008})},\
  \Eprint {https://arxiv.org/abs/https://doi.org/10.1080/00018730802564254}
  {https://doi.org/10.1080/00018730802564254} \BibitemShut {NoStop}%
\bibitem [{\citenamefont {Horv\'ath}\ \emph {et~al.}(2019)\citenamefont
  {Horv\'ath}, \citenamefont {Lovas}, \citenamefont {Kormos}, \citenamefont
  {Tak\'acs},\ and\ \citenamefont {Zar\'and}}]{Horvath2019}%
  \BibitemOpen
  \bibfield  {author} {\bibinfo {author} {\bibfnamefont {D.~X.}\ \bibnamefont
  {Horv\'ath}}, \bibinfo {author} {\bibfnamefont {I.}~\bibnamefont {Lovas}},
  \bibinfo {author} {\bibfnamefont {M.}~\bibnamefont {Kormos}}, \bibinfo
  {author} {\bibfnamefont {G.}~\bibnamefont {Tak\'acs}},\ and\ \bibinfo
  {author} {\bibfnamefont {G.}~\bibnamefont {Zar\'and}},\ }\bibfield  {title}
  {\bibinfo {title} {Nonequilibrium time evolution and rephasing in the quantum
  sine-gordon model},\ }\href {https://doi.org/10.1103/PhysRevA.100.013613}
  {\bibfield  {journal} {\bibinfo  {journal} {Phys. Rev. A}\ }\textbf {\bibinfo
  {volume} {100}},\ \bibinfo {pages} {013613} (\bibinfo {year}
  {2019})}\BibitemShut {NoStop}%
\bibitem [{\citenamefont {Mennemann}\ \emph {et~al.}(2021)\citenamefont
  {Mennemann}, \citenamefont {Mazets}, \citenamefont {Pigneur}, \citenamefont
  {Stimming}, \citenamefont {Mauser}, \citenamefont {Schmiedmayer},\ and\
  \citenamefont {Erne}}]{Mennemann2021}%
  \BibitemOpen
  \bibfield  {author} {\bibinfo {author} {\bibfnamefont {J.-F.}\ \bibnamefont
  {Mennemann}}, \bibinfo {author} {\bibfnamefont {I.~E.}\ \bibnamefont
  {Mazets}}, \bibinfo {author} {\bibfnamefont {M.}~\bibnamefont {Pigneur}},
  \bibinfo {author} {\bibfnamefont {H.~P.}\ \bibnamefont {Stimming}}, \bibinfo
  {author} {\bibfnamefont {N.~J.}\ \bibnamefont {Mauser}}, \bibinfo {author}
  {\bibfnamefont {J.}~\bibnamefont {Schmiedmayer}},\ and\ \bibinfo {author}
  {\bibfnamefont {S.}~\bibnamefont {Erne}},\ }\bibfield  {title} {\bibinfo
  {title} {Relaxation in an extended bosonic josephson junction},\ }\href
  {https://doi.org/10.1103/PhysRevResearch.3.023197} {\bibfield  {journal}
  {\bibinfo  {journal} {Phys. Rev. Res.}\ }\textbf {\bibinfo {volume} {3}},\
  \bibinfo {pages} {023197} (\bibinfo {year} {2021})}\BibitemShut {NoStop}%
\bibitem [{\citenamefont {Olshanii}(1998)}]{Olshanii1998}%
  \BibitemOpen
  \bibfield  {author} {\bibinfo {author} {\bibfnamefont {M.}~\bibnamefont
  {Olshanii}},\ }\bibfield  {title} {\bibinfo {title} {Atomic scattering in the
  presence of an external confinement and a gas of impenetrable bosons},\
  }\href {https://doi.org/10.1103/PhysRevLett.81.938} {\bibfield  {journal}
  {\bibinfo  {journal} {Phys. Rev. Lett.}\ }\textbf {\bibinfo {volume} {81}},\
  \bibinfo {pages} {938} (\bibinfo {year} {1998})}\BibitemShut {NoStop}%
\bibitem [{\citenamefont {Tajik}\ \emph {et~al.}(2022)\citenamefont {Tajik},
  \citenamefont {Gluza}, \citenamefont {Sebe}, \citenamefont {Schüttelkopf},
  \citenamefont {Cataldini}, \citenamefont {Sabino}, \citenamefont {Møller},
  \citenamefont {Ji}, \citenamefont {Erne}, \citenamefont {Guarnieri},
  \citenamefont {Sotiriadis}, \citenamefont {Eisert},\ and\ \citenamefont
  {Schmiedmayer}}]{tajik2022}%
  \BibitemOpen
  \bibfield  {author} {\bibinfo {author} {\bibfnamefont {M.}~\bibnamefont
  {Tajik}}, \bibinfo {author} {\bibfnamefont {M.}~\bibnamefont {Gluza}},
  \bibinfo {author} {\bibfnamefont {N.}~\bibnamefont {Sebe}}, \bibinfo {author}
  {\bibfnamefont {P.}~\bibnamefont {Schüttelkopf}}, \bibinfo {author}
  {\bibfnamefont {F.}~\bibnamefont {Cataldini}}, \bibinfo {author}
  {\bibfnamefont {J.}~\bibnamefont {Sabino}}, \bibinfo {author} {\bibfnamefont
  {F.}~\bibnamefont {Møller}}, \bibinfo {author} {\bibfnamefont {S.-C.}\
  \bibnamefont {Ji}}, \bibinfo {author} {\bibfnamefont {S.}~\bibnamefont
  {Erne}}, \bibinfo {author} {\bibfnamefont {G.}~\bibnamefont {Guarnieri}},
  \bibinfo {author} {\bibfnamefont {S.}~\bibnamefont {Sotiriadis}}, \bibinfo
  {author} {\bibfnamefont {J.}~\bibnamefont {Eisert}},\ and\ \bibinfo {author}
  {\bibfnamefont {J.}~\bibnamefont {Schmiedmayer}},\ }\href@noop {} {\bibinfo
  {title} {Experimental observation of curved light-cones in a quantum field
  simulator}} (\bibinfo {year} {2022}),\ \Eprint
  {https://arxiv.org/abs/2209.09132} {arXiv:2209.09132 [cond-mat.quant-gas]}
  \BibitemShut {NoStop}%
\bibitem [{\citenamefont {Kuhnert}\ \emph {et~al.}(2013)\citenamefont
  {Kuhnert}, \citenamefont {Geiger}, \citenamefont {Langen}, \citenamefont
  {Gring}, \citenamefont {Rauer}, \citenamefont {Kitagawa}, \citenamefont
  {Demler}, \citenamefont {Adu~Smith},\ and\ \citenamefont
  {Schmiedmayer}}]{Kuhnert2013}%
  \BibitemOpen
  \bibfield  {author} {\bibinfo {author} {\bibfnamefont {M.}~\bibnamefont
  {Kuhnert}}, \bibinfo {author} {\bibfnamefont {R.}~\bibnamefont {Geiger}},
  \bibinfo {author} {\bibfnamefont {T.}~\bibnamefont {Langen}}, \bibinfo
  {author} {\bibfnamefont {M.}~\bibnamefont {Gring}}, \bibinfo {author}
  {\bibfnamefont {B.}~\bibnamefont {Rauer}}, \bibinfo {author} {\bibfnamefont
  {T.}~\bibnamefont {Kitagawa}}, \bibinfo {author} {\bibfnamefont
  {E.}~\bibnamefont {Demler}}, \bibinfo {author} {\bibfnamefont
  {D.}~\bibnamefont {Adu~Smith}},\ and\ \bibinfo {author} {\bibfnamefont
  {J.}~\bibnamefont {Schmiedmayer}},\ }\bibfield  {title} {\bibinfo {title}
  {Multimode dynamics and emergence of a characteristic length scale in a
  one-dimensional quantum system},\ }\href
  {https://doi.org/10.1103/PhysRevLett.110.090405} {\bibfield  {journal}
  {\bibinfo  {journal} {Phys. Rev. Lett.}\ }\textbf {\bibinfo {volume} {110}},\
  \bibinfo {pages} {090405} (\bibinfo {year} {2013})}\BibitemShut {NoStop}%
\bibitem [{\citenamefont {Schweigler}(2019)}]{schweigler2019}%
  \BibitemOpen
  \bibfield  {author} {\bibinfo {author} {\bibfnamefont {T.}~\bibnamefont
  {Schweigler}},\ }\href@noop {} {\bibinfo {title} {Correlations and dynamics
  of tunnel-coupled one-dimensional bose gases}} (\bibinfo {year} {2019}),\
  \Eprint {https://arxiv.org/abs/1908.00422} {arXiv:1908.00422
  [cond-mat.quant-gas]} \BibitemShut {NoStop}%
\bibitem [{\citenamefont {Kasper}\ \emph {et~al.}(2020)\citenamefont {Kasper},
  \citenamefont {Marino}, \citenamefont {Ji}, \citenamefont {Gritsev},
  \citenamefont {Schmiedmayer},\ and\ \citenamefont {Demler}}]{Kasper2020}%
  \BibitemOpen
  \bibfield  {author} {\bibinfo {author} {\bibfnamefont {V.}~\bibnamefont
  {Kasper}}, \bibinfo {author} {\bibfnamefont {J.}~\bibnamefont {Marino}},
  \bibinfo {author} {\bibfnamefont {S.}~\bibnamefont {Ji}}, \bibinfo {author}
  {\bibfnamefont {V.}~\bibnamefont {Gritsev}}, \bibinfo {author} {\bibfnamefont
  {J.}~\bibnamefont {Schmiedmayer}},\ and\ \bibinfo {author} {\bibfnamefont
  {E.}~\bibnamefont {Demler}},\ }\bibfield  {title} {\bibinfo {title}
  {Simulating a quantum commensurate-incommensurate phase transition using two
  raman-coupled one-dimensional condensates},\ }\href
  {https://doi.org/10.1103/PhysRevB.101.224102} {\bibfield  {journal} {\bibinfo
   {journal} {Phys. Rev. B}\ }\textbf {\bibinfo {volume} {101}},\ \bibinfo
  {pages} {224102} (\bibinfo {year} {2020})}\BibitemShut {NoStop}%
\bibitem [{\citenamefont {Wybo}\ \emph {et~al.}(2022)\citenamefont {Wybo},
  \citenamefont {Knap},\ and\ \citenamefont {Bastianello}}]{Wybo2022}%
  \BibitemOpen
  \bibfield  {author} {\bibinfo {author} {\bibfnamefont {E.}~\bibnamefont
  {Wybo}}, \bibinfo {author} {\bibfnamefont {M.}~\bibnamefont {Knap}},\ and\
  \bibinfo {author} {\bibfnamefont {A.}~\bibnamefont {Bastianello}},\
  }\bibfield  {title} {\bibinfo {title} {Quantum sine-gordon dynamics in
  coupled spin chains},\ }\href {https://doi.org/10.1103/PhysRevB.106.075102}
  {\bibfield  {journal} {\bibinfo  {journal} {Phys. Rev. B}\ }\textbf {\bibinfo
  {volume} {106}},\ \bibinfo {pages} {075102} (\bibinfo {year}
  {2022})}\BibitemShut {NoStop}%
\bibitem [{\citenamefont {Wybo}\ \emph {et~al.}(2023)\citenamefont {Wybo},
  \citenamefont {Bastianello}, \citenamefont {Aidelsburger}, \citenamefont
  {Bloch},\ and\ \citenamefont {Knap}}]{wybo2023}%
  \BibitemOpen
  \bibfield  {author} {\bibinfo {author} {\bibfnamefont {E.}~\bibnamefont
  {Wybo}}, \bibinfo {author} {\bibfnamefont {A.}~\bibnamefont {Bastianello}},
  \bibinfo {author} {\bibfnamefont {M.}~\bibnamefont {Aidelsburger}}, \bibinfo
  {author} {\bibfnamefont {I.}~\bibnamefont {Bloch}},\ and\ \bibinfo {author}
  {\bibfnamefont {M.}~\bibnamefont {Knap}},\ }\href@noop {} {\bibinfo {title}
  {Preparing and analyzing solitons in the sine-gordon model with quantum gas
  microscopes}} (\bibinfo {year} {2023}),\ \Eprint
  {https://arxiv.org/abs/2303.16221} {arXiv:2303.16221 [cond-mat.quant-gas]}
  \BibitemShut {NoStop}%
\bibitem [{\citenamefont {Friedman}\ \emph {et~al.}(2020)\citenamefont
  {Friedman}, \citenamefont {Gopalakrishnan},\ and\ \citenamefont
  {Vasseur}}]{Friedman2020}%
  \BibitemOpen
  \bibfield  {author} {\bibinfo {author} {\bibfnamefont {A.~J.}\ \bibnamefont
  {Friedman}}, \bibinfo {author} {\bibfnamefont {S.}~\bibnamefont
  {Gopalakrishnan}},\ and\ \bibinfo {author} {\bibfnamefont {R.}~\bibnamefont
  {Vasseur}},\ }\bibfield  {title} {\bibinfo {title} {Diffusive hydrodynamics
  from integrability breaking},\ }\href
  {https://doi.org/10.1103/PhysRevB.101.180302} {\bibfield  {journal} {\bibinfo
   {journal} {Phys. Rev. B}\ }\textbf {\bibinfo {volume} {101}},\ \bibinfo
  {pages} {180302} (\bibinfo {year} {2020})}\BibitemShut {NoStop}%
\bibitem [{\citenamefont {Lopez-Piqueres}\ \emph {et~al.}(2021)\citenamefont
  {Lopez-Piqueres}, \citenamefont {Ware}, \citenamefont {Gopalakrishnan},\ and\
  \citenamefont {Vasseur}}]{Piqueres2021}%
  \BibitemOpen
  \bibfield  {author} {\bibinfo {author} {\bibfnamefont {J.}~\bibnamefont
  {Lopez-Piqueres}}, \bibinfo {author} {\bibfnamefont {B.}~\bibnamefont
  {Ware}}, \bibinfo {author} {\bibfnamefont {S.}~\bibnamefont
  {Gopalakrishnan}},\ and\ \bibinfo {author} {\bibfnamefont {R.}~\bibnamefont
  {Vasseur}},\ }\bibfield  {title} {\bibinfo {title} {Hydrodynamics of
  nonintegrable systems from a relaxation-time approximation},\ }\href
  {https://doi.org/10.1103/PhysRevB.103.L060302} {\bibfield  {journal}
  {\bibinfo  {journal} {Phys. Rev. B}\ }\textbf {\bibinfo {volume} {103}},\
  \bibinfo {pages} {L060302} (\bibinfo {year} {2021})}\BibitemShut {NoStop}%
\bibitem [{\citenamefont {Durnin}\ \emph {et~al.}(2021)\citenamefont {Durnin},
  \citenamefont {Bhaseen},\ and\ \citenamefont {Doyon}}]{Durnin2021b}%
  \BibitemOpen
  \bibfield  {author} {\bibinfo {author} {\bibfnamefont {J.}~\bibnamefont
  {Durnin}}, \bibinfo {author} {\bibfnamefont {M.~J.}\ \bibnamefont
  {Bhaseen}},\ and\ \bibinfo {author} {\bibfnamefont {B.}~\bibnamefont
  {Doyon}},\ }\bibfield  {title} {\bibinfo {title} {Nonequilibrium dynamics and
  weakly broken integrability},\ }\href
  {https://doi.org/10.1103/PhysRevLett.127.130601} {\bibfield  {journal}
  {\bibinfo  {journal} {Phys. Rev. Lett.}\ }\textbf {\bibinfo {volume} {127}},\
  \bibinfo {pages} {130601} (\bibinfo {year} {2021})}\BibitemShut {NoStop}%
\bibitem [{\citenamefont {Bastianello}\ \emph {et~al.}(2021)\citenamefont
  {Bastianello}, \citenamefont {Luca},\ and\ \citenamefont
  {Vasseur}}]{Bastianello_2021}%
  \BibitemOpen
  \bibfield  {author} {\bibinfo {author} {\bibfnamefont {A.}~\bibnamefont
  {Bastianello}}, \bibinfo {author} {\bibfnamefont {A.~D.}\ \bibnamefont
  {Luca}},\ and\ \bibinfo {author} {\bibfnamefont {R.}~\bibnamefont
  {Vasseur}},\ }\bibfield  {title} {\bibinfo {title} {Hydrodynamics of weak
  integrability breaking},\ }\href {https://doi.org/10.1088/1742-5468/ac26b2}
  {\bibfield  {journal} {\bibinfo  {journal} {Journal of Statistical Mechanics:
  Theory and Experiment}\ }\textbf {\bibinfo {volume} {2021}},\ \bibinfo
  {pages} {114003} (\bibinfo {year} {2021})}\BibitemShut {NoStop}%
\bibitem [{\citenamefont {De~Nardis}\ \emph {et~al.}(2018)\citenamefont
  {De~Nardis}, \citenamefont {Bernard},\ and\ \citenamefont
  {Doyon}}]{denardis2018}%
  \BibitemOpen
  \bibfield  {author} {\bibinfo {author} {\bibfnamefont {J.}~\bibnamefont
  {De~Nardis}}, \bibinfo {author} {\bibfnamefont {D.}~\bibnamefont {Bernard}},\
  and\ \bibinfo {author} {\bibfnamefont {B.}~\bibnamefont {Doyon}},\ }\bibfield
   {title} {\bibinfo {title} {Hydrodynamic diffusion in integrable systems},\
  }\href {https://doi.org/10.1103/PhysRevLett.121.160603} {\bibfield  {journal}
  {\bibinfo  {journal} {Phys. Rev. Lett.}\ }\textbf {\bibinfo {volume} {121}},\
  \bibinfo {pages} {160603} (\bibinfo {year} {2018})}\BibitemShut {NoStop}%
\bibitem [{\citenamefont {Nardis}\ \emph {et~al.}(2019)\citenamefont {Nardis},
  \citenamefont {Bernard},\ and\ \citenamefont {Doyon}}]{denardis2019}%
  \BibitemOpen
  \bibfield  {author} {\bibinfo {author} {\bibfnamefont {J.~D.}\ \bibnamefont
  {Nardis}}, \bibinfo {author} {\bibfnamefont {D.}~\bibnamefont {Bernard}},\
  and\ \bibinfo {author} {\bibfnamefont {B.}~\bibnamefont {Doyon}},\ }\bibfield
   {title} {\bibinfo {title} {{Diffusion in generalized hydrodynamics and
  quasiparticle scattering}},\ }\href
  {https://doi.org/10.21468/SciPostPhys.6.4.049} {\bibfield  {journal}
  {\bibinfo  {journal} {SciPost Phys.}\ }\textbf {\bibinfo {volume} {6}},\
  \bibinfo {pages} {049} (\bibinfo {year} {2019})}\BibitemShut {NoStop}%
\bibitem [{\citenamefont {Del Vecchio Del~Vecchio}\ \emph
  {et~al.}(2023)\citenamefont {Del Vecchio Del~Vecchio}, \citenamefont
  {Kormos}, \citenamefont {Doyon},\ and\ \citenamefont {Bastianello}}]{Zenodo}%
  \BibitemOpen
  \bibfield  {author} {\bibinfo {author} {\bibfnamefont {G.}~\bibnamefont {Del
  Vecchio Del~Vecchio}}, \bibinfo {author} {\bibfnamefont {M.}~\bibnamefont
  {Kormos}}, \bibinfo {author} {\bibfnamefont {B.}~\bibnamefont {Doyon}},\ and\
  \bibinfo {author} {\bibfnamefont {A.}~\bibnamefont {Bastianello}},\ }\href
  {https://doi.org/10.5281/zenodo.7945732} {\bibinfo {title} {{Exact
  large-scale fluctuations of the phase field in the sine-Gordon model}}}
  (\bibinfo {year} {2023})\BibitemShut {NoStop}%
\end{thebibliography}%

\clearpage
\onecolumngrid
\newpage

\setcounter{equation}{0}  
\setcounter{figure}{0}
\setcounter{page}{1}
\setcounter{section}{0}    
\renewcommand\thesection{\arabic{section}}    
\renewcommand\thesubsection{\arabic{subsection}}    
\renewcommand{\thetable}{S\arabic{table}}
\renewcommand{\theequation}{S\arabic{equation}}
\renewcommand{\thefigure}{S\arabic{figure}}
\setcounter{secnumdepth}{2}  

\begin{center}
{\Large \textbf{Supplementary Material}}\\ \ \\
{\large \textbf{\titleinfo}}
\ \\ \ \\
Giuseppe Del Vecchio Del Vecchio, M\'arton Kormos, Benjamin Doyon, Alvise Bastianello
\end{center}
\bigskip
\bigskip

In the Supplementary Material, we collect more technical consideration supporting the findings discussed in the paper. More precisely, the content of each section is organized as follows

\begin{enumerate}
\item Section \ref{sec_SY}: A short overview of the phenomenological approach of Sachdev and Young to correlations and fluctuations of the order parameter \cite{Sachdev1997}, later adapted to sine-Gordon by Damle and Sachdev \cite{DamleSachdev2005}, is presented.
\item Section \ref{sec_BFT}: We present the rudiments of thermodynamics in integrable models and Ballistic Fluctuation Theory, discussing the main ideas and sketching the derivation of the results, which we later apply to sine-Gordon.
\item Section \ref{sec_SG_thermo}: We discuss the exact thermodynamic description and BFT of the sine-Gordon field theory, discussing special limits and comparing our results with the semiclassical approach of Section \ref{sec_SY}.
\item Section \ref{sec_numerics}: We overview the numerical techniques used in the Letter.
\item Section \ref{sec_exp}: Further details on the experimental realization of sine-Gordon with coupled condensates are given. We carefully discuss the phase-tomography protocol.
\end{enumerate}

\section{Phenomenological approaches to phase fluctuations}
\label{sec_SY}

Building on precedent work by Sachdev and Young \cite{Sachdev1997}, in Ref. \cite{DamleSachdev2005}, Damle and Sachdev proposed a simple heuristic method to compute phase fluctuations in sine-Gordon which we now briefly overview, and later compare with our exact result based on integrability. We leave the detailed calculations to the original reference, here we provide the physical arguments used in the derivation and the final results. The physically-motivated assumptions are:
\begin{enumerate}
    \item At low temperatures, the system is well approximated as a dilute gas of kinks and antikinks. These are seen as a gas of classical solitons distributed with Maxwell-Boltzmann statistics $\sim e^{-\beta \epsilon_K(\theta)}$ with $\epsilon_K(\theta)=Mc^2\cosh\theta$. At low temperature, the relativistic dispersion law is replaced with the low momentum expansion, becoming Galilean. In computing thermodynamics, the effects of breathers and interactions among excitations are neglected. Hence, (anti)kinks are independently distributed: the number of (anti)kinks comprised in an interval of size $L$ obeys Poisson statistics.
    Likewise, the velocity of kinks and antikinks is the bare velocity of relativistic particles $v_K(\theta)=c\tanh\theta$. These considerations determine the initial conditions $t=0$ for this ``gas of solitons".
    \item (Anti)kinks are solely responsible for changes in the phase, making it jump in units of $2\pi$. Hence, computing the phase difference between the edges of a space interval $[0,x]$ amounts to count how many kinks and antikinks are contained in this interval $\phi(x)-\phi(0)=2\pi (N_K-N_{\bar{K}})$.
  \item Computing unequal time correlations requires evolving in time the initial field configuration. This is done as follows: (anti)kinks in isolation evolve as free particles with relativistic velocity $c\tanh\theta$. When they scatter, they undergo nontrivial scattering. Damle and Sachdev considered two scenarios for scattering: in the first case, building on the generic behavior of low-energy scattering processes, they assume purely reflective scattering. In the other case, motivated by the so called ``reflectionless points" of the quantum sine-Gordon, they assumed purely transmissive scattering. It can be envisioned that the approach can be extended to the generic case, giving at each scattering event a probabilistic outcome based on the reflection and transmission amplitudes. This, of course, does not take into consideration potential spatial shifts due to interactions, contrary to what is done, within various models, in the literature on what is normally referred to as soliton gases. Full account of spatial shifts, along with the coherence effects arising from diagonalization of the scattering, would lead to GHD.
\end{enumerate}

Based on these assumptions, Damle and Sachdev computed the correlation function of the vertex operator $\mathcal{V}_{\lambda}(t,x)=\langle e^{i\lambda \phi(t,x)-i\lambda \phi(0,0)} \rangle$ obtaining
\be\label{demle_transmissive}
\mathcal{V}_{\lambda}(t,x)\big|_{\text{transmissive}}=C\exp\left[-2 \sin^2(\pi \lambda)(q_r+q_l)\right]
\ee
and 
\begin{multline}
\mathcal{V}_{\lambda}(t,x)\big|_{\text{reflective}}=C e^{-q_r-q_l}\big[U_0(2iq_r\Theta,2i\sqrt{q_lq_r})+U_0(2iq_l\Theta,2i\sqrt{q_lq_r})-iU_1(2iq_r\Theta,2i\sqrt{q_lq_r})+\\-iU_1(2iq_l\Theta,2i\sqrt{q_lq_r})-I_0(2\sqrt{q_l q_r}\big]
\end{multline}
with $I_0$ the modified Bessel function, $U_{0,1}$ the Lommel functions and the (unimportant) constant $C$ related to the expectation value of $e^{i\lambda \phi}$ on the ground state. We moreover define $\Theta=\cos(2\pi \lambda)$ and
\begin{align}
q_l&=2\int_{c\tanh\theta>x/t}\frac{\dd \theta}{2\pi}Mc \cosh\theta e^{-\beta \epsilon(\theta)} (ct\tanh\theta-x)\,,\\
q_r&=2\int_{c\tanh \theta<x/t}\frac{\dd \theta}{2\pi} Mc\cosh\theta e^{-\beta \epsilon(\theta)} (x-c t\tanh\theta)\,,
\end{align}

The asymptotic behavior at large times of the correlation functions in the two scenarios is very different. In the purely transmissive case, kinks spread ballistically in the system and their trajectories are not affected by other excitations. Note that the transmissive solution can be written as
\begin{equation}
\label{DS_trans}
\mathcal{V}_{\lambda}(t,x)\big|_{\text{transmissive}}=C\exp\left[-2 \sin^2(\pi \lambda) \int\frac{\dd \theta}{2\pi} Mc\cosh\theta\, 2e^{-\beta \epsilon(\theta)} |x-t c \tanh\theta|\right]\,.
\end{equation}
For space-like separations, $|x/t|>c,$ the absolute value can be dropped and the $t$-dependent part vanishes because it is an odd function, leading to
\begin{equation}
\mathcal{V}_{\lambda}(t,x)=C e^{-2 \sin^2(\pi \lambda) n |x|}\,, \qquad \qquad |x/t|>c\,,
\end{equation}
where $n=\int\frac{\dd \theta}{2\pi} Mc\cosh\theta \,2e^{-\beta \epsilon(\theta)}$ is the total density of kinks.

The same result holds for space-like separations in the purely reflective case, which is consistent with a light cone effect: outside the light cone, the nature of collisions is immaterial. 

This is not the case for time-like separations. In the purely reflective case, the ballistic trajectory of a traveling kink is suddenly stopped whenever an antikink is met: as a matter of fact, within this picture, at large scales (anti)kinks do not propagate ballistically in the system, but rather diffusively.
In a generic situation, where both the reflection and transmission channels are possible, one could attempt a hybrid calculation where (anti)kinks are randomly transmitted or reflected with the proper amplitudes, computed from the (exactly known \cite{ZAMOLODCHIKOV1979253}) two-body scattering matrices. Within a naive semiclassical picture, it is tempting to consider the reflection/transmission probability in different scattering as uncorrelated events. In this approximation, diffusion will be asymptotically dominant at late times. In contrast, our exact result derived from Ballistic Fluctuation Theory always has a dominant ballistic component (albeit with highly-renormalized velocities) even when reflective scattering are possible. Physically, this stems from the integrability of the model, which ensures ballistic transport, together with the fact that different scattering cannot be approximated as independent events, and coherence plays a pivotal role.

\section{Overview of Thermodynamic Bethe Ansatz and Ballistic Fluctuation Theory}
\label{sec_BFT}

Ballistic Fluctuation Theory \cite{Doyon2020, 10.21468/SciPostPhys.8.1.007} emerges from the combination of Generalised Hydrodynamics at Euler scale and Large deviation theory. A convenient application is to integrable models where it allows the calculation of the leading exponential decay of \emph{twist fields} as extensively discussed in \cite{delvecchio2022}. Here we can make use of this general framework to describe phase-fluctuations in the sine-Gordon model by building on the connection between topological charge and phase fluctuations. Before getting into the details of sine-Gordon, we provide a short recap of the main ideas and discuss its general consequences for the sine-Gordon application. For the sake of simplicity, in this paragraph we consider the example of a Bethe-Ansatz quantum integrable model featuring only a particle species (as for example Lieb-Liniger or sinh-Gordon): in sine-Gordon, the formulas are generalized to describe the more complex particle spectrum featuring kinks, antikinks and breathers, but retain the same structure as in this minimal example.

Let us then assume only one species of excitation is present, labeled by the rapidity $\theta$ and with energy and momentum $\epsilon(\theta)$ and $p(\theta)$ respectively. Upon scattering, these particles experience a scattering shift $\varphi(\theta)$: all these quantities are exactly known in sine-Gordon.
We assume the state is described by a (generalized) Gibbs Ensemble $e^{-\sum_j \beta_j Q_j}$ where $Q_j$ are the conserved local charges and $\beta_j$ the generalized inverse temperatures. The charges act simply on multi-particle states as $Q_j\ket{\{\theta_a\}_{i=1}^N}=\sum_{a=1}^Nh_{j}(\theta_a)\ket{\{\theta_a\}_{a=1}^N}$ where $h_j$ is the so-called one-particle eigenvalues and will be important in the following discussion.
It is also useful to define (generalized) free energies $A_{\{\beta_j\}_j}$: for a system of length $L$, one defines $A_{\{\beta_j\}_j}=-\lim_{L\to \infty} \frac{1}{L}\log\left[{\rm Tr}\, e^{-\sum_j \beta_j Q_j}\right]$.
In integrable models, free energies can be exactly computed within the framework of Thermodynamic Bethe Ansatz \cite{takahashi2005thermodynamics}. In integrable systems with fermionic scattering matrices, the state occupancy or filling fraction $\vartheta(\theta)=1/(1+e^{\varepsilon(\theta)})$ (in classical systems, the Fermi-Dirac statistics is replaced with Maxwell-Boltzmann or Rayleigh-Jeans distributions) on Generalized Gibbs Ensembles is determined by a set of integral equations
\be
\varepsilon(\theta)=\sum_j \beta_j h_j(\theta)+\int \frac{\dd \theta'}{2\pi} \varphi(\theta-\theta')\log\left(1+e^{-\varepsilon(\theta')}\right)\, .
\ee
Then, the free energy is
\be\label{eq_free_energy}
A_{\{\beta_j\}_j}=-\int\frac{\dd\theta}{2\pi}\partial_\theta p \log(1+e^{-\varepsilon(\theta)})\, .
\ee

Equal-time charge fluctuations can be easily related to thermodynamic quantities: our task is now computing charge fluctuations in a given interval, the possibility of including unequal times will be described later on. Let us assume we pick a given charge $Q_{i^*}=\int \dd x\, q_{i^*}(x)$ and the associated current $j_{i^*}$ and, for example, are interested in the fluctuations over an interval $\int_0^\ell \dd x \, q_{i^*}(x)$.
By definition, the Full Counting Statistics $F(\lambda)$ is computed as
\be
e^{\ell F(\lambda)}=\frac{\text{Tr}\left[ e^{-\lambda \int_0^\ell \dd x \, q_{i^*}(x)}e^{-\sum_{j}\beta_j Q_j}\right]}{\text{Tr}\left[ e^{-\sum_{j}\beta_j Q_j}\right]}\asymp \exp\left[-\ell \left(A_{\{\beta_j\}_j}\Big|_{\beta_{i^*}\to \beta_{i^*}+\lambda}-A_{\{\beta_j\}_j}\right)\right]\, ,
\ee
where in the last passage we take advantage of extensivity and notice that the insertion of $e^{-\lambda \int_0^\ell \dd x \, q_{i^*}(x)}$ has the effect of shifting the generalized inverse temperature of the charge $Q_{i^*}$ within the interval $[0,\ell]$.

Equal-time fluctuations do not contain any input from transport coefficients, which are instead crucial when different times are considered. This case requires Ballistic Fluctuation Theory to be properly handled.
One starts with the definition of the generating function (2) or, even better, one considers the derivative of the FCS with respect to the spectral parameter $\lambda$
\be\label{S8}
\ell \partial_\lambda F_\alpha(\lambda)=\int_0^1\dd s\, \frac{\text{Tr}\left[(\dot{t}_s j(x_s,t_s)-\dot{x}q(t_s,x_s))e^{\lambda \int_0^1 \dd s\,\big(\dot t_s j(x_s,t_s)-\dot x_s q(x_s,t_s)\big)}e^{-\sum_j \beta_j Q_j}\right]}{\text{Tr}\left[e^{\lambda \int_0^1 \dd s\,\big(\dot t_s j(x_s,t_s)-\dot x_s q(x_s,t_s)\big)}e^{-\sum_j \beta_j Q_j}\right]}\, .
\ee
where $Q_i^{*}\equiv Q$ with density $q_i^{*}\equiv q$ and current $j_i^{*}\equiv j$ (we omit writing explicitly the label of the charge we are focusing on). Here $q(x_s,t_s)$ and $j(x_s,t_s)$ are the charge density and current evolved along a path with endpoints $(x_0,t_0)=(0,0)$ and $(x_1,t_1)=\ell (c^{-1}\sin\alpha,\cos\alpha)$. The final result is independent from the chosen path, so we fix $(x_s,t_s)=\ell (s\sin\alpha,s c^{-1}\cos\alpha)$.
Two observations are now crucial:
\begin{enumerate}
    \item In the limit of large $\ell$, for most values of the $s$ parametrization the point $(t_s,x_s)$ is very far from the boundaries. Therefore, at the price of neglecting subextensive terms, one can replace $e^{\lambda \int_0^1 \dd s\,\big(\dot t_s j(t_s,x_s)-\dot x_s q(t_s,x_s)\big)}\to e^{W(\lambda,\alpha)}\equiv e^{\lambda \int_{-\infty}^{\infty} \dd y\,\big(c^{-1}\cos\alpha j(y\sin\alpha,y c^{-1}\cos\alpha)-\sin\alpha q(y\sin\alpha,y c^{-1}\cos\alpha)\big)}$. The integrand in Eq. \eqref{S8} thus become translational invariant and the $\ell$ extensive scale becomes apparent.
    \item The insertion of the deformation $e^{W(\lambda,\alpha)}$ in the Generalized Gibbs Ensemble is equivalent to a deformation of the GGE itself. Namely, for any local observables $O(x)$, there exists a set of generalized effective temperatures $\beta_j(\lambda,\alpha)$ such that $\text{Tr}\left[O(x) e^{W(\lambda,\alpha)} e^{-\sum_j \beta_j Q_j}\right]=\text{Tr}\left[O(x)  e^{-\sum_j \beta_j(\lambda,\alpha) Q_j}\right]$. This is a highly nontrivial fact whose proof we leave to the literature \cite{Doyon2020}.
\end{enumerate}
Combining the two observations above and integrating Eq. \eqref{S8}, the integral expression for the Full Counting Statistics (Eq. (3) in the main text) is recovered
\begin{equation}\label{eq_integrated_flow}
    F_\alpha(\lambda)
    = \int_0^\lambda \dd \lambda'\,\big( c^{-1}\cos\alpha \,\mathtt j_{\lambda'} - \sin\alpha\, \mathtt q_{\lambda'}\big),
\end{equation}
where  $\mathtt q_\lambda = \braket{q}_\lambda$ and $\mathtt j_\lambda = \braket{j}_\lambda$ are evaluated on the $\lambda$-dependent GGE. Expectation values of charges and currents \cite{Doyon2016,Bertini2016} are easily computable on a known GGE,
\be\label{eq_currentexp}
\mathtt q_\lambda=\int \dd\theta\, h(\theta)\rho_{\lambda}(\theta)\hspace{2pc}\mathtt j_\lambda=\int \dd\theta\, h(\theta)v^{\text{eff}}_{\lambda}(\theta)\rho_{\lambda}(\theta)\, ,
\ee
where for every test function $\tau(\theta)$, one defines the dressing operation $\tau(\theta)\to \tau^\cdr(\theta)$ through the integral equation $\tau^{\cdr}(\theta)=\tau(\theta)-\int \frac{\dd\theta'}{2\pi} \varphi(\theta-\theta')\vartheta(\theta')\tau^{\cdr}(\theta)$ and the effective velocity is $v^{\text{eff}}(\theta)=(\partial_\theta \epsilon)^{\text{dr}}/(\partial_\theta p)^{\text{dr}}$. In Eq.\eqref{eq_currentexp} we add a $\lambda-$label to the velocity to stress its parametric $\lambda-$dependence inherited from the flowing GGE.

The last task is now finding an equation to determine the $\lambda-$dependent GGE. This can be done from the conserved charge correlations: let us consider the expectation value of the density of a conserved charge on a $\lambda-$dependent GGE $\langle q_i(0,0)\rangle$, the response function upon infinitesimal changes of the effective energy $\varepsilon(\theta, \lambda)\to \varepsilon(\theta,\lambda+\delta \lambda)$ can be easily computed, but on the other hand a comparison with the right hand side of Eq. \eqref{S8} makes evident that $\partial_\lambda \langle q_i(0,0)\rangle=\int_{-\infty}^{\infty} \dd y\,\langle q_i(0,0)\big(c^{-1}\cos\alpha j(y\sin\alpha,y c^{-1}\cos\alpha)-\sin\alpha q(y\sin\alpha,y c^{-1}\cos\alpha)\big)\rangle_{\text{connected}}$. Connected correlation functions at large separations will dominate the integral and the latter are computable within hydrodynamics \cite{DeNardis_2022}. This allows one to find a close set of equations for the $\lambda-$flow of the GGE, which finally takes a very simple form
\begin{equation}
    \p_\lambda \varepsilon(\theta,\lambda) =  \sign(c\tan\alpha - v^{\text{eff}}(\theta,\lambda))h^\cdr(\theta,\lambda)\quad\,,\quad \varepsilon(\theta,0) = \varepsilon(\theta)\, .\label{eq:flow_equation}
\end{equation}
The dressing operations in Eq. \eqref{eq:flow_equation} are taken on the $\lambda-$dependent GGE and thus the dressed charge and effective velocity has a parametric $\lambda-$dependence, preventing an easy analytical solution. A notational remark is important: when no $\lambda-$dependence appears, quantities are evaluated at $\lambda=0$. Nonetheless, the flow equations are easy to integrate numerically, using standard methods to handle integral equations of integrable models, see Section \ref{sec_numerics}.
By expanding the Full Counting Statistics in powers of $\lambda$ one can systematically compute the cumulants: the flow equations are particularly suited for this, since such a power expansion eventually amounts to a $\lambda-$perturbative solution of the differential equation Eq. \eqref{eq:flow_equation}. In principle, close analytical expressions for the cumulants can be obtained, although the derivation quickly becomes cumbersome and for highest cumulants a numerical integration of the flow equation is preferred. In particular, the second cumulant reads
\be
c_2= \int\dd\theta\,  \rho(\theta)f(\theta)|v^\text{eff}(\theta)\cos\alpha-\sin\alpha|[h^{\cdr}(\theta)]^2\, .
\ee
The third cumulant vanished for parity-invariant states (like thermal states we focus on) but its expression is reported in \cite{Doyon2020}. The next non-trivial cumulant here is the fourth and on arbitrary GGEs and space-time rays reads
\begin{align}
    c_4 = \int\dd \theta (v^{\text{eff}}\cos\alpha - \sin\alpha)f \rho&\Big\{(h^{\cdr})^4 s \hat{f}\Tilde{f}+3 s [[(sf(h^{\cdr})^2]^\cdr]^2 + 4 h^{\cdr}s[f \Tilde{f}(h^{\cdr})^3]^{\cdr}+ 6 \Tilde{f}(h^{\cdr})^2[sf(h^{\cdr})^2]^\cdr
    \nonumber\\
    &+ 12 h^{\cdr}s[sfh^{\cdr}(sf(h^{\cdr})^2)^{\cdr}]^{\cdr}\Big\}
\end{align}
where $\rho$ is the GGE density and
\begin{equation}
    \hat{f} = -\frac{d}{d\epsilon}\log(f\Tilde{f}) - 3 f \quad , \quad \Tilde{f}=-\frac{d}{d\epsilon}\log(f) - 2 f \quad , \quad s = \sign(v^{\text{eff}}\cos\alpha - \sin\alpha)
\end{equation}
where $f(\theta)$ is a statistical factor (see Ref. \cite{Bastianello2018} for an account). In the present cases, $f = 1-\vartheta$ for quantum sine-Gordon and $f=1$ for classical sine-Gordon. In the classical case, the expressions are slightly modified to account for a more convenient definition of the filling functions and dressing operations, as we will discuss.
When multiple species of particles are present, one has to additionally sum over all particle types amounting to $\int \dd\theta\to \sum_a\int \dd\theta_a$.
In the following, we apply this formalism to the sine-Gordon field theory and describe phase fluctuations.

\section{The thermodynamics of the sine-Gordon model}
\label{sec_SG_thermo}

The exact thermodynamics of the sine-Gordon field theory is built on the knowledge of the exact spectrum and scattering matrix, derived in Ref. \cite{ZAMOLODCHIKOV1979253}. In this section, we recall the rudiments of Thermodynamic Bethe Ansatz and its application to BFT. As the theory is relativistic, the energy and momentum of a particle with mass $m$ are $\epsilon(\theta)=mc^2\cosh\theta$ and $p(\theta)=mc^2\sinh\theta$, respectively.

As recalled in the main text, the interactions determine the species of the breathers according to the parameter $\xi =\frac{8\pi}{cg^2}\left(1-\frac{8\pi}{cg^2}\right)^{-1}$ according to the mass law $ m_n = 2M \sin(\pi n \xi / 2)$, where $ n=1,\dots, \lceil\xi^{-1}\rceil$.
In contrast, kink and antikink are always present in the specturm and have mass $M$ \cite{ZAMOLODCHIKOV1995}: kink, antikink and breathers exhaust the particle content of the model.
Upon scattering, breathers have transmissive scattering with a scattering matrix
\begin{align}
    S_{n,m}(\theta) = &\frac{\sinh(\theta) + i\sin((n+m)\pi \xi/2 )}{\sinh(\theta) - i\sin((n+m)\pi \xi/2 )}\frac{\sinh(\theta) + i\sin(|n-m|\pi \xi/2 )}{\sinh(\theta) - i\sin(|n-m|\pi \xi/2 )}
    \nonumber
    \\
    &\times\prod_{k=1}^{\min(n,m)-1}\frac{\sin^2((|n-m|+2k)\pi \xi/4 - i\theta /2)\cos^2((n+m-2k)\pi \xi/4 + i\theta /2)}{\sin^2((|n-m|+2k)\pi \xi/4 + i\theta /2)\cos^2((n+m-2k)\pi \xi/4 - i\theta /2)}
\end{align}
where $\theta$ is the difference of the rapidities of the scattering particles.
Also (anti)kink are transmitted upon scattering with breathers, with scattering matrix
\begin{equation}
    S_n(\theta) = \frac{\sinh(\theta) + i \cos(n \pi \xi / 2)}{\sinh(\theta) - i \cos(n \pi \xi / 2)}\prod_{k=1}^{n-1}\frac{\sin^2((n-2k)\pi\xi/4 - \pi/4 + i \theta / 2)}{\sin^2((n-2k)\pi\xi/4 - \pi/4 - i \theta / 2)}\quad .
\end{equation}
In contrast, scattering of topological excitations is far richer, since both transmission and reflection are possible. in this case, the scattering matrix is an actual $2\times 2$ matrix with entries

\begin{equation}
    S(\theta_{ij})=\begin{pmatrix}
        &S_0(\theta) &0 &0 &0 \\
        &0 &S_T(\theta) &S_R(\theta) &0 \\
        &0 &S_R(\theta) &S_T(\theta) &0 \\
        &0 &0 &0 &S_0(\theta) \\
\end{pmatrix}\label{eq:scattering_matrix_quantum_sine_gordon}
\end{equation}
where $S_T(\theta)=\frac{\sinh (\xi^{-1}\theta)}{\sinh((i\pi-\theta)\xi^{-1})}S_0(\theta)$ and $S_R(\theta)=i\frac{\sin(\pi\xi^{-1})}{\sinh((i\pi-\theta)\xi^{-1})}S_0(\theta)$ weight the transmission and reflection channel respectively, where 
\be\label{eq_S0}
S_0(\theta)=-\exp\left[-i\int_0^\infty \frac{\dd t}{t} \frac{\sinh(\pi t(1-\xi)/2)}{\sinh(\pi\xi t/2)\cosh(\pi t/2)}\sin(\theta t)\right]\,.
\ee

For general values of the interaction $\xi$, the solution of thermodynamics passes through the Nested Bethe Ansatz, here for the sake of simplicity we focus on certain regimes where simplifications occur, while the BFT's prediction is of broader applicability.

\subsection{The quantum sine-Gordon at the reflectionless points}

When special values of the interactions are considered $\xi=1/(N+1)$ with $N\in \mathbb{N}$, the reflection channel of the kink-antikink scattering vanishes for any rapidity $S_R(\theta)=0$ and the scattering becomes purely transmissive. In this case, the thermodynamics is derived by the standard Thermodynamic Bethe Ansatz \cite{takahashi2005thermodynamics} machinery. To this hand, together with the root density characterizing the particle density of kink, antikink and breathers for any rapidity, one also introduces the filling fraction $\vartheta$ quantify the relative mode occupation. Conveniently, one parametrize the filling functions through the effective energy as $\vartheta(\theta)=1/(1+e^{\varepsilon(\theta)})$. Kink, antikink and breathers are described by their own effective energy found as the solution of the following integral equations

\be
\varepsilon_K(\theta)=\beta Mc^2\cosh\theta+\int \frac{\dd\theta'}{2\pi}\varphi(\theta-\theta')\left(\log(1+e^{-\varepsilon_K(\theta')})+\log(1+e^{-\varepsilon_{\bar{K}}(\theta')})\right)+\sum_{n=1}^{N}\int \frac{\dd \theta'}{2\pi}\varphi_n(\theta-\theta')\log(1+e^{-\varepsilon_n(\theta')})\, ,
\ee
\be
\varepsilon_n(\theta)=\beta m_nc^2\cosh\theta+\int \frac{\dd\theta'}{2\pi}\varphi_n(\theta-\theta')\left(\log(1+e^{-\varepsilon_K(\theta')})+\log(1+e^{-\varepsilon_{\bar{K}}(\theta')})\right)+\sum_{n'=1}^{N}\int \frac{\dd \theta'}{2\pi}\varphi_{n,n}(\theta-\theta')\log(1+e^{-\varepsilon_n(\theta')})
\ee
The integral equation for the effective energy of antikinks $\varepsilon_{\bar{K}}$ is analogous to the kinks' one and thus is omitted. Above, the scattering shifts $\varphi$ are defined as the logarithm derivatives of the scattering matrices $\varphi_{n,n'}(\theta)=i\partial_\theta \log S_{n,n'}(\theta)$, $\varphi_{n}(\theta)=i\partial_\theta \log S_{n}(\theta)$, and $\varphi(\theta)=i\partial_\theta \log S_0(\theta)$. Notably, at the reflectionless points the scattering matrix $S_0$ is greatly simplified $\log S_0(\theta)=i\pi (N+1)+\sum_{j=1}^{N}\log\left[\frac{\exp\left[\theta-i\pi j/(N+1)\right]+1}{\exp\theta+\exp[-i \pi j/(N+1)]}\right]$.
The above integral equations can be solved by standard numerical methods upon properly discretizing the integrals. We provide a short commented Mathematica notebook on Zenodo \cite{Zenodo}.
From the filling functions, one can then recover the density of particles. To this end, it is useful to define the dressing operation. For any triplet of test functions we can define the dressed operation $\{\tau_K(\theta), \tau_{\bar{K}}(\theta), \tau_n(\theta)\}\to\{\tau^\cdr_K(\theta),\tau^\cdr_{\bar{K}}(\theta),\tau^\cdr_n(\theta)\}$ by solving
\be
\tau^\cdr_n(\theta)=\tau_n(\theta)-\int \frac{\dd \theta'}{2\pi}\varphi_n(\theta-\theta')[\vartheta_K(\theta)\tau^\cdr_K(\theta)+\vartheta_{\bar{K}}(\theta)\tau^\cdr_{\bar{K}}(\theta)]-\sum_{n'=1}^{N}\int \frac{\dd\theta'}{2\pi} \varphi_{n,n'}(\theta-\theta')\vartheta_{n'}(\theta')\tau_{n'}^\cdr(\theta')\label{eq:dressing_qbreather}
\ee
\be
 \tau^\cdr_K(\theta)=\tau_K(\theta)-\int \frac{\dd \theta'}{2\pi}\varphi(\theta-\theta')(\vartheta_K(\theta)\tau^\cdr_K(\theta)+\vartheta_{\bar{K}}(\theta)\tau^\cdr_{\bar{K}}(\theta))-\sum_{n=1}^{N}\int \frac{\dd\theta'}{2\pi} \varphi_{n}(\theta-\theta')\vartheta_{n}(\theta')\tau_{n}^\cdr(\theta')\, .\label{eq:dressing_qkink}
\ee
and an analogous equation holds for antikinks. The root densities are then proportional to the filling fraction times the dressed derivative of the momentum $\rho=\frac{1}{2\pi}\vartheta (\partial_\theta p)^\cdr$.

\ \\
\textbf{The Ballistic Fluctuation Theory and its low temperature regime.---} We now apply the general BFT framework discussed in Section \ref{sec_BFT} to the topological charge in sine-Gordon, focusing on the reflectionless point.
Therefore, in place of a generic charge eigenvalue to be plugged in the flow equation \eqref{eq:flow_equation}, we use the topological charge. Its eigenvalue if defined as being $Z_K=+1$ for kinks, $Z_{\bar{K}}=-1$ for antikinks and zero over all the breathers $Z_n=0$.
The flow equation and cumulants must be generalized to include the effect of kinks, antikinks and breathers, but this simply amounts to sum over the different contributions.
We notice that, as it follows from a quick inspection of the dressing equations, in the case where the populations of kinks and antikinks are evenly matched $\rho_K(\theta)=\rho_{\bar{K}}(\theta)$, the dressing operation is ineffective on the topological charge $\{Z_K^\cdr(\theta),Z_{\bar{K}}^\cdr(\theta),Z_n^\cdr(\theta)\}=\{Z_K(\theta),Z_{\bar{K}}(\theta),Z_n(\theta)\}$, significantly simplifying the expressions for the cumulants. Instead, in the computation of the flow equation \eqref{eq:flow_equation} the flow parameter $\lambda$ explicitly biases the kink-antikink population, resulting in a non-trivial dressing of the topological charge.
We provide a Mathematica Notebook on Zenodo \cite{Zenodo} with the numerical solution of the thermodynamics and the computation of $c_2$ and $c_4$ from the explicit formulas.

\bigskip
\emph{The low-temperature limit.---}
It is instructive to analyze the low-temperature limit of the BFT prediction, to be compared with the result by Damle and Sachdev discussed in Section \ref{sec_SY}. In this limit, the flow equation greatly simplifies and it can be exactly integrated.
Let us thus assume the temperature is very low compared to the soliton mass scale $\beta c^2 M\gg 1$: this does not necessarily mean the temperature is small compared with the lightest breather $\beta c^2 m_1$ since, depending on the interaction, $m_1$ can be much smaller than $M$. Indeed, this is the case when approaching the semiclassical limit.
Since (anti)kink populations are exponentially suppressed in the low-temperature regime, their effect on dressing operations can be neglected: in this approximation, the dressed topological charge becomes identical to the bare one also along the flow. Furthermore, in the flow equations, the bias in the topological charge affects only (anti)kink leaving the breather populations unscathed. Hence, even when non-trivial, dressing of quantities (such as energy, momentum and their derivative) becomes flow-independent. Gathering these considerations, one simply integrates the flow equation
\be
\varepsilon_{K,\lambda}(\theta) =  \sign(c\tan\alpha - v_K^{\text{eff}}(\theta))\lambda+\varepsilon_K(\theta)\, ,\hspace{2pc}\varepsilon_{\bar{K},\lambda}(\theta) =  \sign(c\tan\alpha - v_{\bar{K}}^{\text{eff}}(\theta))\lambda+\varepsilon_{\bar{K}}(\theta)\, ,\hspace{2pc}\varepsilon_{n,\lambda}(\theta)=\varepsilon_{n}(\theta)
\ee
Above, quantities without a $\lambda-$dependence are computed on the unbiased thermal state. From the above, in the low temperature approximation we find $\rho_{\lambda,K}(\theta)=\rho_K(\theta)e^{\sign(c\tan\alpha - v_K^{\text{eff}}(\theta))\lambda}$ and $\rho_{\lambda,\bar{K}}(\theta)=\rho_{\bar{K}}(\theta)e^{-\sign(c\tan\alpha - v_{\bar{K}}^{\text{eff}}(\theta))\lambda}$, while the breather roots do not flow. We plug this last piece of information in Eq. \eqref{eq_integrated_flow} and perform the integral. For simplicity, we assume are considering unbiased GGEs where the kink and antikink populations are equal, obtaining the remarkably simple expression
\be\label{eq_FlowT}
F_\alpha(\lambda)\simeq c^{-1}\cos\alpha (\cosh \lambda-1)\int \dd\theta |v_K^{\text{eff}}(\theta)-c\tanh\alpha |\rho_K(\theta)\hspace{2pc} \text{At low temperature } \beta c^2 M\gg 1
\ee
This simple solution should be eventually compared with the phenomenological approach of Damle and Sachdev in the transmissive regime \eqref{DS_trans}. This requires a last passage to analytically continue the flow parameter to imaginary values $\lambda\to i\lambda$ and, as we discuss in the main text, obtain the correlation function of the vertex operator. Upon a $2\pi$ rescaling of $\lambda$ due to our different normalization of the correlator, we formally find the same expression of Damle and Sachdev with the minimal modification of capturing the effects of the background breather excitations by replacing the bare velocity and population of kinks with their dressed counterparts. If, furthermore, one also assumes $\beta c^2m_1\gg 1$, dressing effects can be entirely neglected and Eq. \eqref{DS_trans} is exactly recovered.

\subsection{The sine-Gordon model in the semiclassical regime}

The semiclassical regime is relevant both for practical numerical benchmark and for experimental applications, as we discuss in the main text. The exact thermodynamics of the classical sine-Gordon model has been derived only very recently \cite{koch2023exact}: we leave to the original reference a detailed discussion and report the main formulas and considerations. The semiclassical regime can be seen as a proper limiting case of the quantum model, more specifically one introduces an effective Planck constant $\hbar$ and simultaneously rescale the interaction $g_{\text{quantum}}^2=\hbar g_{\text{classical}}^2$ and temperature $\beta_{\text{quantum}}=\hbar \beta_{\text{classical}}$, while keeping the fixed the product $\beta_{\text{quantum}}M_{\text{quantum}}=\beta_{\text{classical}}M_{\text{classical}}$. In this limit, one defines the continuum classical spectral parameter $\sigma\in [0,1]$ through the correspondence $\sigma\leftrightarrow n \hbar/\sm$ with $\sm=\frac{8\pi}{cg^2_{\text{classical}}}$.
We used this scaling to compare the quantum and classical cumulants in Figure 3 of the main text.

Below, we will solely focus on classical quantities and so drop the ``classical" and ``quantum" label, always referring to classical couplings.

In the semiclassical limit, the reflection component of the kink-antikink scattering matrix vanishes $S_R\to 0$ and scattering becomes purely transmissive, resulting in similar equations to those reported in the previous section.
Nonetheless, a few important differences arise: the naively-obtained semiclassical limit of the filling fractions is singular for small $\sigma$, but this singularity is balanced by the dressed momentum derivative, eventually giving finite root densities and observables.
To avoid dealing with fictitious singularities, it is useful to properly redefine effective energies, dressing and fillings.
The correct dictionary stems from the derivation of the classical Thermodynamics Bethe Ansatz of Ref. \cite{koch2023exact}. Here, we report the result and adapt the notation for our purposes: we use a tilde for the nonsingular parametrization. The parametrization of filling functions in terms of effective energies is
\be
\vartheta_K(\theta)=e^{-\varepsilon_K(\theta)}\,,\hspace{2pc}\vartheta_{\bar{K}}(\theta)=e^{-\varepsilon_{\bar{K}}(\theta)}\,,\hspace{2pc}\vartheta_\sigma(\theta)=e^{-\varepsilon_\sigma(\theta)}\, .
\ee
Nonsingular fillings and regular effective energies are then introduced as
\be
\tilde{\vartheta}_K(\theta)=e^{-\tilde{\varepsilon}_K(\theta)}=\vartheta_K(\theta)\,, \hspace{2pc}\tilde{\vartheta}_{\bar{K}}(\theta)=e^{-\tilde{\varepsilon}_{\bar{K}}(\theta)}=\vartheta_{\bar{K}}(\theta)\, ,\hspace{2pc}\tilde{\vartheta}_\sigma=e^{-\sigma^2\tilde{\varepsilon}_\sigma(\theta) }=(\sm \sigma)^2\vartheta_\sigma(\theta)\, .
\ee

The integral equations determining the effective energy on thermal states best expressed in the new parametrization

\be\label{eq_tbaB}
\nonumber \sigma \tilde{\varepsilon}_\sigma(\theta)=-2+\beta c^2 \frac{m_\sigma}{ \sigma}\cosh\theta +\frac{1}{\sigma}\int \frac{\dd\theta'}{2\pi} \varphi_\sigma(\theta-\theta')(e^{-\tilde{\varepsilon}_K}+e^{-\tilde{\varepsilon}_{\bar{K}}})+\frac{1}{\sigma}\int \frac{\dd\theta'}{2\pi}\int_{0}^{1}\dd \sigma'\,\varphi_{\sigma,\sigma'}(\theta-\theta')\frac{e^{-( \sigma')^2\tilde{\varepsilon}_{\sigma'}(\theta')}-1}{\sm(\sigma')^2}\, ,
\ee

\be\label{eq_tbaK}
\nonumber \tilde{\varepsilon}_K(\theta)=\log \sm-1+\beta M c^2\cosh\theta +\int \frac{\dd\theta'}{2\pi} \varphi(\theta-\theta')(e^{-\tilde{\varepsilon}_K}+e^{-\tilde{\varepsilon}_{\bar{K}}})
+\int \frac{\dd\theta'}{2\pi}\int_{0}^{1}\dd \sigma  \varphi_\sigma(\theta-\theta')\frac{e^{- \sigma^2\tilde{\varepsilon}_\sigma(\theta')}-1}{\sm\sigma^2}\, .
\ee
Above, the classical breather-breather scattering shift is \be
\label{eq:classical_kernel}
\varphi_{\sigma, \sigma'}(\theta)=\frac{16}{cg^2}\log \left(\frac{[\cosh(\theta)-\cos((\sigma+\sigma')\pi/2)][\cosh(\theta)+\cos((\sigma-\sigma')\pi/2)]}{[\cosh(\theta)-\cos((\sigma-\sigma')\pi/2)][\cosh(\theta)+\cos((\sigma+\sigma')\pi/2)]}\right)\, .
\ee
The remaining scattering shifts can be recovered as $\lim_{\sigma'\to 1}\varphi_{\sigma,\sigma'}(\theta)=2\varphi_\sigma(\theta)$ and $\lim_{\sigma\to 1}\varphi_{\sigma}(\theta)=2\varphi(\theta)$, where $\varphi_\sigma(\theta)$ is the breather-kink scattering shift. Consistently, also dressing equations should be conveniently redefined to remove the spurious singularities. To this end, we define a new dressing operation using bold labels $\{\tau_K,\tau_{\bar{K}},\tau_\sigma\}\to \{\tau_K^{\textbf{\cdr}},\tau_{\bar{K}}^{\textbf{\cdr}},\tau_\sigma^{\textbf{\cdr}}\}$ such that
\be\label{eq_dressing_breather}
\sigma\tau^{\textbf{\cdr}}_{ \sigma}(\theta)=\frac{\tau_{\sigma}(\theta)}{\sigma}-\frac{1}{\sigma}\int \frac{\dd\theta'}{2\pi}\varphi_{ \sigma}(\theta-\theta')[\tilde{\vartheta}_K(\theta')\tau^{\textbf{\cdr}}_K(\theta')+\tilde{\vartheta}_{\bar{K}}(\theta')\tau^{\textbf{\cdr}}_{\bar{K}}(\theta')]
-\frac{1}{\sigma}\int_{0}^1 \frac{\dd\sigma'}{\sm} \int \frac{d\theta'}{2\pi} \varphi_{\sigma,\sigma'}(\theta-\theta')\tilde{\vartheta}_{\sigma'}(\theta')\tau^{\textbf{\cdr}}_{\sigma'}(\theta')\, ,
\ee
\be
\tau^{\textbf{\cdr}}_K(\theta)=\tau_K(\theta)- \int \frac{\dd\theta'}{2\pi} \varphi(\theta-\theta')[\tilde{\vartheta}_K(\theta')\tau^{\textbf{\cdr}}_K(\theta')+\tilde{\vartheta}_{\bar{K}}(\theta')\tau^{\textbf{\cdr}}_{\bar{K}}(\theta')]
-\int_{0}^1 \frac{\dd\sigma}{\sm} \int \frac{\dd\theta'}{2\pi} \varphi_{\sigma}(\theta-\theta')\tilde{\vartheta}_{\sigma'}(\theta)\tau^{\textbf{\cdr}}_{\sigma}(\theta')\, .
\ee

Passing from the standard dressing and the new parametrization, the following identities hold \cite{koch2023exact}: $\tau_K^{\textbf{\cdr}}(\theta)=\tau_K^{\cdr}(\theta)$, $\tau_{\bar{K}}^{\textbf{\cdr}}(\theta)=\tau_{\bar{K}}^{\cdr}(\theta)$, and $\tau_\sigma^{\textbf{\cdr}}(\theta)=\sigma^2\tau_\sigma^{\cdr}(\theta)$.

\bigskip
\bigskip

\textbf{The Ballistic Fluctuation Theory formulas.---} The general formulas of the Ballistic Fluctuation Theory are readily extended to the classical case. The only caveat when summing over the breathers is integrating with the proper measure $\int_0^1 \dd \sigma \sm (...)$ where the $\sm$ parameter explicitly appears. Of course, the standard filling function, effective energies and and dressing should be used in the general formulas of Section \ref{sec_BFT}: one can explicitly check the singularities for $\sigma\to 0$ are canceled out in the final expressions. We provide a Mathematica Notebook on Zenodo \cite{Zenodo} with the numerical solution of the thermodynamics, of the flow equations and as well as the computation of $c_2$ and $c_4$ from the explicit formulas.
The analysis of the low-temperature limit of the flow equation follows the same steps as the reflectionless point, resulting in the same final expression for the FCS \eqref{eq_FlowT}, provided the classical TBA and dressing are used. In the classical case, the breathers' spectrum is gapless and thus dressing effects due to breathers cannot be neglected even at low temperatures.

\section{Numerical methods}
\label{sec_numerics}

In this section, we present a short overview of the numerical methods we used.
\bigskip

\textbf{Solution of the TBA equations. ---} The TBA equations characterizing thermal ensembles, as well as the flow equations for the generating function and the expression for the cumulants can be solved by standard numerical routines once the integral equations have been properly discretized. In the quantum case, approximating integrals by the midpoint rule is sufficient, albeit more efficient discretizations can be envisaged. The classical case requires extra care, due to the singular nature of the scattering kernels: a stable discretization strategy is discussed in Ref. \cite{koch2023exact}, the interested reader can refer to that. A commented Mathematica notebook with working example is provided on Zenodo \cite{Zenodo}.
\bigskip

\textbf{The transfer matrix approach. ---}
An efficient method to compute equal-time expectation values and correlation function on classical one dimensional systems at equilibrium is provided by the Transfer Matrix method \cite{Scalapino1972,Castin2000}.
The key observation is regarding the classical partition function as a fictitious quantum mechanical system.
For example, let us consider the problem of computing  the expectation value of two possibly different phase-dependent observables placed at different positions $\langle \mathcal{O}_1(\phi(0))\mathcal{O}_2(\phi(x))\rangle$, assuming a system of length $L$ with periodic boundary conditions.
The classical thermal expectation value is
\begin{multline}
\langle \mathcal{O}_1(\phi(0)) \mathcal{O}_2(\phi(x))\rangle=\frac{\int \mathcal{D}\phi\,  \mathcal{O}_1(\phi(0)) \mathcal{O}_2(\phi(x)) e^{-\beta \int_{-L/2}^{L/2}\dd y \frac{1}{2 g^2}(\partial_y\phi)^2+\frac{m^2c^2}{g^2}(1-\cos(\phi))}}{\int \mathcal{D} \phi e^{-\beta \int_{-L/2}^{L/2}\dd y \frac{1}{2 g^2}(\partial_y\phi)^2+\frac{m^2c^2}{g^2}(1-\cos(\phi))}}=\\
=\frac{\text{Tr}[ e^{-\frac{L}{2} \hat{H}_\text{eff}}\mathcal{O}_1(\phi) e^{-x \hat{H}_\text{eff}} \mathcal{O}_2(\phi)e^{-(\frac{L}{2}-x) \hat{H}_\text{eff}}]}{\text{Tr}[ e^{-L \hat{H}_\text{eff}}]}\, .
\end{multline}
The last equality is obtained by regarding the path integral as the propagator in imaginary time of an effective quantum problem for a particle with ``position" $\phi$ and with the effective Hamiltonian
\be\label{eq_Heff}
\hat{H}_\text{eff}=-\frac{g^2}{2\beta}\partial^2_\phi+\frac{m^2c^2}{g^2}(1-\cos(\phi))\, .
\ee

In the thermodynamic limit $L\to \infty$, the trace is projected over the ground state $|0\rangle$ of the effective Hamiltonian and one obtains
\be\label{eq_corrq}
\langle \mathcal{O}_1(\phi(0)) \mathcal{O}_2(\phi(x))\rangle
=\frac{\langle 0|\mathcal{O}_1(\phi) e^{-x \hat{H}_\text{eff}} \mathcal{O}_2(\phi)|0\rangle}{e^{-x E_{GS}}}\, .
\ee
With $E_{GS}$ the ground state energy of the fictitious quantum problem. The quantum Hamiltonian \eqref{eq_Heff} is then discretized and numerically diagonalized, giving access to Eq. \eqref{eq_corrq}.
Some technical complications in the discretization arise as a consequence of the non compact nature of the target space $\phi$, which in principle may take values on the whole real axis. 
To this end, we proceed as follows: we enforce a finite domain $\phi\in[-\pi n_c,\pi n_c]$ with periodic boundary conditions and $n_c$ sufficiently large, then the values of $\phi$ are discretized on an uniform grid and the second derivative in the effective Hamiltonian is approximated with a finite-difference increment. We will further comment on the choice of $n_c$ later on.
Then, we take advantage of the translational invariance $\phi\to \phi+2\pi$ of the original problem by restricting $\mathcal{O}_1(\phi)$ to have support only in the fundamental Brillouin zone. Hence
\be\label{eq_cell}
\mathcal{O}_1(\phi)\to\tilde{\mathcal{O}}_1(\phi)=\begin{cases} \mathcal{O}_1(\phi)& \text{if }-\pi<\phi<\pi\\ 0 &\text{otherwise}\end{cases}\, .
\ee
In contrast, $\mathcal{O}_2(\phi)$ is left unaffected: one can finally compute cumulants and full counting statistics. Cumulants are obtained starting with the connected correlation functions and expanding in products of powers of the field $\langle [\phi(0)-\phi(x)]^n\rangle=\sum_{j=0}^n \binom{n}{j} (-1)^{n-j}\langle\phi^j(0)\phi^{n-j}(x)\rangle$, and by setting $\mathcal{O}_1(\phi)\to \phi^j $ and $\mathcal{O}_2(\phi)\to \phi^{n-j} $. Notice that each term in the sum $\langle \phi^j(0)\phi^{n-j}(x)\rangle$ is not invariant for $\phi\to \phi+2\pi$ and it would thus be affected by a different choice of the fundamental cell \eqref{eq_cell}, but this is not the case once each term has been resummed to give $\langle [\phi(0)-\phi(x)]^n\rangle$.
Similarly, one can recover the full counting statistics by its very definition. By calling $P_x(\delta \phi)$ the probability that the field jumps of a factor $\delta \phi$ after a distance $x$, one can write
\be
P_x(\delta \phi)=\int_{-\pi}^{\pi} \frac{\dd\phi}{2\pi}\, \langle \delta(\phi(0)-\phi) \delta(\phi(x)-(\phi+\delta \phi))\rangle\, .
\ee
Where we took advantage of the $2\pi$ periodicity. For each (discretized) value of $\phi$ the integrand can be then computed according to Eq. \eqref{eq_corrq} and the probability is finally recovered by summing over the different terms.
We wish finally to comment on the role of the cutoff $n_c$ in the discretization: for any fixed value of $n_c$, one necessarily has the bound $[\phi(0)-\phi(x)]^n< (2 n_c\pi)^n$, while we know the exact connected cumulant linearly grows upon increasing the distance between points. Thus, assuming odd cumulants are vanishing because of symmetry, one expects $[\phi(0)-\phi(x)]^{2n}\propto |x|^n$. In practice, one needs to adjust the cutoff $n_c$ based on the cumulant to be computed and the space separation: higher cumulants and larger distances require larger values of $n_c$ to attain convergence.
An alternative implementation taking advantage of the periodicity of the effective Hamiltonian \eqref{eq_Heff} may have been built using Bloch wave functions and restricting to diagonalize Hamiltonians in the units cells, but then summing over the Bloch wave vectors.

A commented Mathematica notebook on the Transfer Matrix approach with working examples is provided on Zenodo \cite{Zenodo}.
\bigskip

\begin{figure}[t!]
\centering
\includegraphics[width=0.95\textwidth]{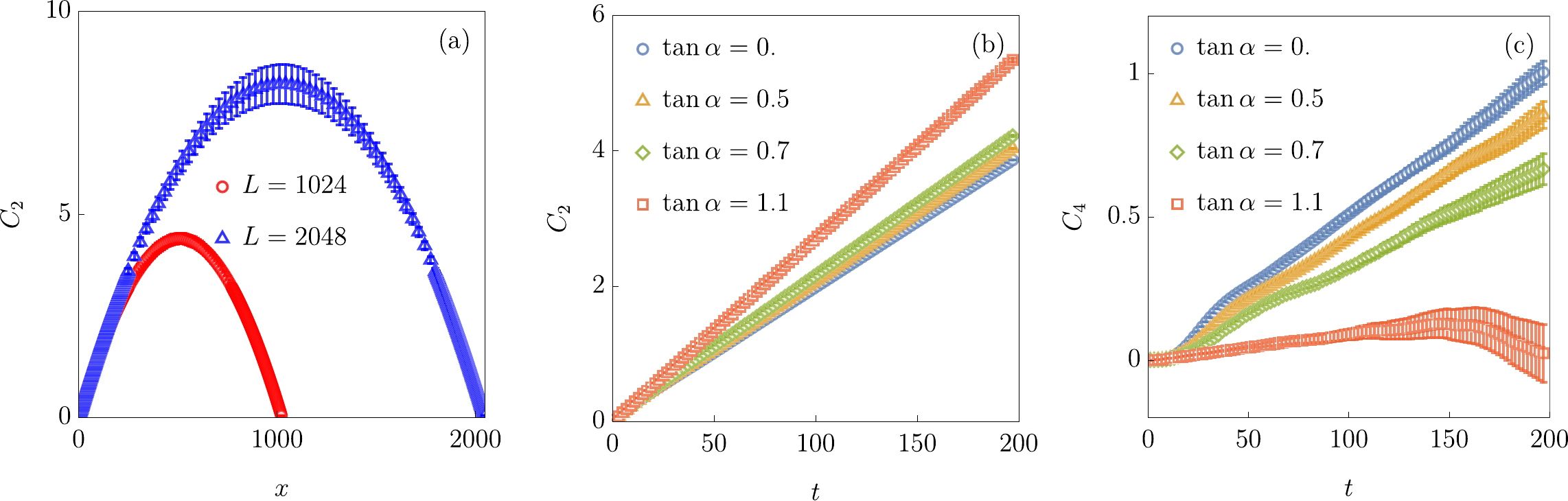}
	\caption{\textbf{Further details on Monte Carlo data.---} (a) To emphasize the role of finite volume and periodic boundary conditions, we show the second cumulant $C_2(x)=\frac{1}{(2\pi)^2}\langle \phi(0)\phi(x)\rangle_c$ at equal times for two different volume realizations. As an example, we choose $g=c=\beta=1$ and $m=0.25$. The second cumulant cannot grow forever and reaches a maximum peak in the center of the system: the linear growth predicted by BFT is realized at large distances compared with the microscopic correlation lengths, but much smaller than the system's size $L$. To reduce finite size effects within a fixed window $[0,\ell_\text{max}]$ with $\ell_\text{max}$ the maximum separation between the two points, we extract the scaling factors $c_n$ by taking three system sizes of $L=1024$, $L=2048$ and $L=4096$ respectively (the latter not shown in (a)) and extrapolate to infinite volume assuming corrections scale as $1/L$. The fourth cumulant at finite system (not shown) experiences stronger finite size corrections and larger uncertainty due to fluctuations.
 (b) \& (c) We show a typical example of the second and fourth cumulants for space-time separation, plotted as a function of time for different rays $\tan\alpha=x/(tc)$ and already extrapolated to infinite size. As an example, we choose $g=c=\beta=1$ and $m=0.25$. The second cumulant (b) shows a clear linear growth with no appreciable corrections, while the fourth cumulant $C_4$ (c) shows finite time corrections in the form of oscillations superimposed on the linear growth. $C_4$ is also more sensitive to finite-size corrections, as it is evident from the curve $\tanh \alpha=1.1$ that, after an initial linear growth, suddenly bends downward with increased uncertainty: there, the extrapolation to infinite size fails and one can trust the curve $\tanh \alpha=1.1$ up to $t\simeq 100$ at most. This issue could be solved by exploring even larger sizes, but the computational time needed for the Monte Carlo to reach convergence becomes prohibitively long.
	}
	\label{fig_appMC}
\end{figure}

\textbf{Monte Carlo simulations. ---}
Monte Carlo samples thermal distribution through a suitable random walk in the phase space \cite{Metropolis1953,Hasting1970}. These are standard techniques so we provide only a short overview.
The field is discretized on a uniform grid with lattice spacing $a$: eventually, the choice of the lattice spacing depends on the temperature and mass scale.
The classical Hamiltonian is discretized as
\be\label{eq_disH}
H[\Pi,\phi]=a\sum_j \frac{c_j^2 g_j^2}{2}\Pi_j+\frac{1}{2 g_j^2 a^2}(\phi_{j+1}-\phi_j)^2-\frac{c_j^2 m^2_j}{g_j^2}\cos(\phi_j)\, ,
\ee
with canonical Poisson brackets $\{\phi_j,\Pi_{j'}\}=\delta_{j,j'}/a$. Above, we promote the couplings to be (weakly) inhomogeneous functions to accommodate for the simulations of the inhomogeneous sine-Gordon stemming from the coupled quasicondensates, see Section \ref{sec_exp}.
On thermal ensemble, the Hamiltonian decouples in the phase and the momentum contributions, hence the two fields can be sampled independently. The distribution of $\Pi_j$ is generated by taking advantage that is Gaussian and independently distributed on the sites, while the phase thermal distribution is sampled by Monte Carlo methods. More specifically, one proposes local updates $\phi_j\to\phi'_j= \phi_j+\delta \phi_j$ where $\delta \phi_j$ is Gaussianly distributed: the move is accepted with probability $p=\exp(-\beta H[\phi'])/\exp(-\beta H[\phi])$, where $H[\phi]$ is the phase-dependent part of the discretized Hamiltonian \eqref{eq_disH}. The variance of the local updates is chosen in such a way the acceptance ratio is approximately $0.5$.

After sufficiently many steps the Monte Carlo converges and starts exploring the thermal distribution, then sampling begins. When unequal time correlations need to be computed, the initial field configurations drawn from the Monte Carlo are evolved according to the deterministic equation of motion derived from the SG Hamiltonian. We use the following discretization \cite{DeLuca2016}
\begin{multline}
\phi_j(t+\dd t)+\phi_j(t-\dd t)-2\phi_j(t)= c^2_j g^2_j\frac{1}{4}\left[\left(\frac{4}{g^2_j}+\frac{1}{g^2_{j+1}}-\frac{1}{g^2_{j-1}}\right)\phi_{j+1}(t)+\left(\frac{4}{g^2_j}+\frac{1}{g^2_{j-1}}-\frac{1}{g^2_{j+1}}\right)\phi_{j-1}(t)\right)+\\-2 c^2_j\phi_j(t) \frac{\dd t^2}{a^2}+ c^4_j m_j^2\sin \phi_j(t)\, .
\end{multline}
We experience this discretization to be stable by checking the energy conservation: corrections remain bounded with time and are decreased upon improving the discretization.
Further details on the discretization used in Fig. 3 and Fig. 3 are provided below.

\begin{enumerate}[i]
\item \emph{Monte Carlo details for Fig. 3}: with the choice $\beta=g=c=1$, we experience convergence can be attained for relatively large lattice spacing. In this figure, we choose $a=0.5$ and $\dd t=10^{-3}$, working with periodic boundary conditions. The major difficulty is due to the large system size needed to extract the scaling of cumulants. Indeed, by considering periodic boundary conditions, the cumulants cannot grow indefinitely. For example, in Fig. \ref{fig_appMC}(a) we show the second cumulant on a finite size, well approximated by an inverted parabola. The scaling behavior $\langle (\phi(x)-\phi(0))^2\rangle \propto |x|$ is recovered only at small distances compared to the overall volume $L$. On the other hand, large separations are needed to attain the scaling regime: this is particularly true for space-time separations outside of the causal light cone $x/t>c$, where an evident oscillatory behavior is superimposed on the linear growth, see Fig. \ref{fig_appMC}. As a compromise, we simulate very large system sizes with $2^{11}$, $2^{12}$ and $2^{13}$ points, extrapolating to infinite volume assuming a $1/L$ scaling. Lastly, the large-separation growth of the cumulants is fitted with a straight line and the slope is extracted.
Extracting the fourth cumulant requires very accurate data: for each chosen volume, we run $100$ independent Monte Carlo sampling collecting at least $1000$ uncorrelated samples for each of them, for a total of $10^5$ samples. The cumulants are independently computed for each Monte Carlo realization, then we choose as the most representative value the average over the $100$ independent realizations and the error bars are taken as the variance. Error bars are propagated to the extrapolated data by linear regression. The data of the so-obtained cumulant, similarly to those shown in Fig. \ref{fig_appMC}, are available on Zenodo \cite{Zenodo}.
\item \emph{Monte Carlo details for Fig. 4 :} Working with realistic experimental parameters (see Section \ref{sec_exp} for discussion) advocated for a smaller lattice spacing, which we chose $a=0.05\text{$\mu$m}$ upon having checked convergence. The inhomogeneous density-profile of the atomic cloud allows for wild phase fluctuations at the edges of the trap, making open boundary conditions the best choice. Due to these largely fluctuating regions, the finite-size problem experienced in the previous point is largely reduced and much smaller system sizes are needed to clearly observe the linear growth of the second cumulant within the experimentally-reachable sizes. In Figure \ref{fig_SM_exp} we provide further data.

\end{enumerate}

\section{Sine-Gordon from coupled condensates}
\label{sec_exp}

In this section, we shortly revisit the emergence of the sine-Gordon field theory as a the low-energy description of coupled-tunnel quasicondensates \cite{Gritsev2007,Schweigler2017}.
We consider two identical one-dimensional atomic clouds in a longitudinal trap $V(x)$. In the absence of further coupling, these two gases are well described by the Hamiltonian
\be\label{eq_LiebLiniger}
 H_{j=\{1,2\}}=\int \dd x\left\{\frac{\hbar^2}{2m}\partial_x\psi^\dagger_j\partial_x\psi_j-\frac{\hbar^2}{m a_{1D}} \psi^\dagger_j\psi^\dagger_j\psi_j\psi_j +V(x)\psi^\dagger_j\psi_j\right\}\, ,
\ee
where $\psi_j$ are canonical bosonic annihilation fields. Here, we focus on repulsive interactions $a_{1D}<0$.
The effective-one dimensional interaction is determined by the three dimensional scattering length $a_{3D}$ and it is largely renormalized by the transverse trap frequency according to \cite{Olshanii1998}
\be
a_{1D}=-\frac{a_\perp}{2}\left(\frac{a_\perp}{a_{3D}}-\mathcal{C}\right)\, ,
\ee
with $\mathcal{C}=1.4603...$, and $a_\perp$ is the perpendicular oscillator length defined as $a_\perp=\sqrt{m/(\hbar \omega_\perp)}$, with $\omega_\perp$ the transverse trap frequency. For small interactions $a_{3D}/a_\perp\ll 1$, the constant $\mathcal{C}$ can be neglected: in the typical configuration of Vienna's experiment \cite{schweigler2019}, ${}^{87}\text{Rb}$ atoms have $a_{3D}=5.2368\times 10^{-3}\text{$\mu$m}$ and the transverse trap frequency is $\omega_\perp\approx 2\pi\times 1.4\text{kHz}$, leading to $a_{3D}/a_\perp\simeq 0.018$.

Atom chips allow for a great tunability of the longitudinal trap $V(x)$ with the possibility of engineer box-like potentials. Therefore, we choose $V(x)=(x/x_0)^6$ as a sixth-order polynomial. 
At low temperature, the atomic profile is well-described by the Thomas-Fermi approximation
\be
n(x)=\sqrt{\frac{m |a_{1D}|}{2\hbar^2}(\mu-V(x))}\, ,
\ee
with $\mu$ the chemical potential. We tune the trap length scale $x_0$ and the chemical potential in such a way $n(x)$ box-shaped with rounded corners and bulk density $40\text{atm/$\mu$m}$, as depicted in Fig. 4(b).

A barrier of adjustable height induces a weak tunneling between the two tubes, leading to the final microscopic dynamics
\be
H=H_1+H_2-t_\perp\int \dd x\, \{ \psi^\dagger_1\psi_2+\psi^\dagger_2\psi_1\}\, .
\ee
The above Hamiltonian is complicated, but it greatly simplifies when focusing on low energies: sine-Gordon lives in this sector, describing the dynamics of the phase-difference between the two condensates. One proceeds as follows. First, one starts with the limit of decoupled condensates $t_\perp=0$: the ground state and the low temperature sector is well-described within bosonization. Hence, one introduces two conjugate fields describing the fluctuations of density and phase of the two condensates
\be\label{eq_densph}
\psi_j\simeq \sqrt{n(x)+\Pi_j(x)} \,e^{i\phi_j(x)}\, ,
\ee
leading to bosonized Hamiltonians. Within the assumption of a smoothly varying potential $V(x)$, its effect enters only as a modulation of the atom density profile, which in turn is translated into a spatial inhomogeneity of the light velocity $c\to c(x)$ and Luttinger parameter $K\to K(x)$ \cite{tajik2022}
\be
H_j=\int \dd x\, \frac{\hbar c(x)}{2}\left(\frac{\pi}{K(x)}\Pi_j^2(x)+\frac{K(x)}{\pi}(\partial_x\phi_j)^2\right)\, .
\ee
In the one-dimensional interacting gas \eqref{eq_LiebLiniger}, the light velocity and Luttinger parameter can be exactly determined from Bethe Ansatz, but in the regime of weak interactions and large atoms number the exact result becomes equivalent to approximating operators with classical fields and straightforwardly replace the density-phase approximation Eq. \eqref{eq_densph} in the microscopic Hamiltonian \eqref{eq_LiebLiniger}, while retaining only the slowest modes. In this case, one simply obtains
\be
c(x)\simeq \frac{\hbar}{m}\sqrt{\frac{2 n(x)}{ |a_\text{1D}|}}
\, ,\hspace{2pc} K(x)\simeq \pi\sqrt{\frac{n(x)|a_\text{1D}|}{2}}\, .
\ee
We will make use of this approximation. One can now reintroduce the tunneling term within bosonization $\int \dd x\, \{\psi^\dagger_1\psi_2+\psi^\dagger_2\psi_1\}\to \int \dd x\,  2 \gamma(x) n(x)\cos(\phi_1(x)-\phi_2(x))$. The coefficient $\gamma(x)$ is added to take care of non-trivial short range renormalizations in the product of the two fields beyond a naive replacement with Eq. \eqref{eq_densph}. However, in the weakly interacting regime these effects can be neglected and one can simply set $\gamma(x)=1$.
The last passage requires changing coordinates in the global Hamiltonian $H$, by considering the symmetric $\Phi(x)=\phi_1(x)+\phi_2(x)$ and antisymmetric $\phi(x)=\phi_1(x)-\phi_2(x)$ degrees of freedom. The two sectors decouple in the bosonized Hamiltonian and, while the symmetric sector remains a gapless Luttinger liquid, in the antisymmetric sector the sine-Gordon Hamiltonian emerges
\be\label{eq_antisymSG}
H_\text{anti-symmetric}=\int \dd x\, \left\{ \frac{\hbar c(x)}{2}\left(\frac{2\pi}{K(x)} \Pi^2(x)+\frac{K(x)}{2\pi}(\partial_x \phi)^2\right)-2 t_\perp n(x)\cos\phi\right\}\, .
\ee
The sine-Gordon couplings in the standard notation are readily obtained by comparison. Interestingly, the Luttinger parameter is directly associated with the renormalized interaction $\xi$, more specifically one has $\xi^{-1}=4K-1$. In turn, $\xi$ fixes the number of different breathers species present in the spectrum and is the knob tuning the quantumness of the theory. For the typical experimental parameters reported above and in the main text, by choosing the bulk density of $40 \text{atm}/\text{$\mu$m}$, one obtains $K\simeq 55$, thus $\xi^{-1}\simeq 220$. Hence, the breather spectrum can be well approximated by a continuum and one expects the semiclassical approximation to be a good description of the model (see also Section \ref{sec_SG_thermo}).
Therefore, to numerically simulate a realistic experimental setup we perform Monte Carlo sampling of the classical sine-Gordon Hamiltonian, the so-extracted phase profiles mimic the outcome of projective measurements.

\bigskip

\textbf{The phase measurement process. ---} In the experimental setup, the phase is extracted from matter-wave interferometry measurements \cite{Schumm2005}. The external three dimensional trap holding in place the two elongated condensates is suddenly switched off and the gas is let free to expand: due to the initial tight confinement in the transverse direction, the momentum of the particles after trap release is very large. Therefore, one can work in the approximation that the expansion happens only within the transverse direction, while longitudinal evolution remains frozen: this ultimately allows for a reconstruction of the position-dependent phase profile. The three-dimensional density profile is thus described by \cite{Kuhnert2013}
\begin{multline}\label{eq_3dD}
n_{3D}(x, \vec{r},t)=|f(\vec{r},t)|^2[\psi^\dagger_1(x)\psi_1(x)+\psi^\dagger_2(x)\psi_2(x)+\psi^\dagger_1(x)\psi_2(x)e^{-i \vec{d}\cdot\vec{r} m/(\hbar t)}+\psi^\dagger_2(x)\psi_1(x)e^{i \vec{d}\cdot\vec{r} m/(\hbar t)}]\simeq\\ |f(\vec{r},t)|^2 2n(x)\left[1+\cos(\phi(x)-\vec{d}\cdot\vec{r} m/(\hbar t))\right]\, ,
\end{multline}
where in the last line one uses the density-phase approximation \eqref{eq_densph}. Above, $x$ refers to the longitudinal spatial coordinate while $\vec{r}$ is the radial direction (perpendicular to the direction of the tubes), $\vec{d}$ is the relative distance of the two tubes. Finally, $f$ is a Gaussian envelope coming from the expansion in plane waves of the transverse oscillator ground state function, the specific form is not needed for our purposes (see however Ref. \cite{Nieuwkerk2018}, also for corrections beyond the weakly-interacting regime).

The three dimensional density is then projected in the plane containing the condensate by integrating in the orthogonal direction. Then the resulting two-dimensional pattern is sliced along the the longitudinal direction $x$ and the oscillating pattern in the remaining orthogonal direction is fitted with the oscillating function, extracting the phase shift $\phi(x)$. This procedure is equivalent to measure independently $n(x)\cos(\phi(x))$ and $n(x)\sin(\phi(x))$, as it is clearly seen by expanding the cosine \eqref{eq_3dD} with the help of trigonometric identities $n(x)\cos(\phi(x)-\vec{d}\cdot\vec{r} m/(\hbar t))=n(x)\cos(\phi(x))\cos(\vec{d}\cdot\vec{r} m/(\hbar t))+n(x)\sin(\phi(x))\sin(\vec{d}\cdot\vec{r} m/(\hbar t))$

In principle, combining these two quantities the phase profile could be exactly recovered: nonetheless, experimental limitations in the form of a finite resolution of the camera are not able to resolve arbitrary small distances, thus inducing coarse graining.
Pixels in the longitudinal directions are equispaced on a grid $\{x_i\}_{i=1}^N$, with approximately $2\text{$\mu$m}$ spacing. The center of each pixel collects signals from its surrounding: in a good approximation, this imperfection can be mimicked by a convolution with a Gaussian with standard deviation $\sigma$ \cite{Kuhnert2013}
\be
[n(x_i)\sin(\phi(x_i))]_\sigma\equiv \int \dd y \, \frac{e^{-\frac{1}{2\sigma^2}(x_i-y)^2}}{\sigma\sqrt{2\pi}} n(y)\sin(\phi(y))\, ,\hspace{2pc} [n(x_i)\cos(\phi(x_i))]_\sigma\equiv \int \dd y \, \frac{e^{-\frac{1}{2\sigma^2}(x_i-y)^2}}{\sigma\sqrt{2\pi}} n(y)\cos(\phi(y))\, .
\ee

\begin{figure}[t!]
\centering
\includegraphics[width=0.95\textwidth]{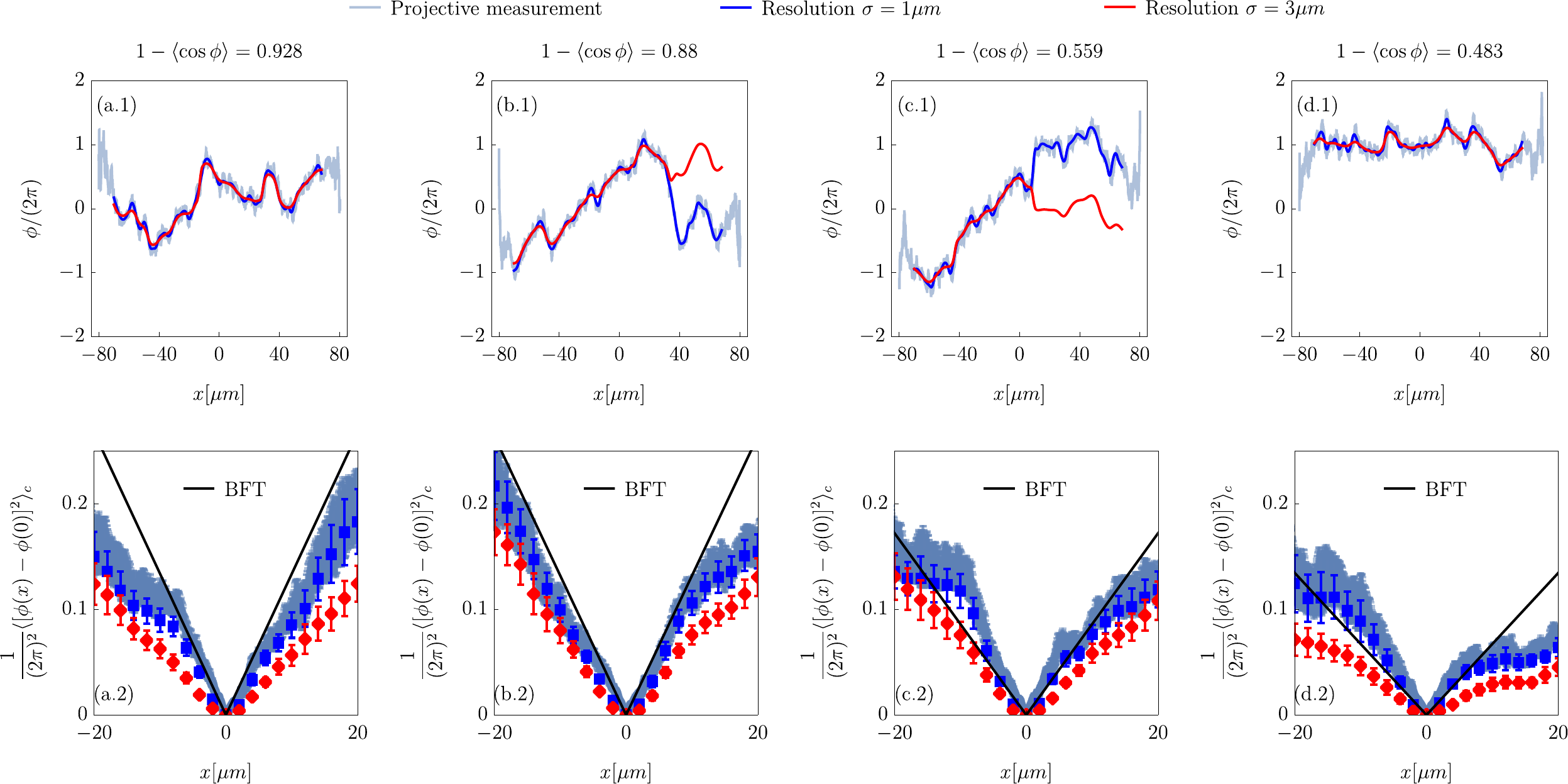}
	\caption{\textbf{Phase measurement and second cumulant for different barriers.---} In this figure, we provide further examples of the phase-measurement process for different tunneling barriers of the two quasicondensates, resulting in different masses of the field theory. We use the same parameters as Fig. 4 of the main text (hence $T=60\text{nK}$, bulk density $40\text{atm/$\mu$m}$), but tune the barrier $t_\perp$ \eqref{eq_antisymSG} to different values. More specifically, we choose $t_\perp=\{0.00351576 , 0.00751331 , 0.0385028 , 0.05\}$ in (a), (b), (c) and (d) respectively, which correspond to the expectation values of the vertex operator (in the bulk) shown in each column. Upper row: example of phase extracted from a single projective measurement with two different resolutions $\sigma$. Bottom row: second cumulant from the trap center obtained by averaging over $100$ independent samples.
	}
	\label{fig_SM_exp}
\end{figure}

From the coarse grain output, the most representative phase field is then extracted. Naively, one would just consider, for example, the arc cosine of the first term upon dividing by the density. However, due to the Gaussian convolution, it does not hold any longer $[n(x_i)\sin(\phi(x_i))]^2_\sigma+[n(x_i)\sin(\phi(x_i))]^2_\sigma\ne n^2(x_i)$. Hence, one needs first to correctly renormalize the coarse grain trigonometric functions, thus defining
\be\label{eq_phasesigma}
[\phi(x_i)]_\sigma= \arccos\left[\frac{[n(x_i)\cos(\phi(x_i))]_\sigma}{\sqrt{[n(x_i)\sin(\phi(x_i))]^2_\sigma+[n(x_i)\cos(\phi(x_i))]^2_\sigma}}\right]\, .
\ee

This identification fixes the phase modulus a sign and a $2\pi$ uncertainty. The sign is resolved by looking at the sign of $[n(x_i)\sin(\phi(x_i))]_\sigma$, leaving a $2\pi-$ambiguity.
The latter is resolved as follows \cite{Kuhnert2013}: one arbitrary assigns a phase comprised between $[-\pi,\pi]$ to a given coordinate $x_{\bar{i}}$, then one scans the system moving on the left and right imposing, as much as possible, the continuity of the phase field. Namely, the sector is chosen in such a way $|[\phi(x_i)]_\sigma-[\phi(x_{i+1})]_\sigma|$ is minimized as much as possible, compatibly with Eq. \eqref{eq_phasesigma}. The reasoning behind this requirement is that large phase jumps will pack large energy, due to the phase-derivative term in sine-Gordon and are thus unlikely in the low-temperature regime.
Nonetheless, kinks are exactly large phase jumps: if the resolution $\sigma$ is not sufficiently narrow and becomes comparable with the kink size, kinks may be undetected.
The phase-profile of a kink at rest is readily obtained by solving the equation of motion as $\phi_K(x)=4\arctan(e^{-mc x})+2\pi$: moving kinks have a reduced length due to Lorentz contraction, but we can use the rest profile as an estimation.
The bare mass scale sets the kink's extent: one should then find the right compromise between having kinks not too big (in such a way they can be placed in the finite volume realized in the experiment) and not too small, otherwise they will be undetected due to the finite resolution of the experiment. In addition, the temperature cannot be too small compared with the kink's mass scale in order to have an appreciable population. 
These considerations led us to the parameter choice shown in Fig. 4, albeit we explore also other temperature regimes (not shown). 
For completeness, for a fixed choice of the background potential, bulk density $n(x)=40\text{atm/$\mu$m}$ and temperature $60\text{nK}$, further data for other choices of the tunneling are shown in Fig. \ref{fig_SM_exp}.

\end{document}